%% file: DESY-12-045.tex
\documentclass[zpreprint,zbstnp]{zeus_paper}

\usepackage[english]{babel}

\input zeus_def.tex

\input definitions.tex

\def\figdir{./}

\begin{document}
\prepnum{{DESY--12--045}}

\title{Inclusive-jet photoproduction at HERA\\
 and determination of {\boldmath $\as$}}

\author{ZEUS Collaboration}
\date{May 2012}

\abstract{
Inclusive-jet cross sections have been measured in the reaction $\eprn$ for
photon virtuality $\q2<1$~\g2\ and $\gp$ centre-of-mass energies in
the region \wrn\ with the ZEUS detector at HERA using an integrated
luminosity of $300$~\pb1. Jets were identified using the $\kt$, 
anti-$\kt$ or SIScone jet algorithms in the
laboratory frame. Single-differential cross sections are presented as
functions of the jet transverse energy, $\etjet$, and pseudorapidity,
$\etajet$, for jets with $\etjet>17$ GeV and $\etar$.
In addition, measurements of double-differential inclusive-jet cross
sections are presented as functions of $\etjet$ in different regions
of $\etajet$.
Next-to-leading-order QCD calculations give a good description of the
measurements, except for jets with low $\etjet$ and high $\etajet$. 
The influence of non-perturbative effects not related to hadronisation
was studied.
Measurements of the ratios of cross
sections using different jet algorithms are also presented; 
the measured ratios are well described by calculations including up to
$\oass$ terms.
Values of $\asz$ were extracted from the measurements and the
energy-scale dependence of the coupling was determined. The value of
$\asz$ extracted from the measurements based on the $\kt$ jet
algorithm is
$\asz= 0.1206^{+0.0023}_{-0.0022}{\rm (exp.)}^{+0.0042}_{-0.0035}{\rm (th.)}$;
the results from the anti-$\kt$ and SIScone algorithms are compatible
with this value and have a similar precision.
}

\makezeustitle

\pagenumbering{Roman}

\begin{center}
{                      \Large  The ZEUS Collaboration              }
\end{center}

{\small

        {\raggedright
H.~Abramowicz$^{45, ai}$, 
I.~Abt$^{35}$, 
L.~Adamczyk$^{13}$, 
M.~Adamus$^{54}$, 
R.~Aggarwal$^{7, c}$, 
S.~Antonelli$^{4}$, 
P.~Antonioli$^{3}$, 
A.~Antonov$^{33}$, 
M.~Arneodo$^{50}$, 
V.~Aushev$^{26, 27, aa}$
% ...
        }

\newpage

\makebox[3em]{$^{1}$}
\begin{minipage}[t]{14cm}
{\it Argonne National Laboratory, Argonne, Illinois 60439-4815, USA}~$^{A}$

\end{minipage}\\
\makebox[3em]{$^{2}$}
\begin{minipage}[t]{14cm}
{\it Andrews University, Berrien Springs, Michigan 49104-0380, USA}

\end{minipage}\\
\makebox[3em]{$^{3}$}
\begin{minipage}[t]{14cm}
{\it INFN Bologna, Bologna, Italy}~$^{B}$

\end{minipage}\\
\makebox[3em]{$^{4}$}
\begin{minipage}[t]{14cm}
{\it University and INFN Bologna, Bologna, Italy}~$^{B}$

\end{minipage}\\
\makebox[3em]{$^{5}$}
\begin{minipage}[t]{14cm}
{\it Physikalisches Institut der Universit\"at Bonn,
Bonn, Germany}~$^{C}$

\end{minipage}\\
\makebox[3em]{$^{6}$}
\begin{minipage}[t]{14cm}
{\it H.H.~Wills Physics Laboratory, University of Bristol,
Bristol, United Kingdom}~$^{D}$

\end{minipage}\\
\makebox[3em]{$^{7}$}
\begin{minipage}[t]{14cm}
{\it Panjab University, Department of Physics, Chandigarh, India}

\end{minipage}\\
\makebox[3em]{$^{8}$}
\begin{minipage}[t]{14cm}
{\it Calabria University,
Physics Department and INFN, Cosenza, Italy}~$^{B}$

\end{minipage}\\
\makebox[3em]{$^{9}$}
\begin{minipage}[t]{14cm}
{\it Institute for Universe and Elementary Particles, Chonnam National University,\\
Kwangju, South Korea}

\end{minipage}\\
\makebox[3em]{$^{10}$}
\begin{minipage}[t]{14cm}
{\it Jabatan Fizik, Universiti Malaya, 50603 Kuala Lumpur, Malaysia}~$^{E}$

\end{minipage}\\
\makebox[3em]{$^{11}$}
\begin{minipage}[t]{14cm}
{\it Nevis Laboratories, Columbia University, Irvington on Hudson,
New York 10027, USA}~$^{F}$

\end{minipage}\\
\makebox[3em]{$^{12}$}
\begin{minipage}[t]{14cm}
{\it The Henryk Niewodniczanski Institute of Nuclear Physics, Polish Academy of \\
Sciences, Krakow, Poland}~$^{G}$

\end{minipage}\\
\makebox[3em]{$^{13}$}
\begin{minipage}[t]{14cm}
{\it AGH-University of Science and Technology, Faculty of Physics and Applied Computer
Science, Krakow, Poland}~$^{H}$

\end{minipage}\\
\makebox[3em]{$^{14}$}
\begin{minipage}[t]{14cm}
{\it Department of Physics, Jagellonian University, Cracow, Poland}

\end{minipage}\\
\makebox[3em]{$^{15}$}
\begin{minipage}[t]{14cm}
{\it Deutsches Elektronen-Synchrotron DESY, Hamburg, Germany}

\end{minipage}\\
\makebox[3em]{$^{16}$}
\begin{minipage}[t]{14cm}
{\it Deutsches Elektronen-Synchrotron DESY, Zeuthen, Germany}

\end{minipage}\\
\makebox[3em]{$^{17}$}
\begin{minipage}[t]{14cm}
{\it INFN Florence, Florence, Italy}~$^{B}$

\end{minipage}\\
\makebox[3em]{$^{18}$}
\begin{minipage}[t]{14cm}
{\it University and INFN Florence, Florence, Italy}~$^{B}$

\end{minipage}\\
\makebox[3em]{$^{19}$}
\begin{minipage}[t]{14cm}
{\it Fakult\"at f\"ur Physik der Universit\"at Freiburg i.Br.,
Freiburg i.Br., Germany}

\end{minipage}\\
\makebox[3em]{$^{20}$}
\begin{minipage}[t]{14cm}
{\it School of Physics and Astronomy, University of Glasgow,
Glasgow, United Kingdom}~$^{D}$

\end{minipage}\\
\makebox[3em]{$^{21}$}
\begin{minipage}[t]{14cm}
{\it Department of Engineering in Management and Finance, Univ. of
the Aegean, Chios, Greece}

\end{minipage}\\
\makebox[3em]{$^{22}$}
\begin{minipage}[t]{14cm}
{\it Hamburg University, Institute of Experimental Physics, Hamburg,
Germany}~$^{I}$

\end{minipage}\\
\makebox[3em]{$^{23}$}
\begin{minipage}[t]{14cm}
{\it Imperial College London, High Energy Nuclear Physics Group,
London, United Kingdom}~$^{D}$

\end{minipage}\\
\makebox[3em]{$^{24}$}
\begin{minipage}[t]{14cm}
{\it Institute of Particle and Nuclear Studies, KEK,
Tsukuba, Japan}~$^{J}$

\end{minipage}\\
\makebox[3em]{$^{25}$}
\begin{minipage}[t]{14cm}
{\it Institute of Physics and Technology of Ministry of Education and
Science of Kazakhstan, Almaty, Kazakhstan}

\end{minipage}\\
\makebox[3em]{$^{26}$}
\begin{minipage}[t]{14cm}
{\it Institute for Nuclear Research, National Academy of Sciences, Kyiv, Ukraine}

\end{minipage}\\
\makebox[3em]{$^{27}$}
\begin{minipage}[t]{14cm}
{\it Department of Nuclear Physics, National Taras Shevchenko University of Kyiv, Kyiv, Ukraine}

\end{minipage}\\
\makebox[3em]{$^{28}$}
\begin{minipage}[t]{14cm}
{\it Kyungpook National University, Center for High Energy Physics, Daegu,
South Korea}~$^{K}$

\end{minipage}\\
\makebox[3em]{$^{29}$}
\begin{minipage}[t]{14cm}
{\it Institut de Physique Nucl\'{e}aire, Universit\'{e} Catholique de Louvain, Louvain-la-Neuve,\\
Belgium}~$^{L}$

\end{minipage}\\
\makebox[3em]{$^{30}$}
\begin{minipage}[t]{14cm}
{\it Departamento de F\'{\i}sica Te\'orica, Universidad Aut\'onoma
de Madrid, Madrid, Spain}~$^{M}$

\end{minipage}\\
\makebox[3em]{$^{31}$}
\begin{minipage}[t]{14cm}
{\it Department of Physics, McGill University,
Montr\'eal, Qu\'ebec, Canada H3A 2T8}~$^{N}$

\end{minipage}\\
\makebox[3em]{$^{32}$}
\begin{minipage}[t]{14cm}
{\it Meiji Gakuin University, Faculty of General Education,
Yokohama, Japan}~$^{J}$

\end{minipage}\\
\makebox[3em]{$^{33}$}
\begin{minipage}[t]{14cm}
{\it Moscow Engineering Physics Institute, Moscow, Russia}~$^{O}$

\end{minipage}\\
\makebox[3em]{$^{34}$}
\begin{minipage}[t]{14cm}
{\it Lomonosov Moscow State University, Skobeltsyn Institute of Nuclear Physics,
Moscow, Russia}~$^{P}$

\end{minipage}\\
\makebox[3em]{$^{35}$}
\begin{minipage}[t]{14cm}
{\it Max-Planck-Institut f\"ur Physik, M\"unchen, Germany}

\end{minipage}\\
\makebox[3em]{$^{36}$}
\begin{minipage}[t]{14cm}
{\it NIKHEF and University of Amsterdam, Amsterdam, Netherlands}~$^{Q}$

\end{minipage}\\
\makebox[3em]{$^{37}$}
\begin{minipage}[t]{14cm}
{\it Physics Department, Ohio State University,
Columbus, Ohio 43210, USA}~$^{A}$

\end{minipage}\\
\makebox[3em]{$^{38}$}
\begin{minipage}[t]{14cm}
{\it Department of Physics, University of Oxford,
Oxford, United Kingdom}~$^{D}$

\end{minipage}\\
\makebox[3em]{$^{39}$}
\begin{minipage}[t]{14cm}
{\it INFN Padova, Padova, Italy}~$^{B}$

\end{minipage}\\
\makebox[3em]{$^{40}$}
\begin{minipage}[t]{14cm}
{\it Dipartimento di Fisica dell' Universit\`a and INFN,
Padova, Italy}~$^{B}$

\end{minipage}\\
\makebox[3em]{$^{41}$}
\begin{minipage}[t]{14cm}
{\it Department of Physics, Pennsylvania State University, University Park,\\
Pennsylvania 16802, USA}~$^{F}$

\end{minipage}\\
\makebox[3em]{$^{42}$}
\begin{minipage}[t]{14cm}
{\it Polytechnic University, Sagamihara, Japan}~$^{J}$

\end{minipage}\\
\makebox[3em]{$^{43}$}
\begin{minipage}[t]{14cm}
{\it Dipartimento di Fisica, Universit\`a 'La Sapienza' and INFN,
Rome, Italy}~$^{B}$

\end{minipage}\\
\makebox[3em]{$^{44}$}
\begin{minipage}[t]{14cm}
{\it Rutherford Appleton Laboratory, Chilton, Didcot, Oxon,
United Kingdom}~$^{D}$

\end{minipage}\\
\makebox[3em]{$^{45}$}
\begin{minipage}[t]{14cm}
{\it Raymond and Beverly Sackler Faculty of Exact Sciences, School of Physics, \\
Tel Aviv University, Tel Aviv, Israel}~$^{R}$

\end{minipage}\\
\makebox[3em]{$^{46}$}
\begin{minipage}[t]{14cm}
{\it Department of Physics, Tokyo Institute of Technology,
Tokyo, Japan}~$^{J}$

\end{minipage}\\
\makebox[3em]{$^{47}$}
\begin{minipage}[t]{14cm}
{\it Department of Physics, University of Tokyo,
Tokyo, Japan}~$^{J}$

\end{minipage}\\
\makebox[3em]{$^{48}$}
\begin{minipage}[t]{14cm}
{\it Tokyo Metropolitan University, Department of Physics,
Tokyo, Japan}~$^{J}$

\end{minipage}\\
\makebox[3em]{$^{49}$}
\begin{minipage}[t]{14cm}
{\it Universit\`a di Torino and INFN, Torino, Italy}~$^{B}$

\end{minipage}\\
\makebox[3em]{$^{50}$}
\begin{minipage}[t]{14cm}
{\it Universit\`a del Piemonte Orientale, Novara, and INFN, Torino,
Italy}~$^{B}$

\end{minipage}\\
\makebox[3em]{$^{51}$}
\begin{minipage}[t]{14cm}
{\it Department of Physics, University of Toronto, Toronto, Ontario,
Canada M5S 1A7}~$^{N}$

\end{minipage}\\
\makebox[3em]{$^{52}$}
\begin{minipage}[t]{14cm}
{\it Physics and Astronomy Department, University College London,
London, United Kingdom}~$^{D}$

\end{minipage}\\
\makebox[3em]{$^{53}$}
\begin{minipage}[t]{14cm}
{\it Faculty of Physics, University of Warsaw, Warsaw, Poland}

\end{minipage}\\
\makebox[3em]{$^{54}$}
\begin{minipage}[t]{14cm}
{\it National Centre for Nuclear Research, Warsaw, Poland}

\end{minipage}\\
\makebox[3em]{$^{55}$}
\begin{minipage}[t]{14cm}
{\it Department of Particle Physics and Astrophysics, Weizmann
Institute, Rehovot, Israel}

\end{minipage}\\
\makebox[3em]{$^{56}$}
\begin{minipage}[t]{14cm}
{\it Department of Physics, University of Wisconsin, Madison,
Wisconsin 53706, USA}~$^{A}$

\end{minipage}\\
\makebox[3em]{$^{57}$}
\begin{minipage}[t]{14cm}
{\it Department of Physics, York University, Ontario, Canada M3J
1P3}~$^{N}$

\end{minipage}\\
\vspace{30em} \pagebreak[4]

\makebox[3ex]{$^{ A}$}
\begin{minipage}[t]{14cm}
 supported by the US Department of Energy\
\end{minipage}\\
\makebox[3ex]{$^{ B}$}
\begin{minipage}[t]{14cm}
 supported by the Italian National Institute for Nuclear Physics (INFN) \
\end{minipage}\\
\makebox[3ex]{$^{ C}$}
\begin{minipage}[t]{14cm}
 supported by the German Federal Ministry for Education and Research (BMBF), under
 contract No. 05 H09PDF\
\end{minipage}\\
\makebox[3ex]{$^{ D}$}
\begin{minipage}[t]{14cm}
 supported by the Science and Technology Facilities Council, UK\
\end{minipage}\\
\makebox[3ex]{$^{ E}$}
\begin{minipage}[t]{14cm}
 supported by an FRGS grant from the Malaysian government\
\end{minipage}\\
\makebox[3ex]{$^{ F}$}
\begin{minipage}[t]{14cm}
 supported by the US National Science Foundation. Any opinion,
 findings and conclusions or recommendations expressed in this material
 are those of the authors and do not necessarily reflect the views of the
 National Science Foundation.\
\end{minipage}\\
\makebox[3ex]{$^{ G}$}
\begin{minipage}[t]{14cm}
 supported by the Polish Ministry of Science and Higher Education as a scientific project No.
 DPN/N188/DESY/2009\
\end{minipage}\\
\makebox[3ex]{$^{ H}$}
\begin{minipage}[t]{14cm}
 supported by the Polish Ministry of Science and Higher Education and its grants
 for Scientific Research\
\end{minipage}\\
\makebox[3ex]{$^{ I}$}
\begin{minipage}[t]{14cm}
 supported by the German Federal Ministry for Education and Research (BMBF), under
 contract No. 05h09GUF, and the SFB 676 of the Deutsche Forschungsgemeinschaft (DFG) \
\end{minipage}\\
\makebox[3ex]{$^{ J}$}
\begin{minipage}[t]{14cm}
 supported by the Japanese Ministry of Education, Culture, Sports, Science and Technology
 (MEXT) and its grants for Scientific Research\
\end{minipage}\\
\makebox[3ex]{$^{ K}$}
\begin{minipage}[t]{14cm}
 supported by the Korean Ministry of Education and Korea Science and Engineering
 Foundation\
\end{minipage}\\
\makebox[3ex]{$^{ L}$}
\begin{minipage}[t]{14cm}
 supported by FNRS and its associated funds (IISN and FRIA) and by an Inter-University
 Attraction Poles Programme subsidised by the Belgian Federal Science Policy Office\
\end{minipage}\\
\makebox[3ex]{$^{ M}$}
\begin{minipage}[t]{14cm}
 supported by the Spanish Ministry of Education and Science through funds provided by
 CICYT\
\end{minipage}\\
\makebox[3ex]{$^{ N}$}
\begin{minipage}[t]{14cm}
 supported by the Natural Sciences and Engineering Research Council of Canada (NSERC) \
\end{minipage}\\
\makebox[3ex]{$^{ O}$}
\begin{minipage}[t]{14cm}
 partially supported by the German Federal Ministry for Education and Research (BMBF)\
\end{minipage}\\
\makebox[3ex]{$^{ P}$}
\begin{minipage}[t]{14cm}
 supported by RF Presidential grant N 4142.2010.2 for Leading Scientific Schools, by the
 Russian Ministry of Education and Science through its grant for Scientific Research on
 High Energy Physics and under contract No.02.740.11.0244 \
\end{minipage}\\
\makebox[3ex]{$^{ Q}$}
\begin{minipage}[t]{14cm}
 supported by the Netherlands Foundation for Research on Matter (FOM)\
\end{minipage}\\
\makebox[3ex]{$^{ R}$}
\begin{minipage}[t]{14cm}
 supported by the Israel Science Foundation\
\end{minipage}\\
\vspace{30em} \pagebreak[4]

\makebox[3ex]{$^{ a}$}
\begin{minipage}[t]{14cm}
now at University of Salerno, Italy\
\end{minipage}\\
\makebox[3ex]{$^{ b}$}
\begin{minipage}[t]{14cm}
now at Queen Mary University of London, United Kingdom\
\end{minipage}\\
\makebox[3ex]{$^{ c}$}
\begin{minipage}[t]{14cm}
also funded by Max Planck Institute for Physics, Munich, Germany\
\end{minipage}\\
\makebox[3ex]{$^{ d}$}
\begin{minipage}[t]{14cm}
also Senior Alexander von Humboldt Research Fellow at Hamburg University,
 Institute of Experimental Physics, Hamburg, Germany\
\end{minipage}\\
\makebox[3ex]{$^{ e}$}
\begin{minipage}[t]{14cm}
also at Cracow University of Technology, Faculty of Physics,
 Mathemathics and Applied Computer Science, Poland\
\end{minipage}\\
\makebox[3ex]{$^{ f}$}
\begin{minipage}[t]{14cm}
supported by the research grant No. 1 P03B 04529 (2005-2008)\
\end{minipage}\\
\makebox[3ex]{$^{ g}$}
\begin{minipage}[t]{14cm}
supported by the Polish National Science Centre, project No. DEC-2011/01/BST2/03643\
\end{minipage}\\
\makebox[3ex]{$^{ h}$}
\begin{minipage}[t]{14cm}
now at Rockefeller University, New York, NY
 10065, USA\
\end{minipage}\\
\makebox[3ex]{$^{ i}$}
\begin{minipage}[t]{14cm}
now at DESY group FS-CFEL-1\
\end{minipage}\\
\makebox[3ex]{$^{ j}$}
\begin{minipage}[t]{14cm}
now at Institute of High Energy Physics, Beijing, China\
\end{minipage}\\
\makebox[3ex]{$^{ k}$}
\begin{minipage}[t]{14cm}
now at DESY group FEB, Hamburg, Germany\
\end{minipage}\\
\makebox[3ex]{$^{ l}$}
\begin{minipage}[t]{14cm}
also at Moscow State University, Russia\
\end{minipage}\\
\makebox[3ex]{$^{ m}$}
\begin{minipage}[t]{14cm}
now at University of Liverpool, United Kingdom\
\end{minipage}\\
\makebox[3ex]{$^{ n}$}
\begin{minipage}[t]{14cm}
now at CERN, Geneva, Switzerland\
\end{minipage}\\
\makebox[3ex]{$^{ o}$}
\begin{minipage}[t]{14cm}
also affiliated with Universtiy College London, UK\
\end{minipage}\\
\makebox[3ex]{$^{ p}$}
\begin{minipage}[t]{14cm}
now at Goldman Sachs, London, UK\
\end{minipage}\\
\makebox[3ex]{$^{ q}$}
\begin{minipage}[t]{14cm}
also at Institute of Theoretical and Experimental Physics, Moscow, Russia\
\end{minipage}\\
\makebox[3ex]{$^{ r}$}
\begin{minipage}[t]{14cm}
also at FPACS, AGH-UST, Cracow, Poland\
\end{minipage}\\
\makebox[3ex]{$^{ s}$}
\begin{minipage}[t]{14cm}
partially supported by Warsaw University, Poland\
\end{minipage}\\
\makebox[3ex]{$^{ t}$}
\begin{minipage}[t]{14cm}
now at Istituto Nucleare di Fisica Nazionale (INFN), Pisa, Italy\
\end{minipage}\\
\makebox[3ex]{$^{ u}$}
\begin{minipage}[t]{14cm}
now at Haase Energie Technik AG, Neum\"unster, Germany\
\end{minipage}\\
\makebox[3ex]{$^{ v}$}
\begin{minipage}[t]{14cm}
now at Department of Physics, University of Bonn, Germany\
\end{minipage}\\
\makebox[3ex]{$^{ w}$}
\begin{minipage}[t]{14cm}
now at Biodiversit\"at und Klimaforschungszentrum (BiK-F), Frankfurt, Germany\
\end{minipage}\\
\makebox[3ex]{$^{ x}$}
\begin{minipage}[t]{14cm}
also affiliated with DESY, Germany\
\end{minipage}\\
\makebox[3ex]{$^{ y}$}
\begin{minipage}[t]{14cm}
also at University of Tokyo, Japan\
\end{minipage}\\
\makebox[3ex]{$^{ z}$}
\begin{minipage}[t]{14cm}
now at Kobe University, Japan\
\end{minipage}\\
\makebox[3ex]{$^{\dagger}$}
\begin{minipage}[t]{14cm}
 deceased \
\end{minipage}\\
\makebox[3ex]{$^{aa}$}
\begin{minipage}[t]{14cm}
supported by DESY, Germany\
\end{minipage}\\
\makebox[3ex]{$^{ab}$}
\begin{minipage}[t]{14cm}
member of National Technical University of Ukraine, Kyiv Polytechnic Institute,
 Kyiv, Ukraine\
\end{minipage}\\
\makebox[3ex]{$^{ac}$}
\begin{minipage}[t]{14cm}
member of National University of Kyiv - Mohyla Academy, Kyiv, Ukraine\
\end{minipage}\\
\makebox[3ex]{$^{ad}$}
\begin{minipage}[t]{14cm}
partly supported by the Russian Foundation for Basic Research, grant 11-02-91345-DFG\_a\
\end{minipage}\\
\makebox[3ex]{$^{ae}$}
\begin{minipage}[t]{14cm}
Alexander von Humboldt Professor; also at DESY and University of Oxford\
\end{minipage}\\
\makebox[3ex]{$^{af}$}
\begin{minipage}[t]{14cm}
STFC Advanced Fellow\
\end{minipage}\\
\makebox[3ex]{$^{ag}$}
\begin{minipage}[t]{14cm}
nee Korcsak-Gorzo\
\end{minipage}\\
\makebox[3ex]{$^{ah}$}
\begin{minipage}[t]{14cm}
This material was based on work supported by the
 National Science Foundation, while working at the Foundation.\
\end{minipage}\\
\makebox[3ex]{$^{ai}$}
\begin{minipage}[t]{14cm}
also at Max Planck Institute for Physics, Munich, Germany, External Scientific Member\
\end{minipage}\\
\makebox[3ex]{$^{aj}$}
\begin{minipage}[t]{14cm}
now at Tokyo Metropolitan University, Japan\
\end{minipage}\\
\makebox[3ex]{$^{ak}$}
\begin{minipage}[t]{14cm}
now at Nihon Institute of Medical Science, Japan\
\end{minipage}\\
\makebox[3ex]{$^{al}$}
\begin{minipage}[t]{14cm}
now at Osaka University, Osaka, Japan\
\end{minipage}\\
\makebox[3ex]{$^{am}$}
\begin{minipage}[t]{14cm}
also at \L\'{o}d\'{z} University, Poland\
\end{minipage}\\
\makebox[3ex]{$^{an}$}
\begin{minipage}[t]{14cm}
member of \L\'{o}d\'{z} University, Poland\
\end{minipage}\\
\makebox[3ex]{$^{ao}$}
\begin{minipage}[t]{14cm}
now at Department of Physics, Stockholm University, Stockholm, Sweden\
\end{minipage}\\
\makebox[3ex]{$^{ap}$}
\begin{minipage}[t]{14cm}
also at Cardinal Stefan Wyszy\'nski University, Warsaw, Poland\
\end{minipage}\\

}

\newpage

\pagenumbering{arabic} 
\pagestyle{plain}

\section{Introduction}
\label{intro}
The study of jet production in $ep$ collisions at HERA has been well
established as a testing ground of perturbative QCD (pQCD). Jet cross
sections provided precise determinations of the strong coupling
constant, $\as$, and its scale dependence. The jet observables used to
test pQCD included 
inclusive-jet~\cite{epj:c19:289,pl:b547:164,pl:b551:226,np:b765:1,pl:b649:12,epj:c65:363,epj:c67:1},
dijet~\cite{pl:b507:70,epj:c19:289,epj:c23:13,np:b765:1,epj:c65:363,epj:c67:1}
and multijet~\cite{pl:b515:17,epj:c44:183,pr:d85:052008,epj:c65:363,epj:c67:1} cross sections
in neutral current (NC) deep inelastic $ep$ scattering (DIS),
inclusive-jet~\cite{pl:b560:7,epj:c29:497},
dijet~\cite{epj:c11:35,epj:c23:615,pl:b531:9,epj:c25:13,pl:b639:21,pr:d76:072011}
and multijet~\cite{pl:b443:394,np:b792:1} cross sections in
photoproduction and the internal structure of jets in
NC~\cite{np:b545:3,pl:b558:41,np:b700:3} and charged
current~\cite{epj:c31:149,pr:d78:032004} DIS. These studies also
demonstrated that the $\kt$ cluster algorithm~\cite{np:b406:187} in
the longitudinally invariant inclusive mode~\cite{pr:d48:3160} results
in the smallest uncertainties in the reconstruction of jets in $ep$
collisions. Jet cross sections in NC DIS~\cite{pl:b547:164} and
photoproduction~\cite{epj:c23:615} were used by
ZEUS~\cite{epj:c42:1} as input in a QCD analysis to extract the parton
distribution functions (PDFs) of the proton; these data helped to
constrain the gluon density at medium- to high-$x$ values, where $x$
is the fraction of the proton momentum carried by the gluon.

The $\kt$ algorithm is well suited for $ep$ collisions and yields
infrared- and collinear-safe cross sections at any order of
pQCD. However, it might not be best suited to reconstruct jets in 
hadron-hadron collisions, such as those at the LHC. In order to
optimise the reconstruction of jet observables in such environments,
new infrared- and collinear-safe jet algorithms were recently
developed, namely the anti-$\kt$~\cite{jhep:04:063}, a
recombination-type jet algorithm, and the ``Seedless Infrared-Safe''
cone (SIScone)~\cite{jhep:05:086} algorithms. Measurements of jet
cross sections in NC DIS using these algorithms were recently
published~\cite{pl:b691:127} and constituted the first measurements
with these new jet algorithms. The results tested the performance of
these jet algorithms with data in a well understood hadron-induced
reaction and it was shown that pQCD calculations with up to four
partons in the final state provide a good description of the
differences between jet algorithms.

Measurements of inclusive-jet cross sections in photoproduction are
presented in this paper. Two types of QCD processes contribute to jet
production in photoproduction; at leading order they can be separated
into~\cite{pl:b79:83,*np:b166:413,*pr:d21:54,*zfp:c6:241,proc:hera:1987:331,*prl:61:275,*prl:61:682,*pr:d39:169,*zfp:c42:657,*pr:d40:2844}
the direct process, in which the photon interacts directly with a
parton in the proton, and the resolved process, in which the photon
acts as a source of partons, one of which interacts with a parton in
the proton. Due to the presence of the resolved processes, the analysis
of jet cross sections in photoproduction with different jet algorithms
provides a test of their performance in a reaction closer to
hadron-hadron interactions than NC DIS.

In this paper, single-differential inclusive-jet cross sections as
functions of the jet transverse energy, $\etjet$, and pseudorapidity,
$\etajet$, are presented based on the $\kt$, anti-$\kt$ and SIScone jet
algorithms. The results based on the anti-$\kt$ and SIScone jet 
algorithms are compared to the measurements based on the $\kt$ via
the ratios of cross sections. In addition, measurements of cross
sections are also presented as functions of $\etjet$ in different
regions of $\etajet$, which have the potential to constrain further
the gluon density at high $x$. Next-to-leading-order (NLO) QCD
calculations using recent parameterisations of the 
proton
and 
photon
PDFs are compared to the measurements. A determination of $\asz$ as well
as of its energy-scale dependence are also presented. The analyses
presented here are based on a data sample with a more than three-fold
increase in statistics with respect to the previous study~\cite{pl:b560:7}.

\section{Experimental set-up}
A detailed description of the ZEUS detector can be found
elsewhere~\cite{pl:b293:465,zeus:1993:bluebook}. A brief outline of
the components most relevant for this analysis is given below.

Charged particles were tracked in the central tracking detector
(CTD)~\cite{nim:a279:290,*npps:b32:181,*nim:a338:254}, which operated
in a magnetic field of $1.43\Tesla$ provided by a thin superconducting
solenoid. The CTD consisted of $72$~cylindrical drift-chamber
layers, organised in nine superlayers covering the
polar-angle\footnote{The ZEUS coordinate system was a right-handed
  Cartesian system, with the $Z$ axis pointing in the proton beam
  direction, referred to as the ``forward direction'', and the $X$
  axis pointing towards the centre of HERA. The coordinate origin
  was at the nominal interaction point.}
region \mbox{$15^\circ<\theta<164^\circ$}. The CTD was complemented by a
silicon microvertex detector (MVD)~\cite{nim:a581:656}, consisting of
three active layers in the barrel and four disks in the forward
region. For CTD-MVD tracks that pass through all nine CTD superlayers,
the momentum resolution was 
$\sigma(p_T)/p_T=0.0029p_T\oplus 0.0081\oplus 0.0012/p_T$, 
with $p_T$ in GeV.

The high-resolution uranium--scintillator calorimeter
(CAL)~\cite{nim:a309:77,*nim:a309:101,*nim:a321:356,*nim:a336:23}
consisted of three parts: the forward (FCAL), the barrel (BCAL) and
the rear (RCAL) calorimeters. Each part was subdivided transversely
into towers and longitudinally into one electromagnetic section (EMC)
and either one (in RCAL) or two (in BCAL and FCAL) hadronic sections
(HAC). The smallest subdivision of the calorimeter was called a
cell. The CAL energy resolutions, as measured under test-beam
conditions, were $\sigma(E)/E=0.18/\sqrt E$ for electrons and 
$\sigma(E)/E=0.35/\sqrt E$ for hadrons, with $E$ in GeV.

The luminosity was measured using the Bethe-Heitler reaction 
$ep\rightarrow e\gamma p$ by a luminosity detector which consisted of
a lead--scintillator calorimeter~\cite{desy-92-066,*zfp:c63:391,*acpp:b32:2025}
and an independent magnetic spectrometer~\cite{nim:a565:572}.
The fractional uncertainty on the measured luminosity was $1.8\%$.

\section{Data selection}
The data were collected during the running period 2005--2007, when HERA
operated with protons of energy $E_p=920$~GeV and electrons or
positrons\footnote{In the following, the term electron will refer to
  both the electron and positron, unless otherwise stated.} of energy
$E_e=27.5$~GeV, at an $ep$ centre-of-mass energy of $\sqrt s=318$ GeV,
and correspond to an integrated luminosity of $299.9\pm 5.4$~\pb1.

A three-level trigger system was used to select events
online~\cite{zeus:1993:bluebook,proc:chep:1992:222}. At the first
level, events were triggered by a coincidence of a regional or
transverse energy sum in the CAL and at least one track from the
interaction point measured in the CTD. At the second level, a total
transverse energy of at least $12$~GeV, excluding the energy in the
eight CAL towers immediately surrounding the forward beampipe, was
required, and cuts on CAL energies and timing were used to suppress
events caused by interactions between the proton beam and residual gas
in the beampipe. At the third level, two different methods were
applied to select the events. The first method selected events with a
total transverse energy of at least $25$~GeV, excluding the energy in
the eight CAL towers immediately surrounding the forward beampipe. For
the second method, a jet algorithm was applied to the CAL cells and
jets were reconstructed using the energies and positions of these
cells; events with at least one jet of $E_T>10$~GeV and $\eta<2.5$
were accepted. Additional requirements based on CAL energies, tracking
and timing were used to suppress further the non-$ep$ background.

Events from collisions between quasi-real photons and protons were
selected offline using similar criteria to those reported in the previous 
ZEUS publication~\cite{pl:b560:7}. The selection criteria applied were:
\begin{itemize}
\item a reconstructed event vertex along the $Z$ axis within 35~cm of the
  nominal interaction point was required;
\item cuts based on tracking information were applied to remove the
  contamination from beam-gas interactions, cosmic-ray showers and
  beam-halo muons;
\item charged current DIS events were rejected by requiring the total
  missing transverse momentum, $\ptmis$, to be small compared to the
  total transverse energy, $E^{\rm tot}_T$, i.e. 
  \mbox{$\ptmis/\sqrt{E^{\rm tot}_T}<2\ \sqrt{\rm GeV}$};
\item any NC DIS event with an identified scattered-electron candidate
  in the CAL was rejected;
\item the events were restricted to $\gamma p$ centre-of-mass
  energies in the region \mbox{\wrn}, where $\wgp=\sqrt{sy}$;
  $y$ is the inelasticity and was estimated as $y_{\rm
  JB}=(E-p_Z)/2E_e$, where $E$ is the total energy measured in the CAL
  and $p_Z$ is the longitudinal component of the total momentum.
\end{itemize}

After these selection criteria were applied, the contamination from
beam-gas interactions, cosmic-ray showers and beam-halo muons was
found to be negligible. The remaining background from NC DIS events
was estimated by Monte Carlo (MC) techniques to be around $1\%$ and
was neglected. The contamination from charged current DIS events was
found to be even smaller. The selected sample consisted of events from
$ep$ interactions with $\q2<1$ \g2, where $\q2$ is the virtuality of
the exchanged photon, and a median $\q2\approx 10^{-4}$~\g2,
estimated using MC techniques.

\section{Jet search}
\label{js}
In photoproduction, jets are usually defined using the transverse-energy
flow in the pseudorapidity-azimuth ($\etaphi$) plane of the
laboratory frame. The procedure to reconstruct jets with the $\kt$
algorithm from an initial list of objects (e.g. final-state partons,
final-state hadrons or energy deposits in the calorimeter) is
described below in some detail. In the following discussion, $E_T^i$
denotes the transverse energy, $\eta^i$ the pseudorapidity and
$\phi^i$ the azimuthal angle of object $i$. For each pair of objects,
the quantity

\begin{equation}
d_{ij}={\rm min}((E_T^i)^2,(E_T^j)^2)\cdot[(\eta^i-\eta^j)^2+(\phi^i-\phi^j)^2]/R^2
\end{equation}
is calculated, where $R$ is the jet radius. For each individual
object, the distance to the beam, $d_i=(E_T^i)^2$, is also
calculated. If, of all the values $\{d_{ij},d_i\}$, $d_{kl}$ is the
smallest, then objects $k$ and $l$ are combined into a single new
object. If, however, $d_k$ is the smallest, then object $k$ is
considered a jet and removed from the sample. The procedure is
repeated until all objects are assigned to jets.

The anti-$\kt$ algorithm is identical to the $\kt$ except for a
modified distance measure,

\begin{equation}
d_{ij}={\rm min}((E_T^i)^{-2},(E_T^j)^{-2})\cdot[(\eta^i-\eta^j)^2+(\phi^i-\phi^j)^2]/R^2,
\end{equation}

and the distance to the beam, which is defined as $d_i=(E_T^i)^{-2}$.

The SIScone algorithm consists of two steps. First, for a given set of
initial objects, all stable cones are identified; cones are classified
as stable by the coincidence of the cone axis with that defined by 
the total momentum of the objects contained in the given cone of
radius $R$ in the $\eta-\phi$ plane. In this procedure, no seed is
used. Stable cones are then discarded if their transverse momentum is
below a given threshold, $p_{t,{\rm min}}$. For each selected stable
cone, the scalar sum of the transverse momentum of the objects
associated to it, $\tilde{p}_t$, is defined. Second, overlapping cones
are identified and subsequently split or merged according to the following
procedure. Two cones are merged if the scalar sum of the transverse 
momentum of the objects shared by the two cones exceeds a certain
fraction $f$ of the lowest-$\tilde{p}_t$ cone; otherwise, two
different cones are considered and the common objects are assigned to
the nearest cone.

For the measurements presented in this paper, the jet radius $R$ was
set to unity and the jet variables were defined according to the
Snowmass convention~\cite{proc:snowmass:1990:134} for all three jet
algorithms. In the application of the SIScone algorithm, the fraction
$f$ was set to $0.75$ and $p_{t,{\rm min}}$ was set to zero.

The $\kt$, anti-$\kt$ and SIScone jet algorithms\footnote{The
  {\sc Fastjet} 2.4.1~\cite{pl:b641:57} package was used.}
were used to reconstruct jets in the hadronic final state from the
energy deposits in the CAL cells. The jets reconstructed from
the CAL cell energies are called calorimetric jets and the variables
associated with them are denoted by $\etcal$, $\etacal$ and
$\phical$. Three samples of events were selected for further analysis,
one for each jet algorithm, which contain at least one jet satisfying
$\etcal>13$~GeV and $\etacr$.

\section{Monte Carlo simulations}
\label{mc}
Samples of events were generated to determine the response of the
detector to jets of hadrons and the correction factors necessary to
obtain the hadron-level jet cross sections. In addition, these samples
were used to estimate hadronisation corrections to the NLO
calculations (see Section~\ref{nlo}).

The MC programs {\sc Pythia}~6.146~\cite{cpc:82:74} and {\sc
  Herwig}~6.504~\cite{cpc:67:465,*jhep:0101:010} were used to generate
resolved and direct photoproduction events. In both generators, the
partonic processes are simulated using leading-order matrix elements,
with the inclusion of initial- and final-state parton
showers. Fragmentation into hadrons was performed using the Lund string
model~\cite{prep:97:31}, as implemented in
{\sc Jetset}~\cite{cpc:39:347,*cpc:43:367} in the case of {\sc Pythia}, and a
cluster model~\cite{np:b238:492} in the case of {\sc Herwig}. The
CTEQ4M~\cite{pr:d55:1280} (GRV-HO~\cite{pr:d45:3986,*pr:d46:1973})
sets were used for the proton (photon) PDFs. Samples of {\sc Pythia}
including multiparton interactions~\cite{pr:d36:2019} ({\sc
  Pythia}-MI) with a minimum transverse momentum for the secondary
scatter, $p_{T,{\rm min}}^{\rm sec}$, of 1, 1.5 or 2~GeV were used to
simulate contributions from non-perturbative effects not related to
hadronisation (NP), such as the underlying event. All
the samples of generated events were passed through the {\sc
  Geant}~3.21-based~\cite{tech:cern-dd-ee-84-1} ZEUS detector- and
trigger-simulation programs~\cite{zeus:1993:bluebook}. They were 
reconstructed and analysed by the same program chain as the data.

The jet search was performed on the MC events using the energy
measured in the CAL cells as described in Section~\ref{js}. In addition,
the three jet algorithms were also applied to the final-state particles
(hadron level) and partons (parton level). The hadron level is defined
by those hadrons with lifetime $\tau\geq 10$~ps and the parton level
is defined as those partons present after the parton-shower procedure.

\section{Transverse-energy and acceptance corrections}
\label{corfac}
The comparison of the reconstructed jet variables for the hadronic and 
the calorimetric jets in MC-simulated events showed that no correction
was needed for the jet pseudorapidity and azimuth. However, $\etcal$
underestimates the corresponding hadronic-jet transverse energy by
$\approx 14\%$ with an r.m.s. of $\approx 10\%$. This underestimation is
mainly due to the energy lost by the particles in the inactive
material in front of the CAL. The transverse-energy corrections to
calorimetric jets, as functions of $\etacal$ and $\etcal$ and
averaged over $\phical$, were determined using the MC events. Further
corrections to the jet transverse energy were applied to the data to
account for differences in the jet energy scale between data and MC
simulations; the method presented
previously~\cite{pl:b531:9,hep-ex-0206036}, which relies on a good 
understanding of the performance of the track reconstruction, was
used to calibrate the absolute energy scale of the jets down to $\pm
1\%$. This calibration was cross checked by means of the
transverse-momentum balance in the CAL between the electron candidate
and the jet in single-jet NC DIS events.

Henceforth, jet variables without subscript refer to the
corrected values. After all these corrections to the jet transverse
energy, events with at least one jet satisfying $\etjet>17$~GeV and
$\etar$ were retained. The number of events and jets in the final data
samples are shown in Table~\ref{tab2} for each jet algorithm. No
events with more than four jets were found in these samples.

The $\etjet$ and $\etajet$ distributions in the data were corrected
for detector effects using bin-by-bin acceptance correction factors
determined using the MC samples. These correction factors take into
account the efficiency of the trigger, the selection criteria and the
purity and efficiency of the jet reconstruction. For this approach to
be valid, the uncorrected distributions of the data must be adequately
described by the MC simulations at the detector level. This condition
was satisfied by both the {\sc Pythia} and {\sc Herwig} MC
samples. The average between the acceptance correction factors obtained
from {\sc Pythia} and {\sc Herwig} was used to correct the data to the
hadron level. The deviations in the results obtained by using either
{\sc Pythia} or {\sc Herwig} to correct the data from their average
were taken to represent systematic uncertainties of the effect of the 
QCD-cascade and hadronisation models in the corrections (see
Section~\ref{expunc}). The acceptance correction factors differed from
unity by typically less than $20\%$. 

\section{Experimental uncertainties}
\label{expunc}
The following sources of systematic uncertainty were considered for
the measured cross sections:
\begin{itemize}
\item the differences in the results obtained by using either
  {\sc Pythia} or {\sc Herwig} to correct the data for detector
  effects. The resulting uncertainty was typically below $\pm 4\%$;
\item the effect of the CAL energy-scale uncertainty on $\wgp$ was
  estimated by varying $y_{\rm JB}$ by $\pm 1\%$ in simulated events.
  The uncertainty in the cross sections was below $\pm 1\%$ at low
  $\etjet$, increasing to $\approx\pm 3\%$ at high $\etjet$;
\item the effect of the uncertainty on the parameterisations of the
  proton and photon PDFs was estimated by using alternative sets of
  PDFs (MRST(c-g)~\cite{epj:c4:463} for the proton and
  AFG-HO~\cite{zp:c64:621} for the photon) in the MC simulation to
  compute the acceptance correction factors. The variation of the
  cross sections was typically smaller than $\pm 1\%$;
\item the uncertainty in the cross sections due to that in the
  simulation of the trigger was found to be negligible.
\end{itemize}

All the above systematic uncertainties were added in quadrature and
are shown in the figures as error bars. The resulting systematic
uncertainty in the cross sections based on the three jet algorithms
was similar and typically below $\pm 5\%$. 

The absolute energy scale
of the calorimetric jets in simulated events was varied by its
uncertainty of $\pm 1\%$ (see Section~\ref{corfac}); the effect of
this variation on the inclusive-jet cross sections was typically $\mp
5\%$ at low $\etjet$, increasing up to $\mp 10\%$ at high $\etjet$. This
uncertainty is fully correlated between measurements in different
bins and is shown separately as a shaded band in the figures. In
addition, there was an overall normalisation uncertainty of $\pm 1.8\%$
from the luminosity determination, which is not included in the
figures and tables.

\section{Next-to-leading-order QCD calculations}
\label{nlo}
\label{thun}
The NLO QCD ($\oass$) calculations used in the analysis presented here
were computed using the program by Klasen, Kleinwort and
Kramer~\cite{epjdirectcbg:c1:1}. The calculations use the
phase-space-slicing method~\cite{kramer:1984:slicing} with an
invariant-mass cut to isolate the singular regions of the phase
space. The number of flavours was set to five and 
the renormalisation ($\mu_R$) and factorisation ($\mu_F$) scales were
chosen to be $\mu_R=\mu_F=\mu=\etjet$. The strong coupling constant
was calculated at two loops with $\Lambda_{\overline{\rm
    MS}}=226$~MeV, corresponding to $\asz=0.118$. The calculations
were performed using the ZEUS-S~\cite{pr:d67:012007} parameterisations
of the proton PDFs and the GRV-HO sets
for the photon PDFs as default\footnote{The {\sc Lhapdf}
  5.7.1~\cite{hep-ph:0508110} package was used.}. The three
jet algorithms were applied to the partons in the events generated by
this program to compute the jet cross-section predictions. At
$\oass$, the parton-level predictions from the $\kt$ and anti-$\kt$
jet algorithms are identical. The predictions from the $\kt$ and
SIScone jet algorithms start to differ at this order.
 
Since the measurements refer to jets of hadrons, whereas the NLO QCD
calculations refer to jets of partons, the predictions were corrected
to the hadron level using the MC models. The multiplicative correction
factor, $C_{\rm had}$, was defined as the ratio of the cross section
for jets of hadrons over that for jets of partons, estimated by using
the MC programs described in Section~\ref{mc}. The mean of the ratios 
obtained with {\sc Pythia} and {\sc Herwig} was taken as the value
of $C_{\rm had}$. Details on the values of $C_{\rm had}$ are presented
in Section~\ref{results}.

The following sources of uncertainty in the theoretical predictions were
considered:
\begin{itemize}
  \item the uncertainty on the NLO QCD calculations due to that on
    the value of $\asz$ used was estimated by repeating the calculations
    using two additional ZEUS-S sets of proton PDFs, for which different
    values of $\asz$ were assumed in the fits. The difference between
    these calculations was used to determine the uncertainty due to
    that on the value of $\asz$~\cite{jp:g26:r27};
  \item the uncertainty on the NLO QCD calculations due to terms
    beyond NLO was estimated by repeating the calculations using
    values of $\mu$ scaled by factors $0.5$ and $2$;
  \item the uncertainty from the modelling of the QCD cascade and
    hadronisation effects was assumed to be half the difference
    between the hadronisation corrections obtained using the {\sc
      Pythia} and {\sc Herwig} models;
  \item the uncertainty on the NLO QCD calculations due to those on the 
    proton PDFs was estimated by repeating the calculations using 22
    additional sets from the ZEUS-S error analysis with $\asz$ fixed
    to the central value; this error analysis takes into account the
    statistical and correlated experimental uncertainties of each data
    set used in the determination of the proton PDFs;
\item the uncertainty on the NLO QCD calculations due to those on
    the photon PDFs was estimated by using alternative sets of
    parameterisations, AFG04~\cite{epj:c44:395} and CJK~\cite{pr:d70:093004}.
\end{itemize}
 
\begin{figure}[p]
\vfill
\setlength{\unitlength}{1.0cm}
\begin{picture} (18.0,15.0)
\put (-0.2,9.0){\centerline{\epsfig{figure=\figdir 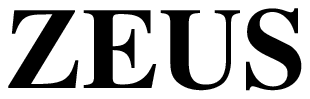,width=12cm}}}
\put (-1.7,8.5){\epsfig{figure=\figdir 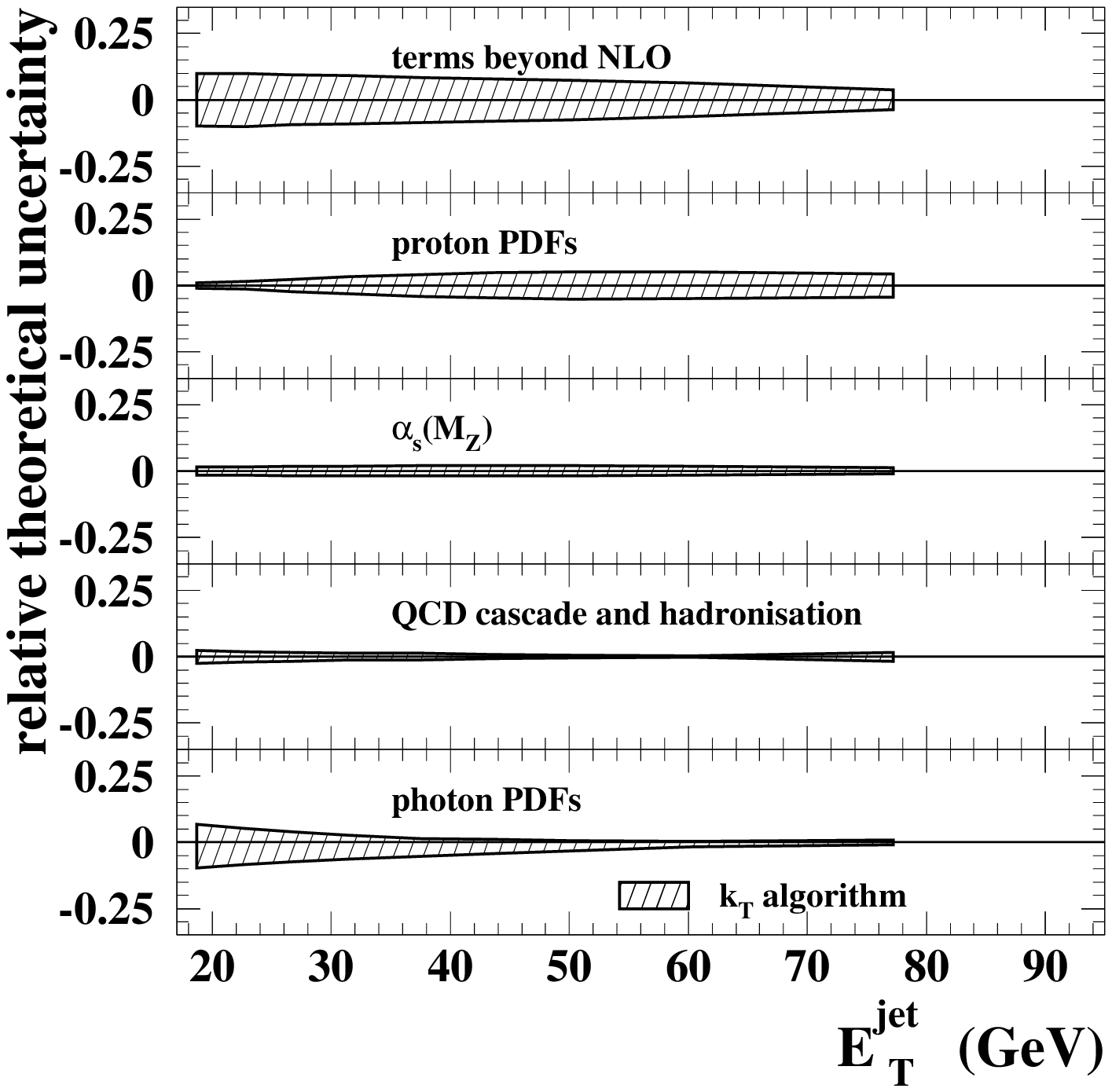,width=12cm}}
\put (5.5,8.5){\epsfig{figure=\figdir 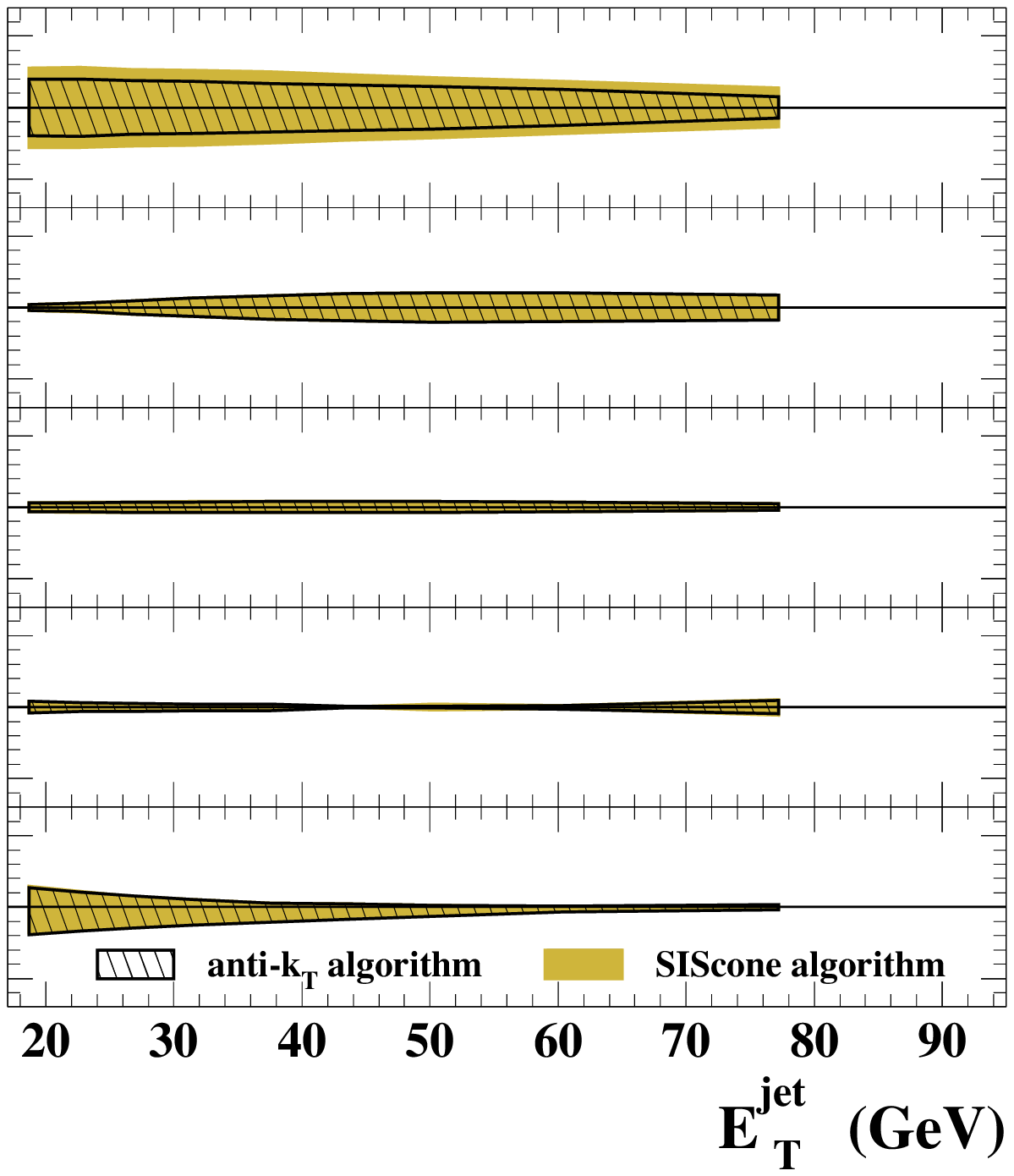,width=12cm}}
\put (-1.7,-1.0){\epsfig{figure=\figdir 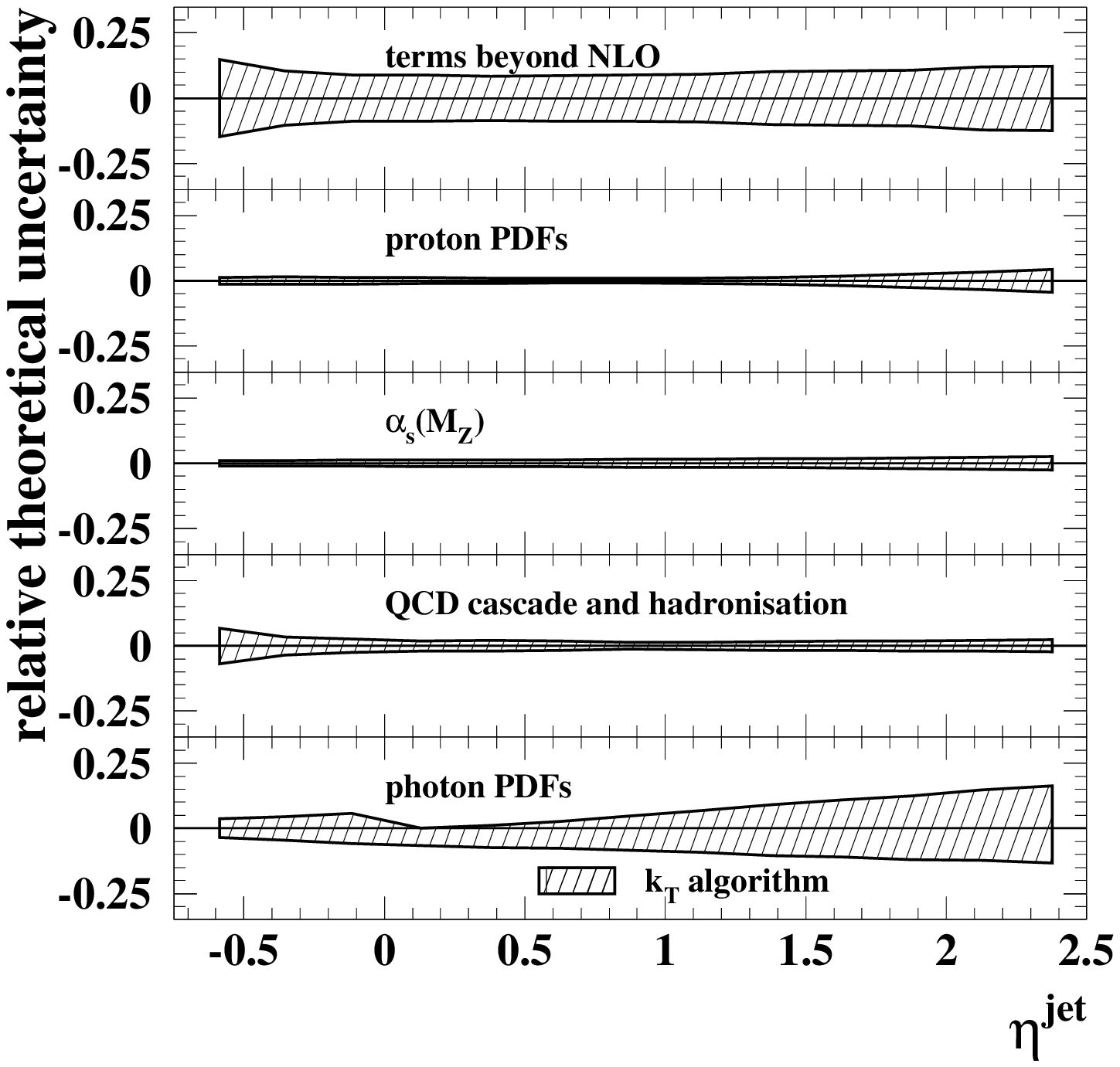,width=12cm}}
\put (5.5,-1.0){\epsfig{figure=\figdir 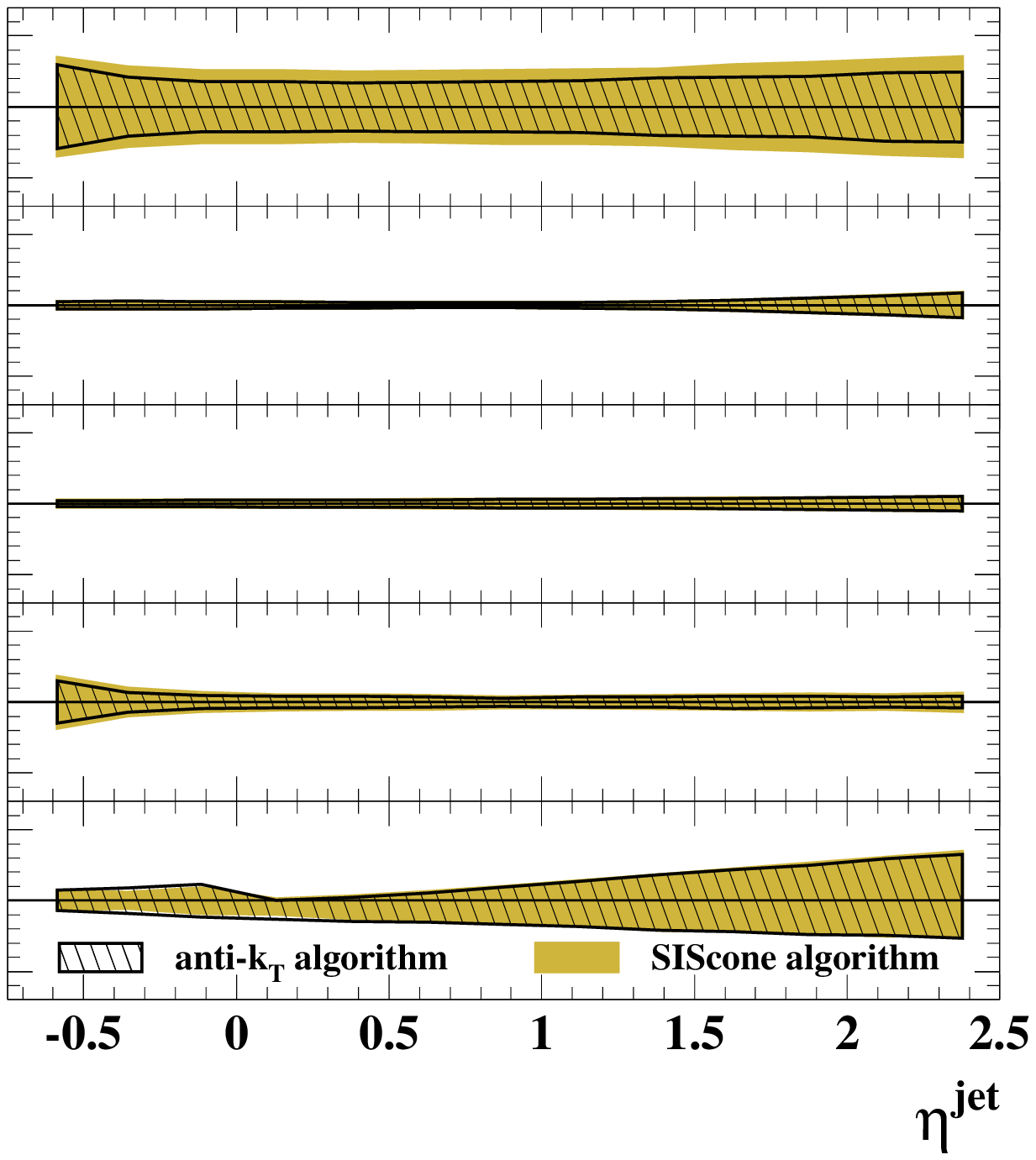,width=12cm}}
\put (4.0,10.0){{\bf\small (a)}}
\put (11.0,10.0){{\bf\small (b)}}
\put (4.0,0.5){{\bf\small (c)}}
\put (11.0,0.5){{\bf\small (d)}}
\end{picture}
\caption
{\it 
Overview of the relative theoretical uncertainties for
inclusive-jet cross sections in photoproduction in the kinematic
region of the measurements as functions of (a,b) $\etjet$ and (c,d)
$\etajet$ for the $\kt$, anti-$\kt$ and SIScone jet algorithms. Shown
are the relative uncertainties induced by the terms beyond NLO, the
proton PDFs, the value of $\asz$, the modelling of the QCD cascade and
hadronisation and the photon PDFs.
}
\label{fig1}

\vfill
\end{figure}

The total theoretical uncertainty was obtained by adding in quadrature
the individual uncertainties listed above. Figure~\ref{fig1} shows an
overview of the relative theoretical uncertainties for the
inclusive-jet cross sections in the kinematic region of the
measurements as functions of $\etajet$ and $\etjet$ and for each jet
algorithm separately. The uncertainty due to higher orders
is somewhat larger for the SIScone than for the $\kt$ and anti-$\kt$
algorithms, whereas the other uncertainties are very similar for the
three jet algorithms. The uncertainty coming from the terms beyond NLO
is dominant in all cases. At high $\etjet$, the proton PDF uncertainty
is of the same order (slightly smaller) as that arising from terms
beyond NLO for the $\kt$ and anti-$\kt$ (SIScone) algorithms. The
uncertainty arising from the photon PDFs at high $\etajet$ becomes
comparable to that coming from higher orders. The uncertainties from
the value of $\asz$ and hadronisation corrections are small.

The samples of {\sc Pythia}-MI described in Section~\ref{mc} were used
to estimate the contribution from non-perturbative effects not related
to hadronisation. Such effects were computed as ratios of the cross
section for jets of hadrons in the {\sc Pythia}-MI samples over that
for the samples of {\sc Pythia}; these ratios are called $C_{\rm NP}$.
The values of $C_{\rm NP}$ depend strongly on
$p_{T,{\rm min}}^{\rm sec}$, the minimum transverse momentum for the
secondary scatter, since the smaller the $p_{T,{\rm min}}^{\rm sec}$ value
is set, the larger the phase space available for production of
secondary interactions and hence the higher the jet rate. Another
feature of such secondary interactions is that, due to the particular
kinematics of HERA, the products of these additional interactions are
expected to be boosted towards the proton direction. Several
predictions including these non-perturbative effects, denoted as
NLO$\otimes$NP, were computed by applying the factors $C_{\rm NP}$,
using the $p_{T,{\rm min}}^{\rm sec}=$ 1, 1.5 and 2~GeV {\sc
  Pythia}-MI samples, to the NLO QCD calculations after hadronisation
corrections.

\section{Results}
\label{results}
Single- and double-differential inclusive-jet cross sections were
measured in the kinematic region given by $\q2<1$~\g2\ and
\wrn. These cross sections include every jet of hadrons with
$\etjet>17$~GeV and $-1<\etajet<2.5$ in each event. The jets were
reconstructed using either the $\kt$, the anti-$\kt$ or the SIScone jet 
algorithms. The $x$ region covered by the measurements was determined
to be $3\cdot 10^{-3}<x<0.95$.

\subsection{Single-differential cross sections}
\begin{figure}[p]
\vfill
\setlength{\unitlength}{1.0cm}
\begin{picture} (18.0,15.0)
\put (0.0,0.0){\centerline{\epsfig{figure=\figdir 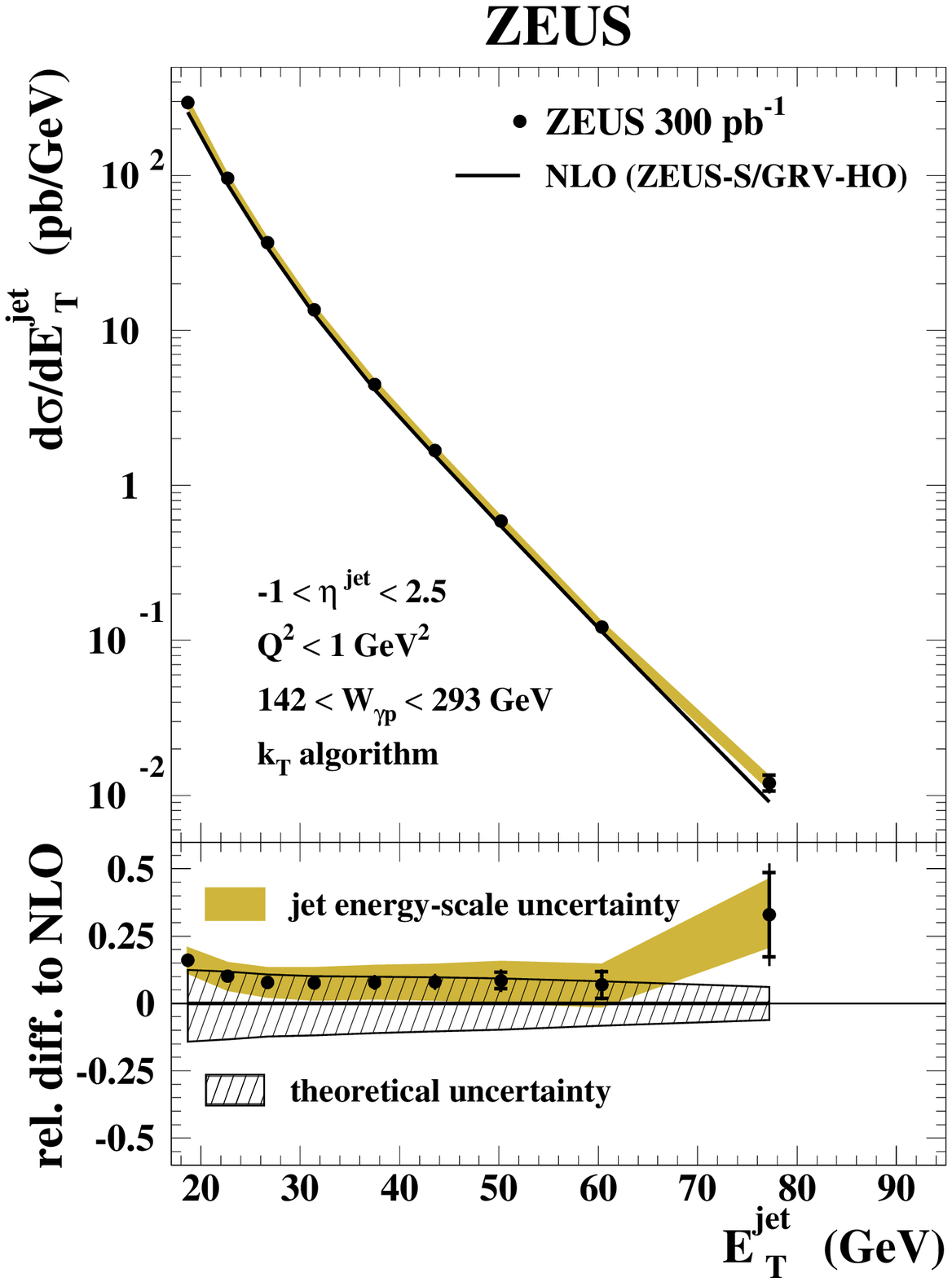,width=15cm}}}
\put (11.7,13.5){{\bf\small (a)}}
\put (11.7,1.8){{\bf\small (b)}}
\end{picture}
\caption
{\it 
(a) The measured differential cross-section $\set$ based on the $\kt$
jet algorithm for inclusive-jet photoproduction with $-1<\etajet<2.5$
(dots) in the kinematic region given by $\q2<1$~\gev$^2$ and
\wrn. The NLO QCD calculation (solid line),
corrected to include hadronisation effects and using the ZEUS-S
(GRV-HO) parameterisations of the proton (photon) PDFs, is also shown. 
(b) The relative difference between the measured $\set$ and the 
NLO QCD calculation (dots). In both figures, the inner error bars
represent the statistical uncertainties; the outer error bars show the
statistical and systematic uncertainties not associated with the
uncertainty in the absolute energy scale of the jets, added in
quadrature; the shaded band displays the uncertainty due to the
absolute energy scale of the jets and the hatched band displays the total
theoretical uncertainty. In some bins, the error bars on the data
points are smaller than the marker size and are therefore not visible.
}
\label{fig2}
\vfill
\end{figure}

\begin{figure}[p]
\vfill
\setlength{\unitlength}{1.0cm}
\begin{picture} (18.0,15.0)
\put (0.0,0.0){\centerline{\epsfig{figure=\figdir 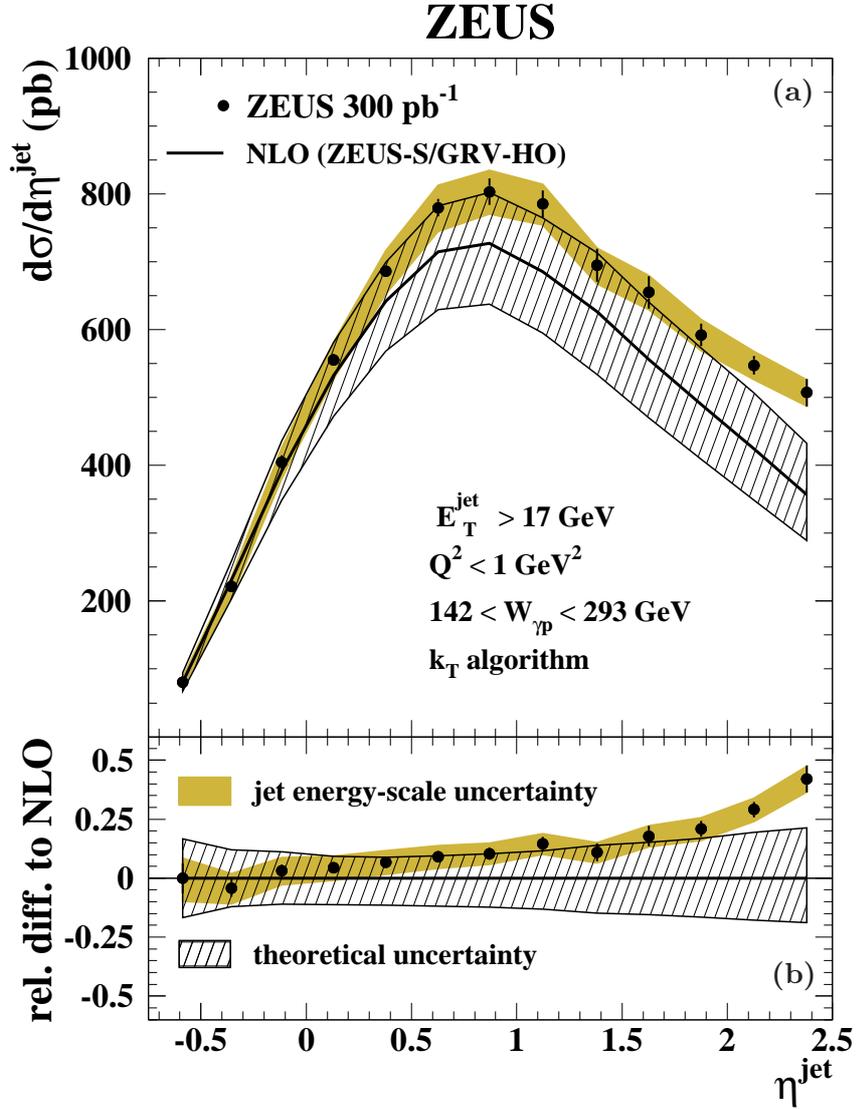,width=15cm}}}
\put (11.7,13.5){{\bf\small (a)}}
\put (11.7,1.8){{\bf\small (b)}}
\end{picture}
\caption
{\it 
(a) The measured differential cross-section $\seta$ based on the $\kt$
jet algorithm for inclusive-jet photoproduction with $\etjet>17$~GeV
(dots) in the kinematic region given by $\q2<1$~\gev$^2$ and \wrn.
(b) The relative difference between the measured $\seta$ and the
NLO QCD calculation (dots).
Other details as in the caption to Fig.~\ref{fig2}. 
}
\label{fig3}
\vfill
\end{figure}

The measurements of the single-differential cross sections based on
the $\kt$ jet algorithm as functions of $\etjet$ and $\etajet$ are
presented in Figs.~\ref{fig2} and \ref{fig3} and Table~\ref{tab3}. In
these and the subsequent figures, each data point is plotted at the
weighted mean of each bin. The measured $\set$ falls by over four
orders of magnitude in the measured range. The measured $\seta$
displays a maximum around $\etajet\approx 1$.

The NLO QCD predictions are compared to the measurements in these
figures. The calculation reproduces the measured $\set$ well.
The measured $\seta$ is well described for
$\etajet\lesssim 2$; however, an excess of data with respect to the
theory is observed for larger $\etajet$ values. Such discrepancies
have already been observed in previous studies of jet
photoproduction~\cite{pr:d76:072011,np:b792:1}. In the study of dijet
production~\cite{pr:d76:072011}, the discrepancies were interpreted as
an inadequacy of the parameterisations of the photon PDFs, which had
been extracted from $\ele$ data at lower scales. The studies of
multijet production~\cite{np:b792:1} showed the need to include
non-perturbative effects not related to hadronisation in the pQCD 
calculations to describe the data.

\begin{figure}[p]
\vfill
\setlength{\unitlength}{1.0cm}
\begin{picture} (18.0,15.0)
\put (0.45,2.0){\centerline{\epsfig{figure=\figdir DESY-12-045_0.eps,width=12cm}}}
\put (-2.0,0.0){\epsfig{figure=\figdir 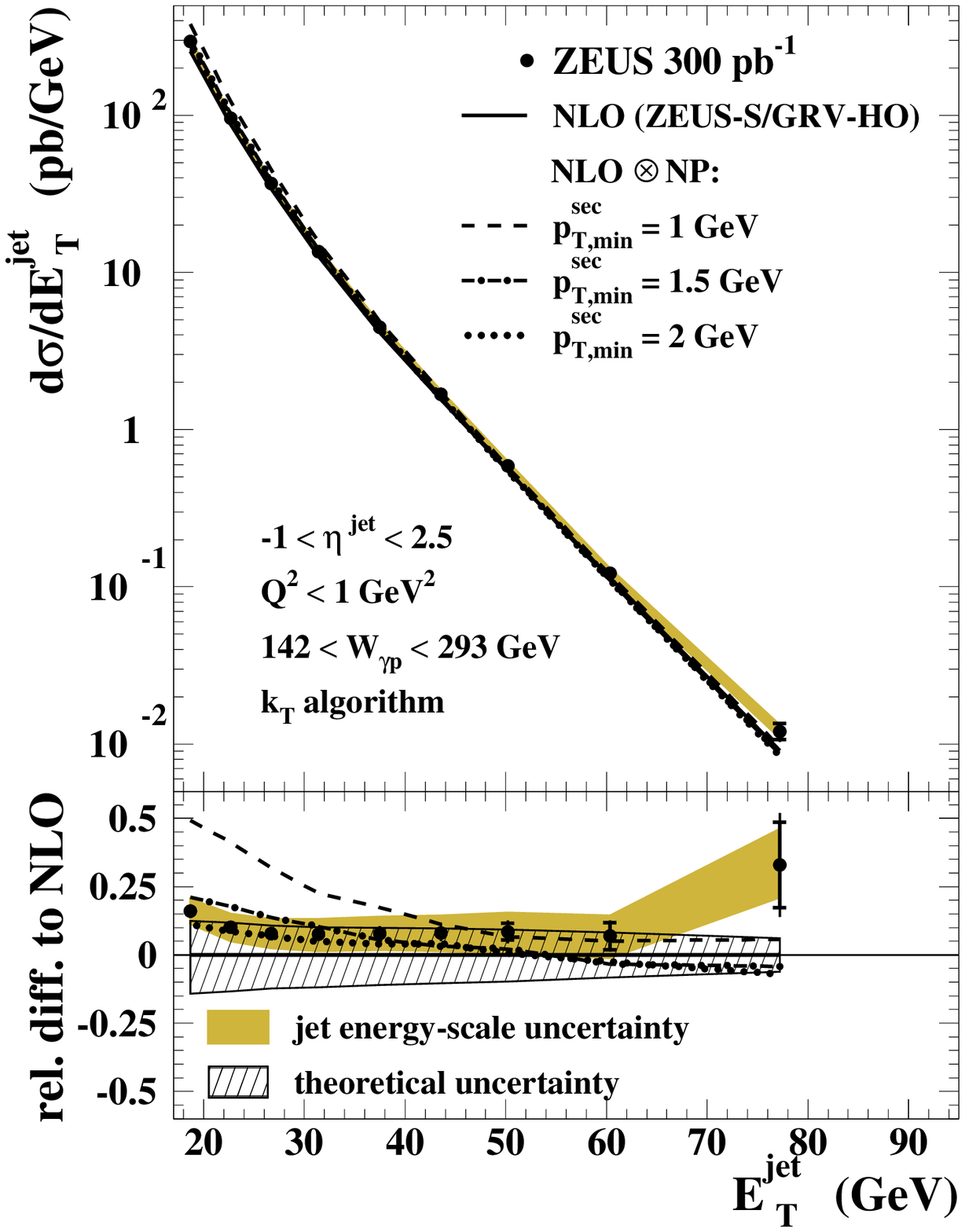,width=12cm}}
\put (7.0,0.0){\epsfig{figure=\figdir 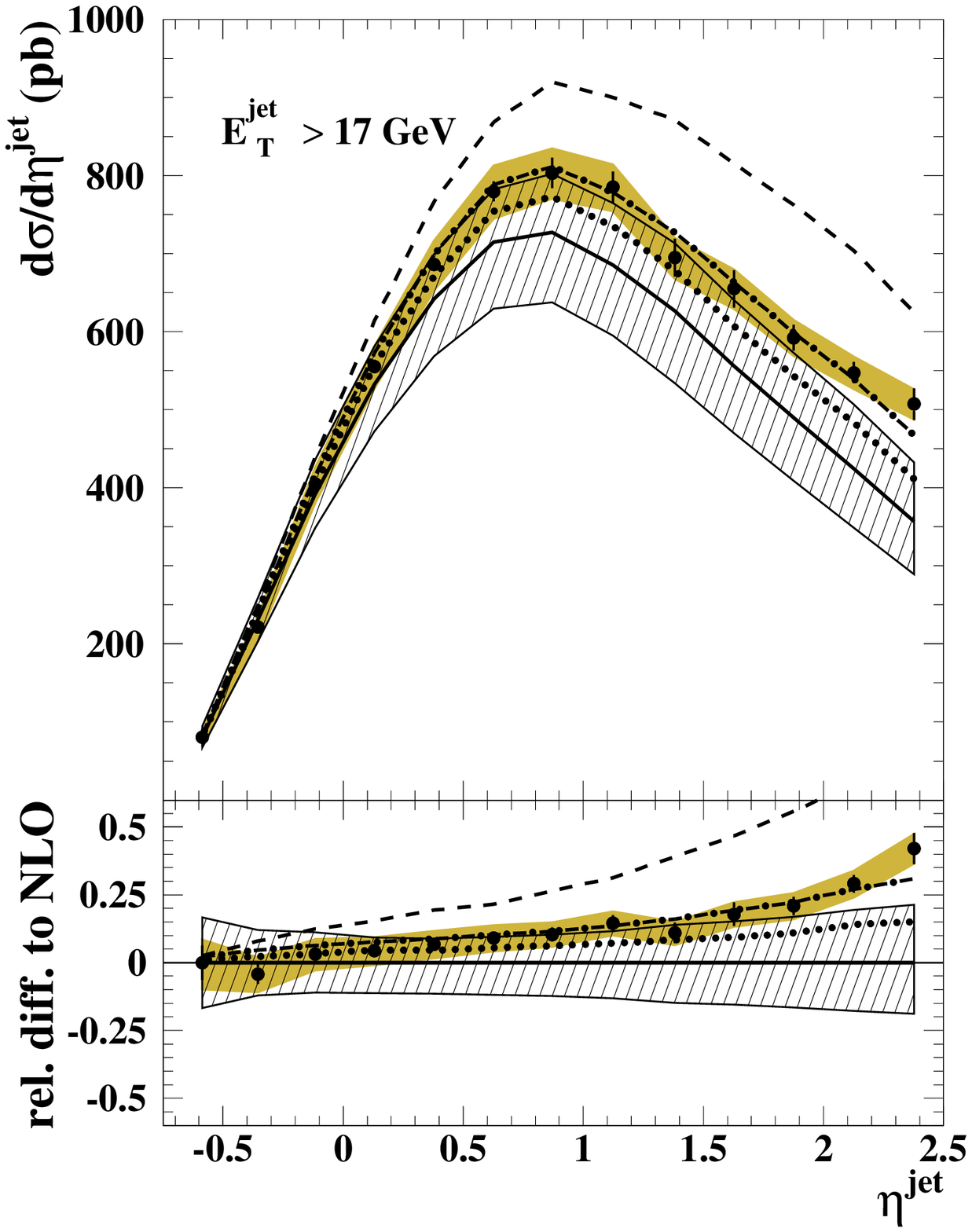,width=12cm}}
\put (6.8,10.7){{\bf\small (a)}}
\put (15.7,10.7){{\bf\small (b)}}
\end{picture}
\caption
{\it 
The measured differential cross-sections (a) $\set$ and (b) $\seta$
based on the $\kt$ jet algorithm for inclusive-jet photoproduction
with $\etjet>17$~GeV and $\etar$ (dots) in the kinematic region given
by $\q2<1$~\gev$^2$ and \wrn. For comparison, the NLO QCD
calculations including an estimation of non-perturbative effects (see
text) are also shown.
Other details as in the caption to Fig.~\ref{fig2}.
}
\label{fig4}
\vfill
\end{figure}

\begin{figure}[p]
\vfill
\setlength{\unitlength}{1.0cm}
\begin{picture} (18.0,15.0)
\put (0.45,2.0){\centerline{\epsfig{figure=\figdir DESY-12-045_0.eps,width=12cm}}}
\put (-2.0,0.0){\epsfig{figure=\figdir 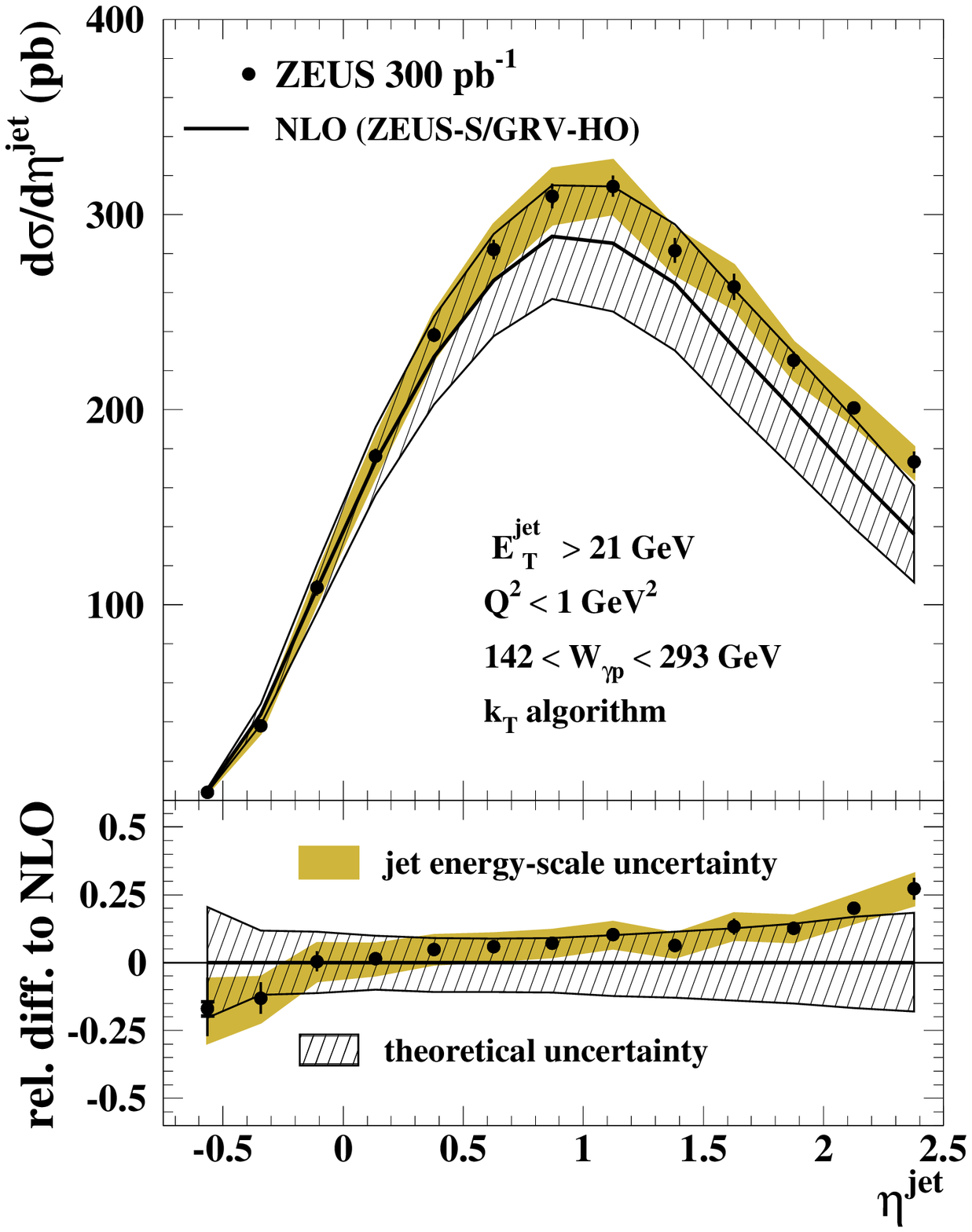,width=12cm}}
\put (7.0,0.0){\epsfig{figure=\figdir 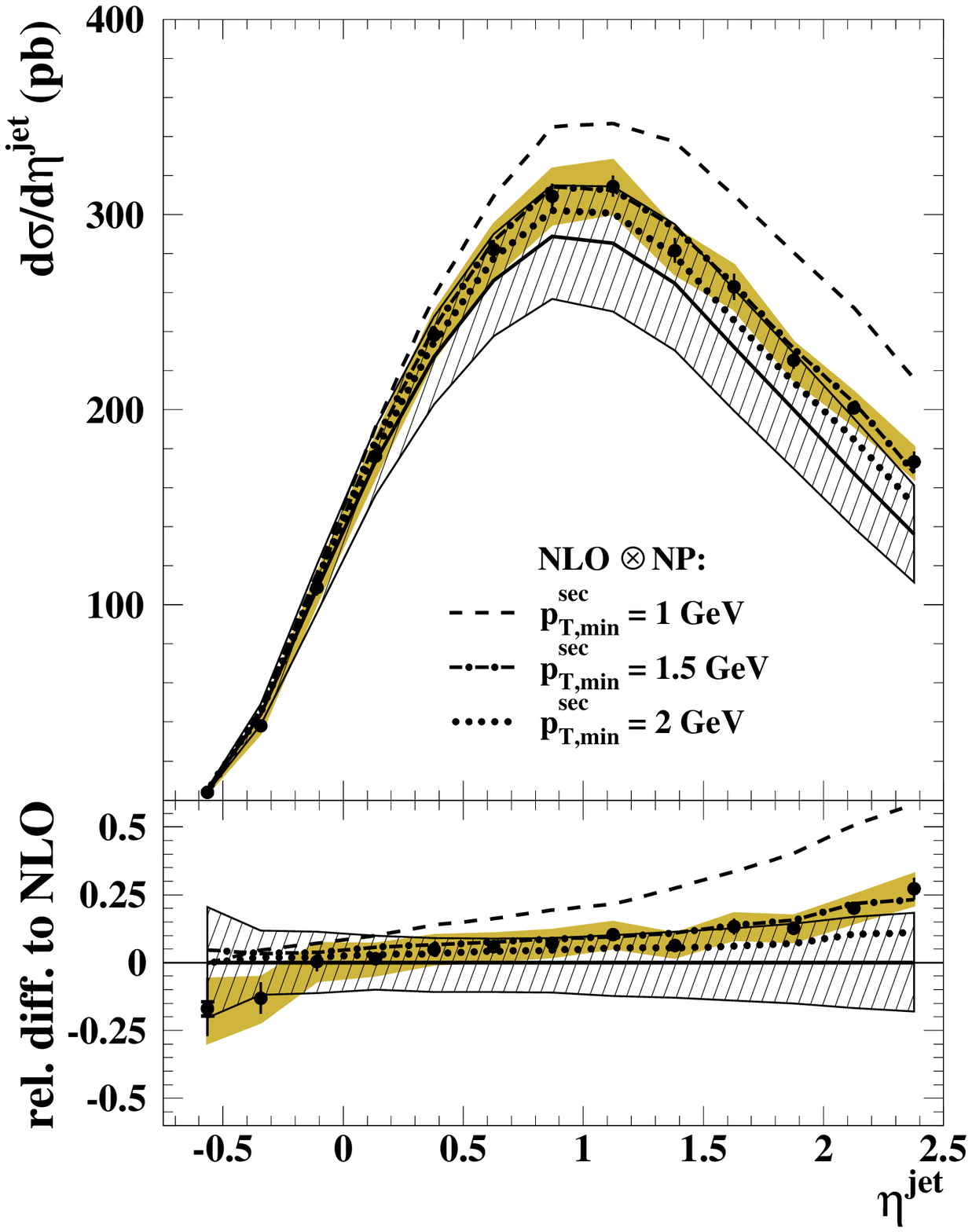,width=12cm}}
\put (6.8,10.7){{\bf\small (a)}}
\put (15.7,10.7){{\bf\small (b)}}
\end{picture}
\caption
{\it 
The measured differential cross-section $\seta$  based on the $\kt$
jet algorithm for inclusive-jet photoproduction with $\etjet>21$~GeV
(dots) in the kinematic region given by $\q2<1$~\gev$^2$ and \wrn.
In (b), the NLO QCD calculations including an estimation of
non-perturbative effects (see text) are also shown.
Other details as in the caption to Fig.~\ref{fig2}.
}
\label{fig5}
\vfill
\end{figure}

The influence of non-perturbative effects not related to hadronisation
in the predictions was investigated by using the NLO$\otimes$NP QCD
calculations (see Section~\ref{nlo}). The comparison of
these calculations to the data is shown in Fig.~\ref{fig4}. It is
observed that the NLO$\otimes$NP QCD calculations predict a larger jet
rate at low $\etjet$ and high $\etajet$, in the region where the NLO
QCD prediction fails to describe the data. The NLO$\otimes$NP QCD
prediction with $p_{T,{\rm min}}^{\rm sec}=1.5$ GeV is closest to the
data. These observations indicate the possible presence of effects
such as the underlying event in the data,
which are not included in the NLO QCD calculation.
These non-perturbative contributions are expected to be unrelated to the
hard scattering and approximately constant with the scale of the
interaction, so that the ratio of this non-perturbative contribution to
the jet transverse energy becomes smaller as $\etjet$ increases, as seen in 
Fig.~\ref{fig4}a. This is supported by the good description of the
data by the NLO QCD calculation for $\etjet>21$ GeV (Fig.~\ref{fig4}a)
and by the inclusive-jet cross section as a function of $\etajet$ for
$\etjet>21$ GeV (see Fig.~\ref{fig5} and Table~\ref{tab4}): the NLO
QCD calculation gives a good description of the data in the whole
measured range; in particular, discrepancies between data and theory
are no longer observed at high values of $\etajet$. In addition, the
differences between the NLO$\otimes$NP predictions with different 
$p_{T,{\rm min}}^{\rm sec}$ values become smaller, as seen in
Fig.~\ref{fig5}b.

\begin{figure}[p]
\vfill
\setlength{\unitlength}{1.0cm}
\begin{picture} (18.0,15.0)
\put (0.45,2.0){\centerline{\epsfig{figure=\figdir DESY-12-045_0.eps,width=12cm}}}
\put (-2.0,0.0){\epsfig{figure=\figdir 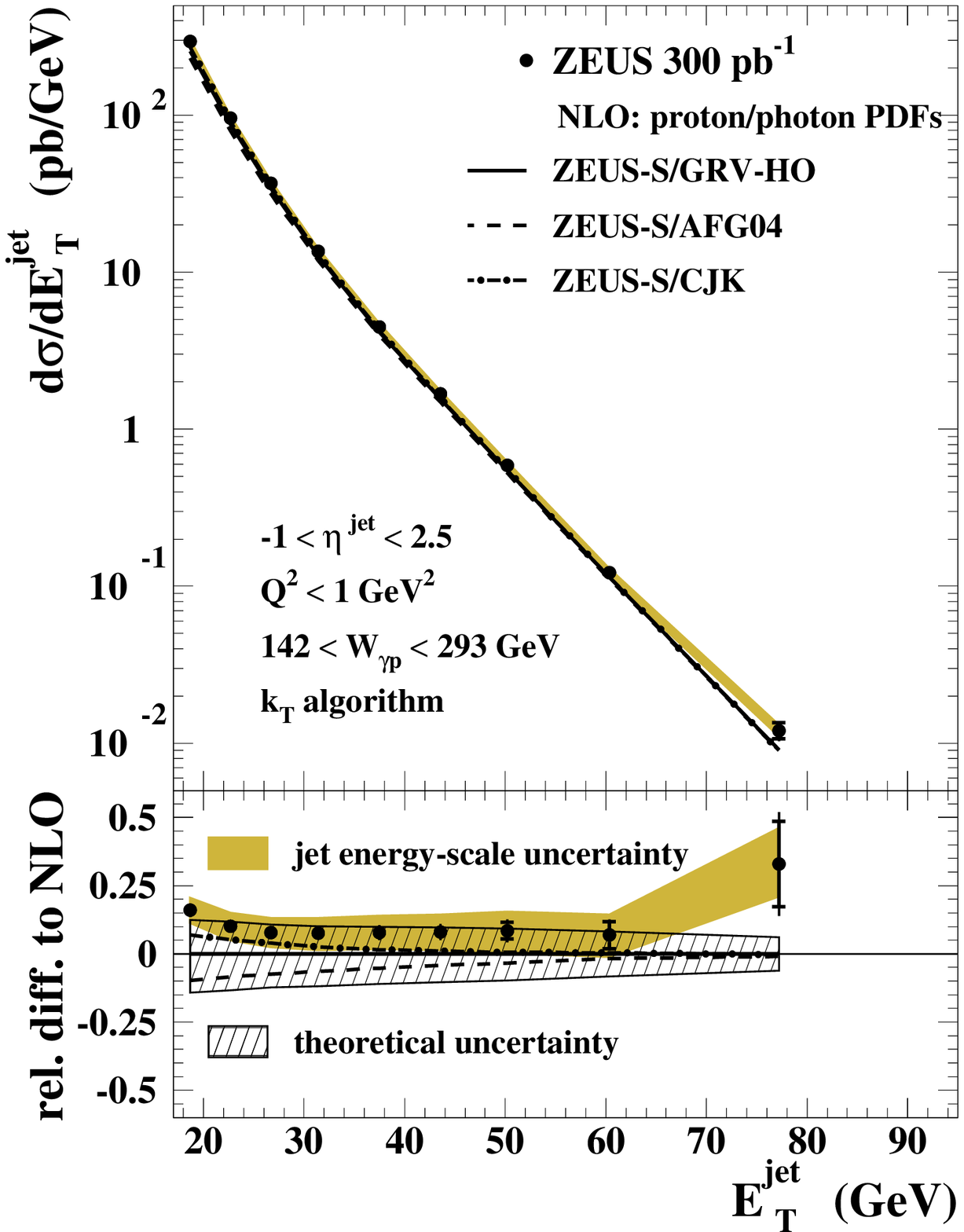,width=12cm}}
\put (7.0,0.0){\epsfig{figure=\figdir 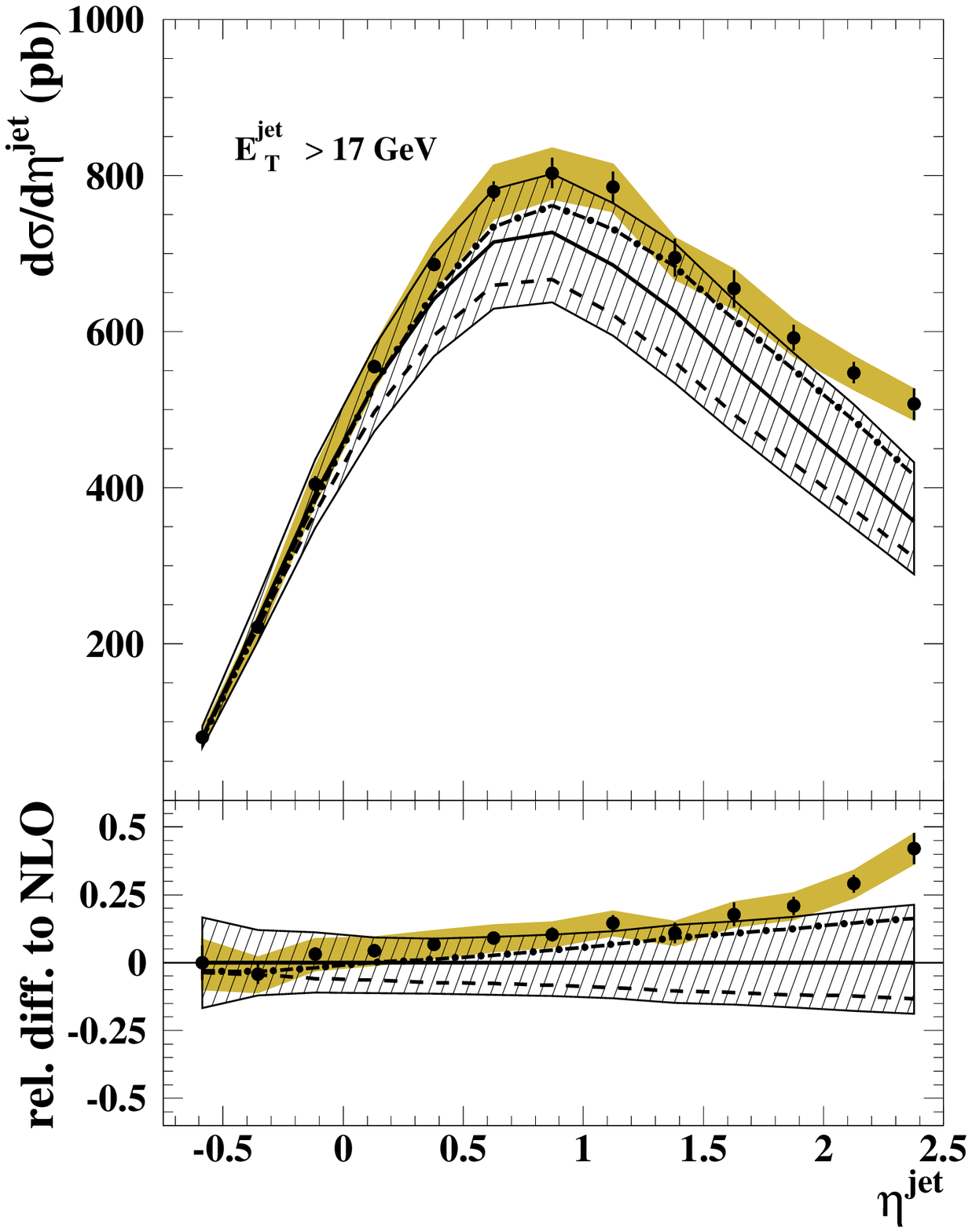,width=12cm}}
\put (6.9,10.7){{\bf\small (a)}}
\put (15.7,10.7){{\bf\small (b)}}
\end{picture}
\caption
{\it 
The measured differential cross-sections (a) $\set$ and (b) $\seta$
based on the $\kt$ jet algorithm for inclusive-jet photoproduction
with $\etjet>17$~GeV and $\etar$ (dots) in the kinematic region given
by $\q2<1$~\gev$^2$ and \wrn. For comparison, the NLO QCD
calculations using different parameterisations of the 
photon PDFs are also shown.
Other details as in the caption to Fig.~\ref{fig2}. 
}
\label{fig6}
\vfill
\end{figure}

The influence of the poorly constrained photon PDFs on the predictions
was investigated by comparing calculations based on different PDF sets
to the data. Figure~\ref{fig6} shows the measurements together with
the NLO QCD predictions using alternatively the AFG04 and CJK sets of
photon PDFs, together with the predictions based on the GRV-HO
set. Some differences are observed between the three predictions,
especially at low $\etjet$ and high $\etajet$. In particular, the
predictions based on AFG04 (CJK) are lower (higher) than those based
on GRV-HO. 

\begin{figure}[p]
\vfill
\setlength{\unitlength}{1.0cm}
\begin{picture} (18.0,15.0)
\put (0.45,2.0){\centerline{\epsfig{figure=\figdir DESY-12-045_0.eps,width=12cm}}}
\put (-2.0,0.0){\epsfig{figure=\figdir 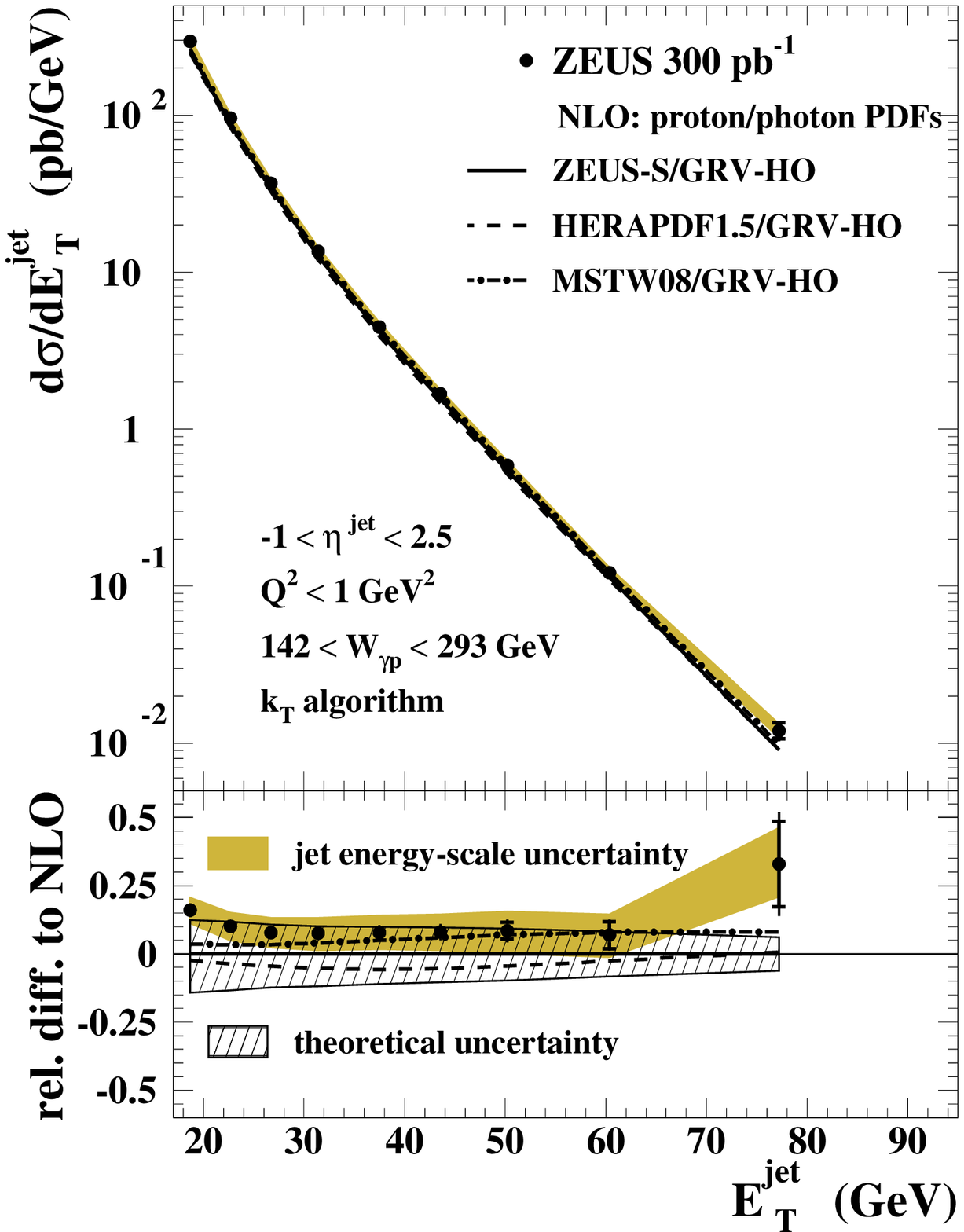,width=12cm}}
\put (7.0,0.0){\epsfig{figure=\figdir 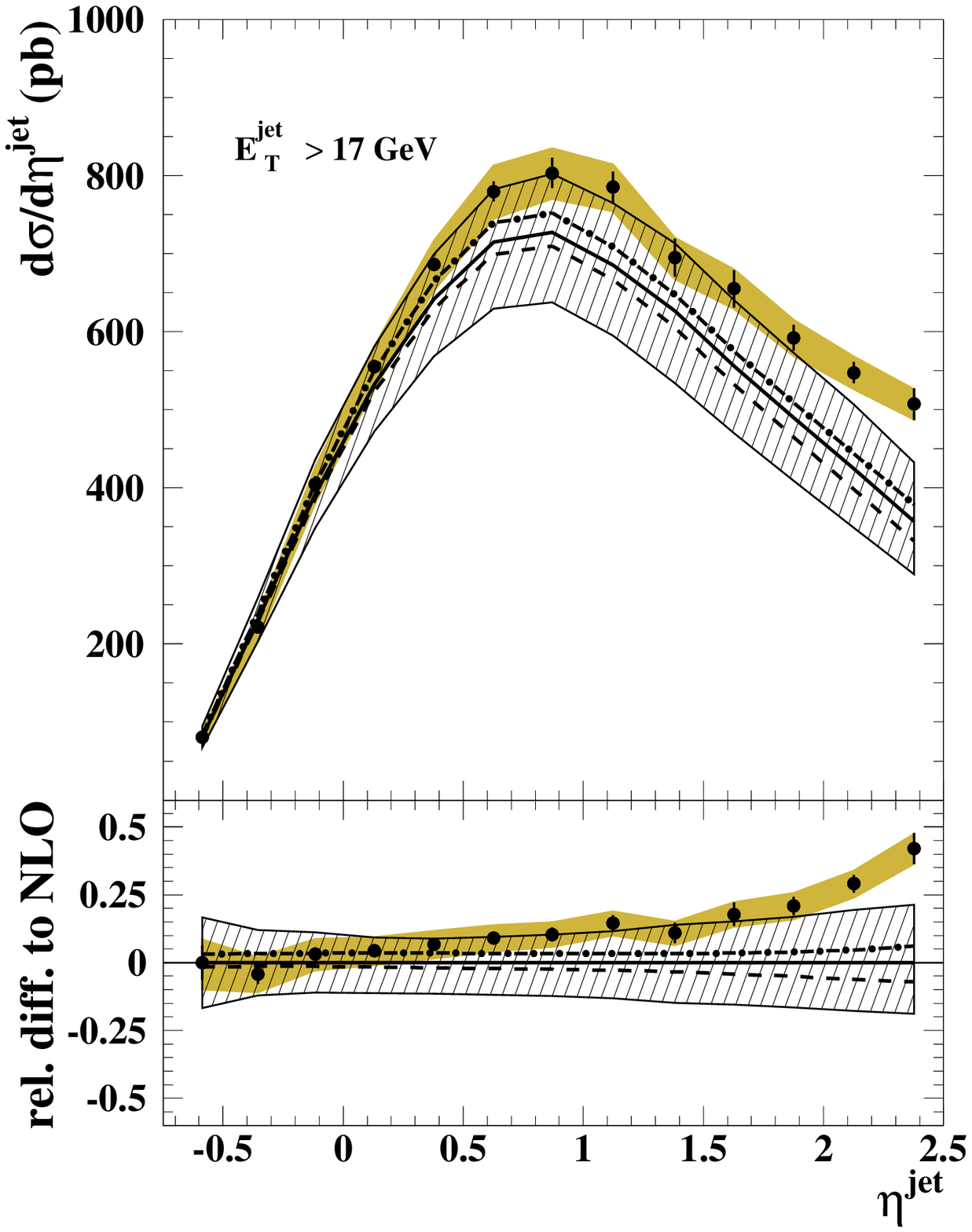,width=12cm}}
\put (6.9,10.7){{\bf\small (a)}}
\put (15.7,10.7){{\bf\small (b)}}
\end{picture}
\caption
{\it 
The measured differential cross-sections (a) $\set$ and (b) $\seta$
based on the $\kt$ jet algorithm for inclusive-jet photoproduction
with $\etjet>17$~GeV and $\etar$ (dots) in the kinematic region given
by $\q2<1$~\gev$^2$ and \wrn. For comparison, the NLO QCD
calculations using different parameterisations of the proton
PDFs are also shown.
Other details as in the caption to
Fig.~\ref{fig2}.
}
\label{fig7}
\vfill
\end{figure}

The influence of the proton PDFs on the predictions was investigated
by comparing calculations based on different PDF sets to the
data. Figure~\ref{fig7} shows the measurements together with 
the NLO QCD predictions using alternatively the
MSTW08~\cite{epj:c63:189} and HERAPDF1.5~\cite{jhep:1001:109} sets of
proton PDFs, together with the predictions based on the ZEUS-S set.
The prediction based on HERAPDF1.5 is lower than that based on
the ZEUS-S set in most of the phase space, whereas the MSTW08
prediction is higher at high $\etjet$. This region of phase space is
not well constrained since the main contribution comes from the
high-$x$ gluon density in the proton.

In summary, the measurements of inclusive-jet cross sections in
photoproduction have the potential to constrain the proton and
the photon PDFs. To study in more detail the sensitivity of the
inclusive-jet cross sections to the proton and photon PDFs and find 
the regions of phase space in which the data can add information to
constrain further these PDFs, double-differential cross sections were
measured and are presented in the next section.

\subsection{Double-differential cross sections}
\begin{figure}[p]
\vfill
\setlength{\unitlength}{1.0cm}
\begin{picture} (18.0,15.0)
\put (0.0,0.5){\centerline{\epsfig{figure=\figdir 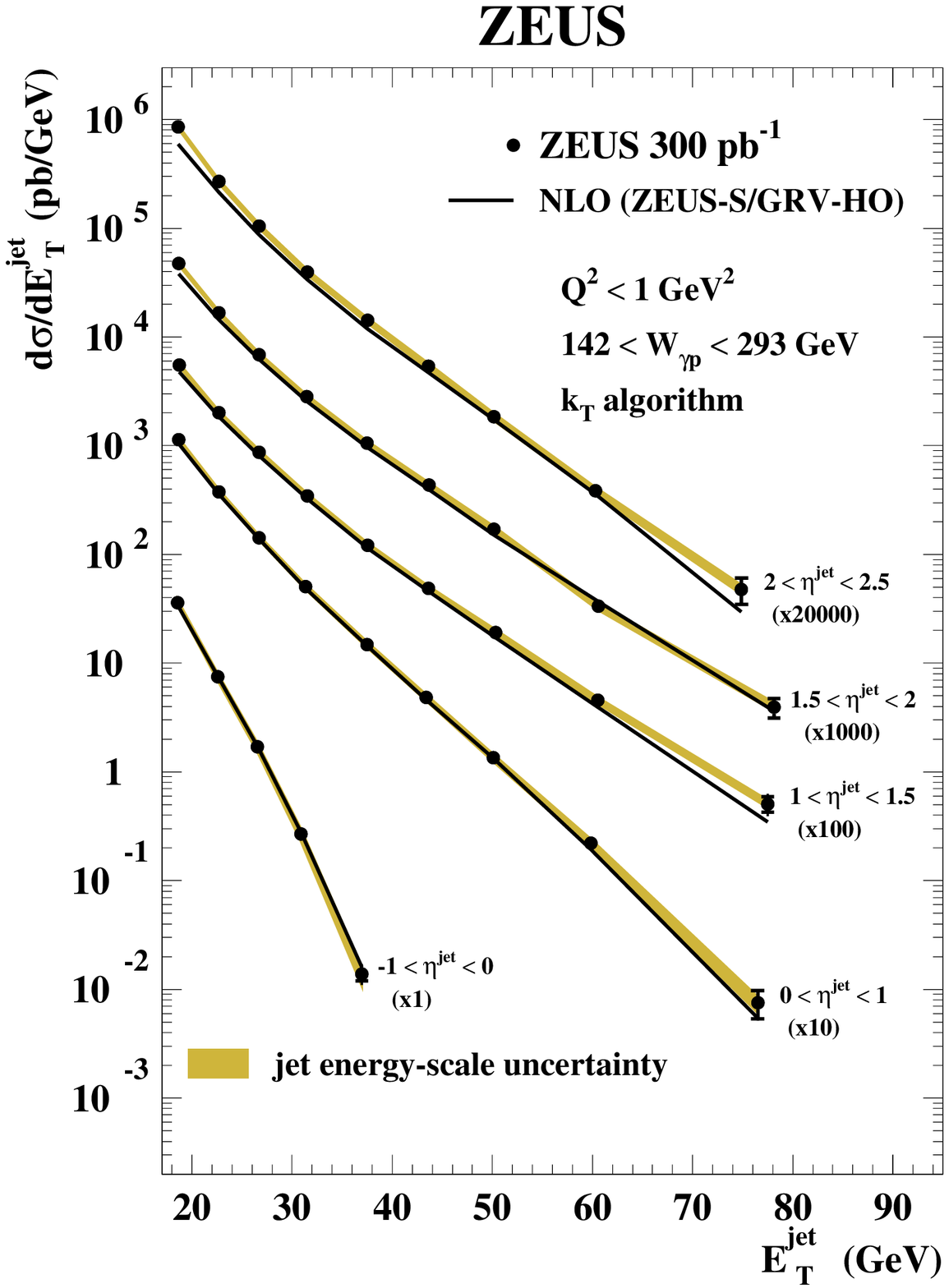,width=15cm}}}
\end{picture}
\vspace{-1.5cm}
\caption
{\it 
The measured differential cross-sections $\set$ based on the $\kt$ jet
algorithm for inclusive-jet photoproduction with $\etjet>17$~GeV in
different regions of $\etajet$ (dots) in the kinematic region given by
$\q2<1$~\gev$^2$ and \wrn. Each cross section has been
multiplied by the scale factor indicated in brackets to aid visibility.
Other details as in the caption to Fig.~\ref{fig2}.
}
\label{fig8}
\vfill
\end{figure}

The measurements of the inclusive-jet cross sections based on the
$\kt$ jet algorithm as functions of $\etjet$ in different regions of
$\etajet$ are presented in Fig.~\ref{fig8} and Tables~\ref{tab5} and
\ref{tab6}. The measured cross sections exhibit a steep fall-off
within the $\etjet$ range considered. The $\etjet$ dependence of the
cross section becomes less steep as $\etajet$ increases.

\begin{figure}[p]
\vfill
\setlength{\unitlength}{1.0cm}
\begin{picture} (18.0,15.0)
\put (0.0,0.5){\centerline{\epsfig{figure=\figdir 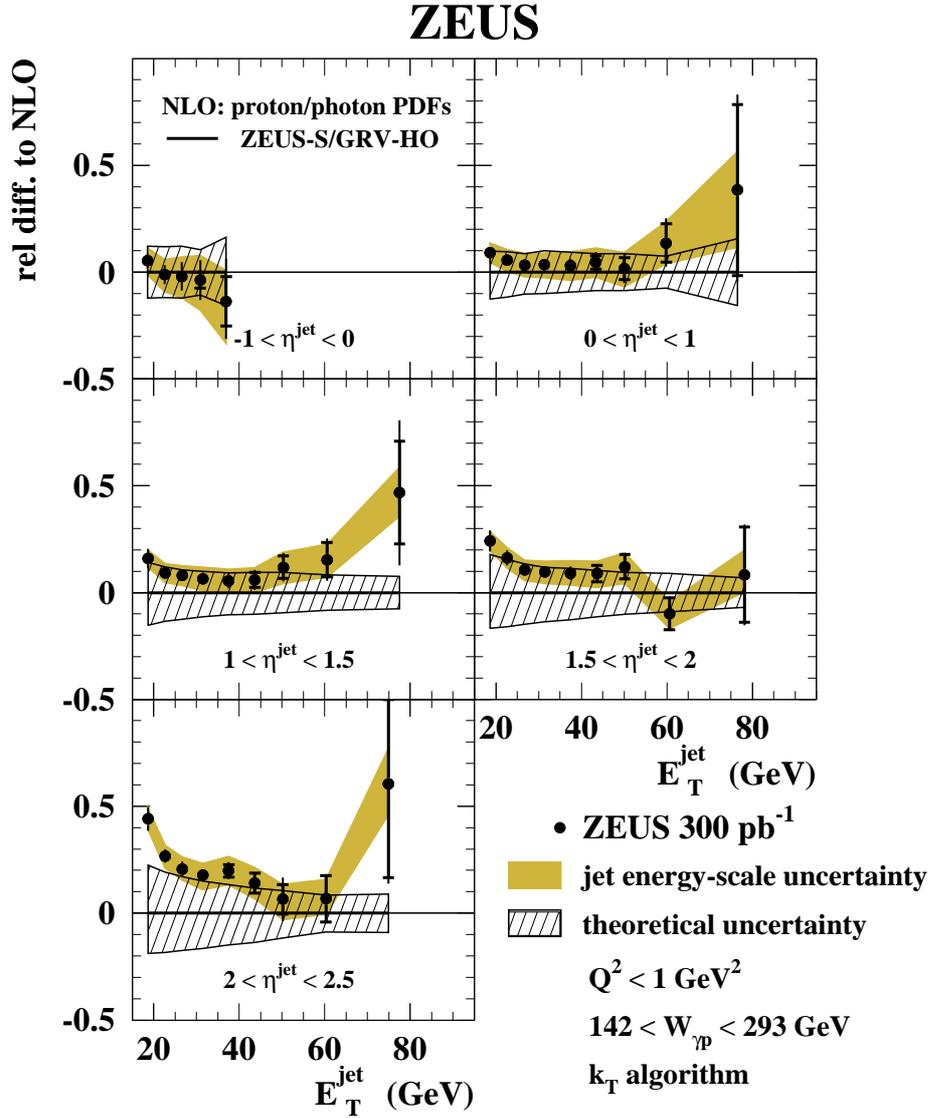,width=15cm}}}
\end{picture}
\vspace{-1.5cm}
\caption
{\it 
The relative differences between the measured differential cross-sections
$\set$ presented in Fig.~\ref{fig8} and the NLO QCD calculations (dots). 
Other details as in the caption to Fig.~\ref{fig2}.
}
\label{fig9}
\vfill
\end{figure}

\begin{figure}[p]
\vfill
\setlength{\unitlength}{1.0cm}
\begin{picture} (18.0,15.0)
\put (0.0,0.5){\centerline{\epsfig{figure=\figdir 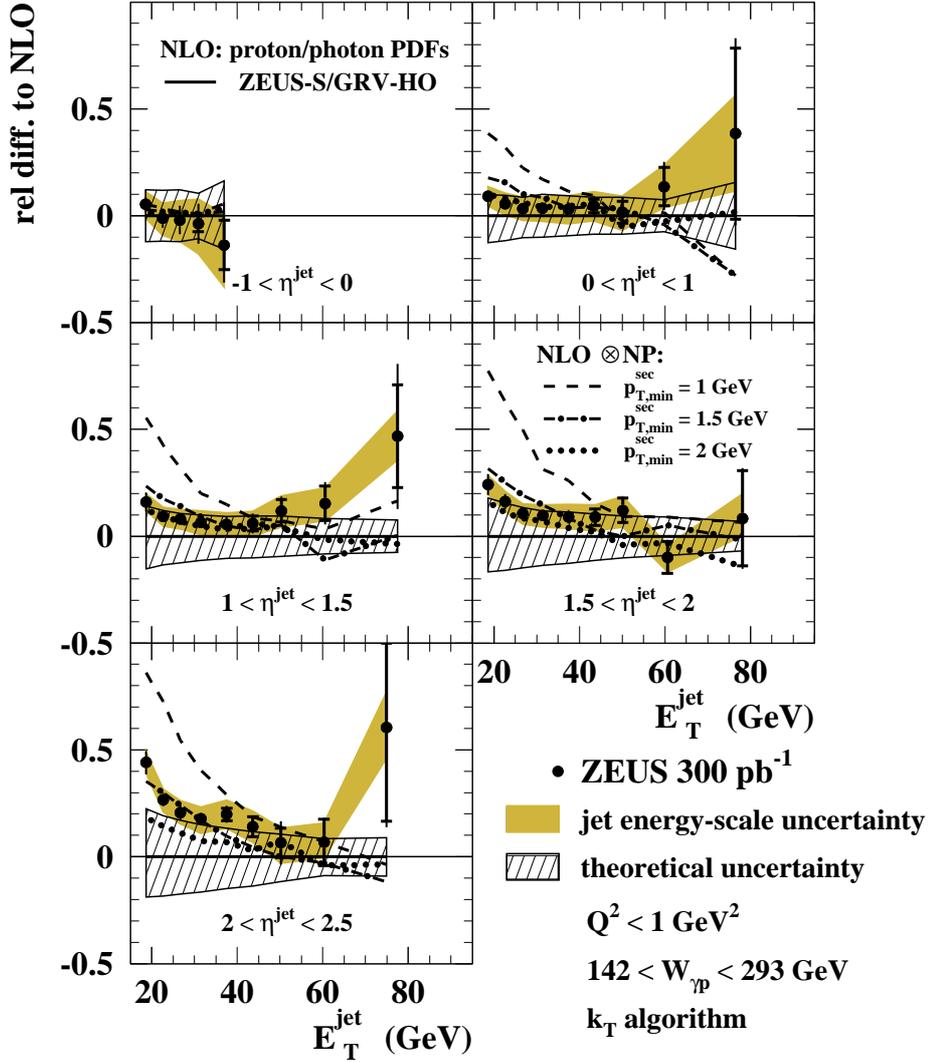,width=15cm}}}
\end{picture}
\vspace{-1.5cm}
\caption
{\it 
The relative differences between the measured differential cross-sections
$\set$ presented in Fig.~\ref{fig8} and the NLO QCD calculations
(dots). The relative differences between the predictions based on
the calculations including an estimation of non-perturbative effects
(see text) and the NLO QCD calculation are also shown.
Other details as in the caption to Fig.~\ref{fig2}.
}
\label{fig10}
\vfill
\end{figure}

The NLO QCD predictions are compared to the measurements in
Fig.~\ref{fig8}. They give a good description of the data, except at
low $\etjet$ and high $\etajet$. Figure~\ref{fig9} shows the relative
difference of the measured differential cross sections to the NLO QCD
calculations. The data are well described by the predictions for
$-1<\etajet<2$ in the whole $\etjet$ range measured.
For the region $2<\etajet<2.5$, where it
is observed that non-perturbative effects not related to hadronisation
(see Fig.~\ref{fig10}) might contribute significantly, the data are well
described only for $\etjet>21$~GeV.

\begin{figure}[p]
\vfill
\setlength{\unitlength}{1.0cm}
\begin{picture} (18.0,15.0)
\put (0.0,0.5){\centerline{\epsfig{figure=\figdir 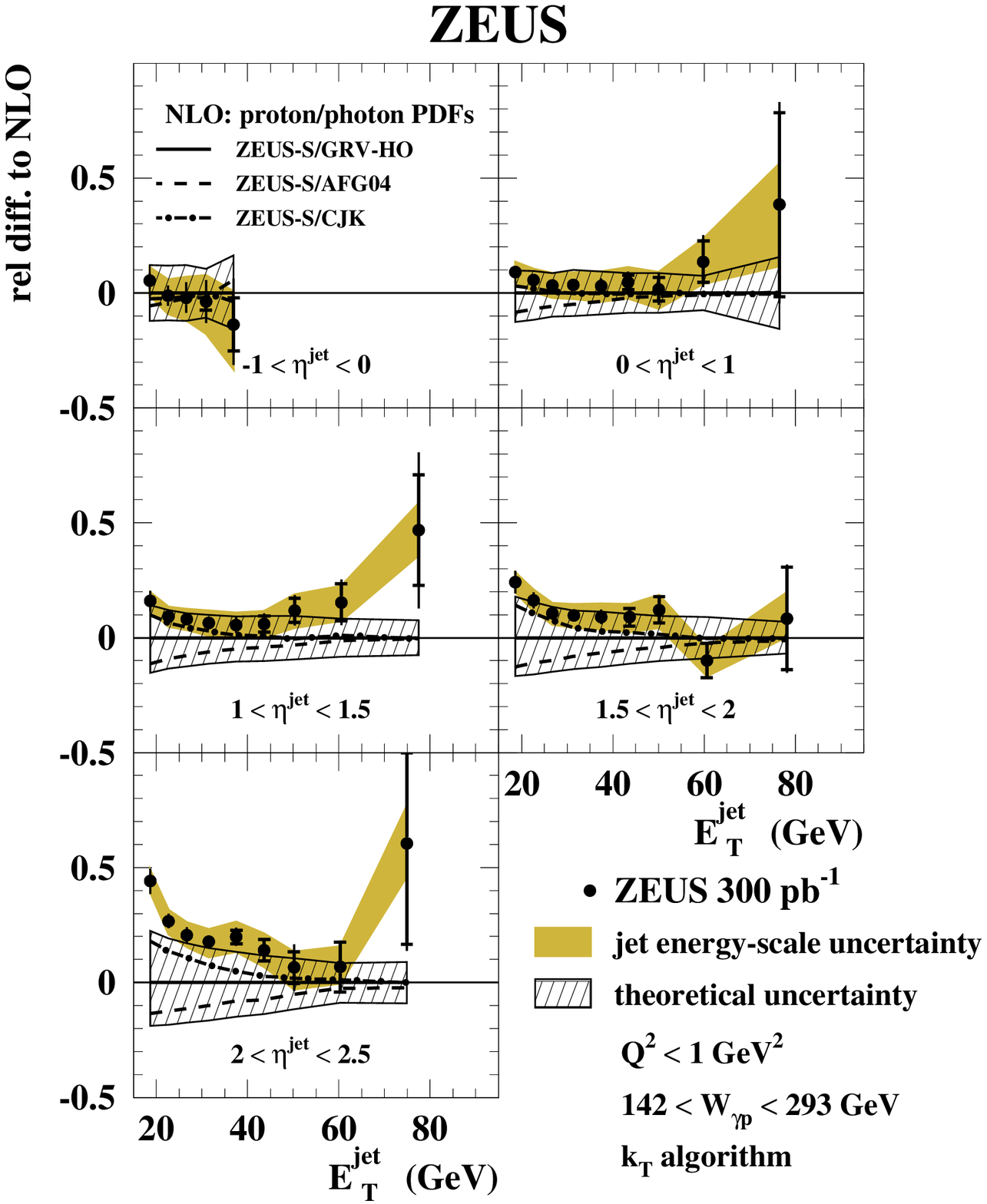,width=15cm}}}
\end{picture}
\vspace{-1.5cm}
\caption
{\it 
The relative differences between the measured differential cross-sections
$\set$ presented in Fig.~\ref{fig8} and the NLO QCD calculations
(dots). The relative differences between the predictions based on
different photon PDFs and that based on the
ZEUS-S/GRV-HO sets are also shown. 
Other details as in the caption to Fig.~\ref{fig2}.
}
\label{fig11}
\vfill
\end{figure}

\begin{figure}[p]
\vfill
\setlength{\unitlength}{1.0cm}
\begin{picture} (18.0,15.0)
\put (0.0,0.5){\centerline{\epsfig{figure=\figdir 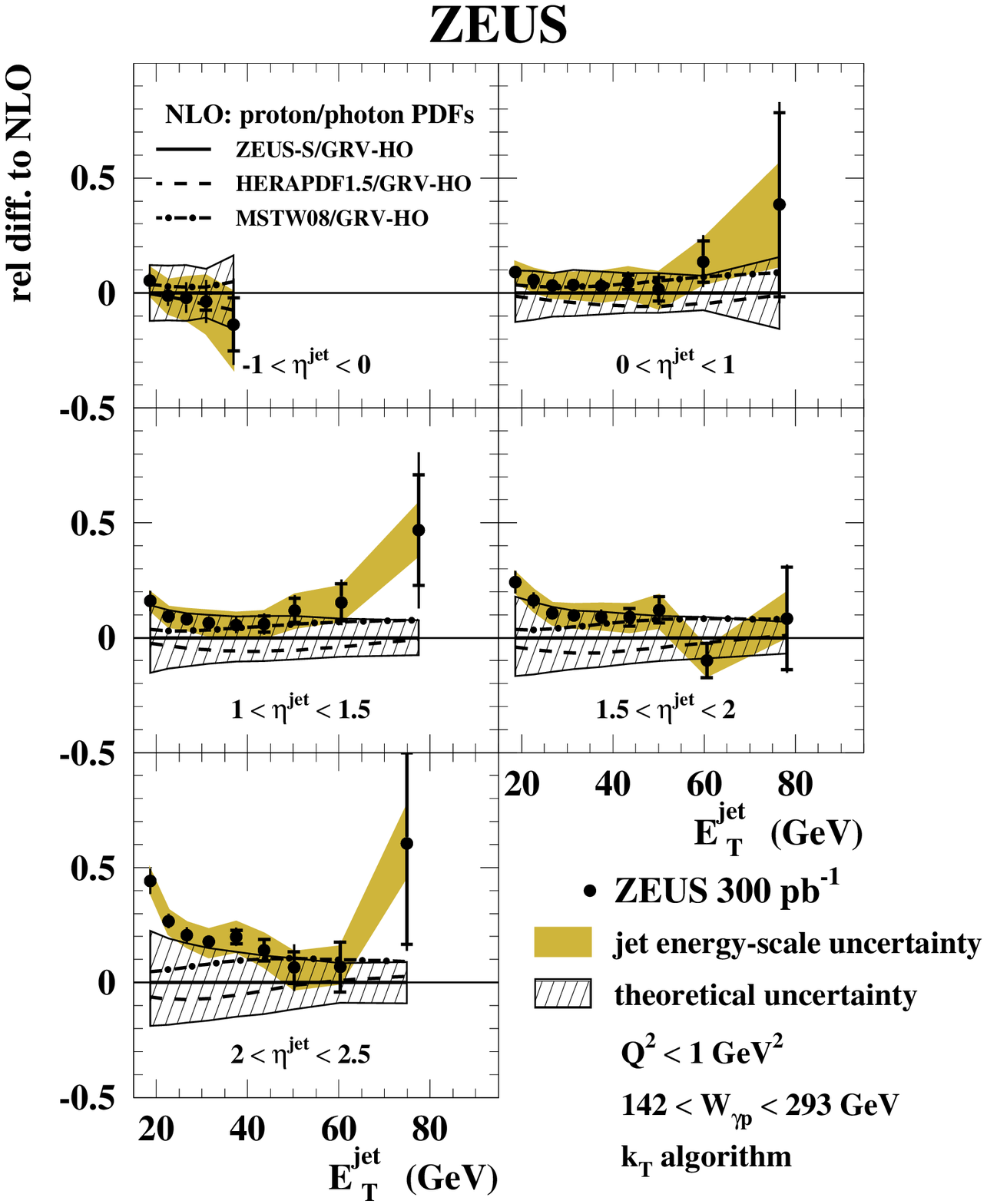,width=15cm}}}
\end{picture}
\vspace{-1.5cm}
\caption
{\it 
The relative differences between the measured differential cross-sections
$\set$ presented in Fig.~\ref{fig8} and the NLO QCD calculations
(dots). The relative differences between the predictions based on
different proton PDFs and that based on the
ZEUS-S/GRV-HO sets are also shown. 
Other details as in the caption to Fig.~\ref{fig2}.
}
\label{fig12}
\vfill
\end{figure}

Figures~\ref{fig11} and \ref{fig12} show the comparison between the
measured cross sections and the predictions based on different photon
and proton PDFs, respectively. As discussed above, differences at low
$\etjet$ and high $\etajet$ are observed between the predictions based
on GRV-HO, AFG04 and CJK. The latter gives predictions closest to the
data, especially in the region $2<\etajet<2.5$. The largest
differences between the predictions based on MSTW08 and ZEUS-S
are observed at high $\etjet$ for $\etajet>1$. The predictions based
on HERAPDF1.5 are lower than those based on ZEUS-S in most of the
phase-space region.

As discussed in Section~\ref{nlo}, the theoretical uncertainties are
dominated by the contribution from higher orders. This uncertainty
decreases as $\etjet$ increases. The contribution from the proton PDF
uncertainty is significant and approximately constant for $\etjet>30$
GeV; at high $\etjet$ values, the proton PDF uncertainty is of the
same order as that coming from higher orders. In these regions, in
which the gluon-induced contribution is still substantial and the
possible presence of non-perturbative effects is expected to be
minimised, the data have the potential to constrain the gluon density
in the proton. The uncertainty coming from the photon PDFs is largest
at low $\etjet$ and high $\etajet$ and approximately of the same order
as that coming from higher-order terms. Therefore, these high-precision
measurements also have the potential to constrain the photon PDFs in these
regions of phase space.

\subsection{Single-differential cross sections based on different  jet
  algorithms}
\begin{figure}[p]
\vfill
\setlength{\unitlength}{1.0cm}
\begin{picture} (18.0,15.0)
\put (0.0,2.5){\centerline{\epsfig{figure=\figdir DESY-12-045_0.eps,width=12cm}}}
\put (-2.0,0.0){\epsfig{figure=\figdir 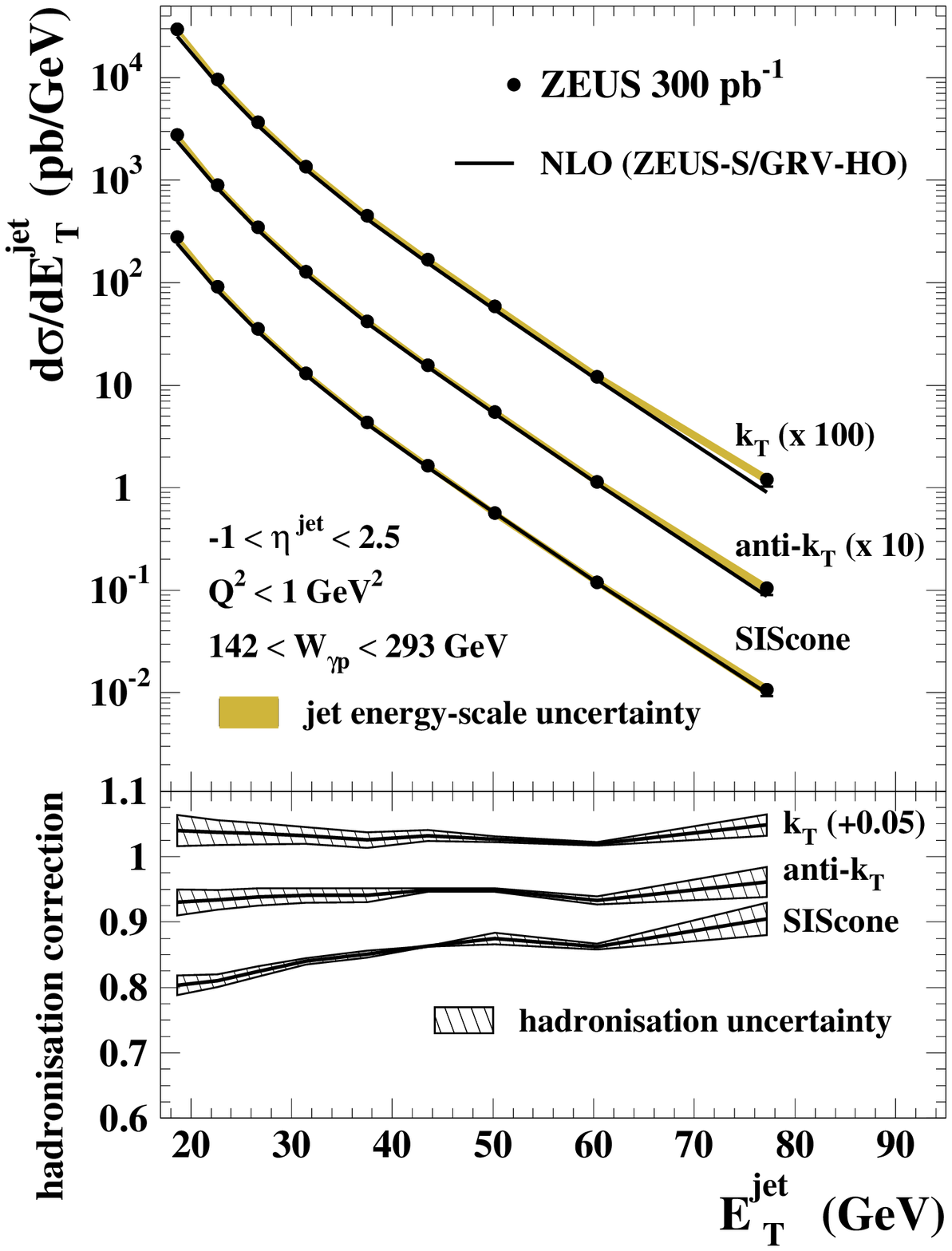,width=12cm}}
\put (7.0,0.0){\epsfig{figure=\figdir 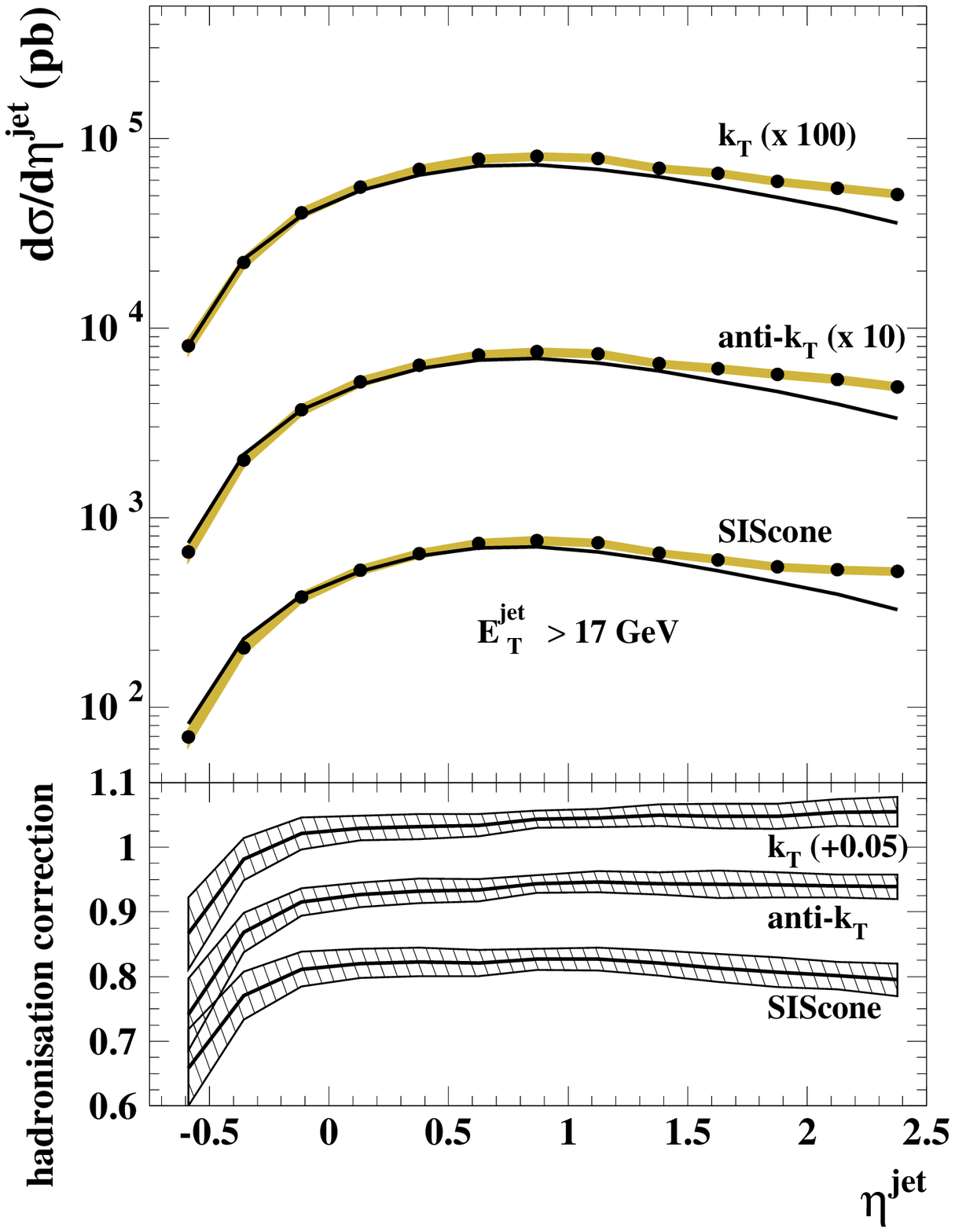,width=12cm}}
\put (6.8,11.3){\bf\small (a)}
\put (15.8,11.3){\bf\small (b)}
\end{picture}
\caption
{\it 
The measured differential cross-sections (a) $\set$ and (b) $\seta$
based on different jet algorithms for inclusive-jet photoproduction
with $\etjet>17$~GeV and $\etar$ (dots) in the kinematic region given
by $\q2<1$~\gev$^2$ and \wrn. The anti-$\kt$ and $\kt$ cross
sections were multiplied by the scale factors indicated in brackets to
aid visibility. The lower part of the figure shows the
hadronisation correction factors applied to the NLO calculations
together with their uncertainty (hatched bands) for each jet
algorithm; the hadronisation correction factor for the $\kt$ algorithm
was shifted by the value indicated in brackets to aid
visibility. Other details as in the caption to Fig.~\ref{fig2}.
}
\label{fig13}
\vfill
\end{figure}

The measurements of the inclusive-jet cross sections based on
the anti-$\kt$ and SIScone jet algorithms as functions of $\etjet$ and
$\etajet$ are presented in Fig.~\ref{fig13}, together with those based
on the $\kt$ algorithm. The measured cross sections are also given in
Tables~\ref{tab7} and \ref{tab8}. The measured $\set$ cross sections
exhibit a steep fall-off of over four orders of magnitude in the
$\etjet$ measured range. The measured $\seta$ cross sections display a
maximum around $\etajet\approx 1$. The measured cross sections using
the three jet algorithms have a similar shape, normalisation and
precision.

\begin{figure}[p]
\vfill
\setlength{\unitlength}{1.0cm}
\begin{picture} (18.0,15.0)
\put (0.0,5.5){\centerline{\epsfig{figure=\figdir DESY-12-045_0.eps,width=12cm}}}
\put (0.0,1.3){\centerline{\epsfig{figure=\figdir 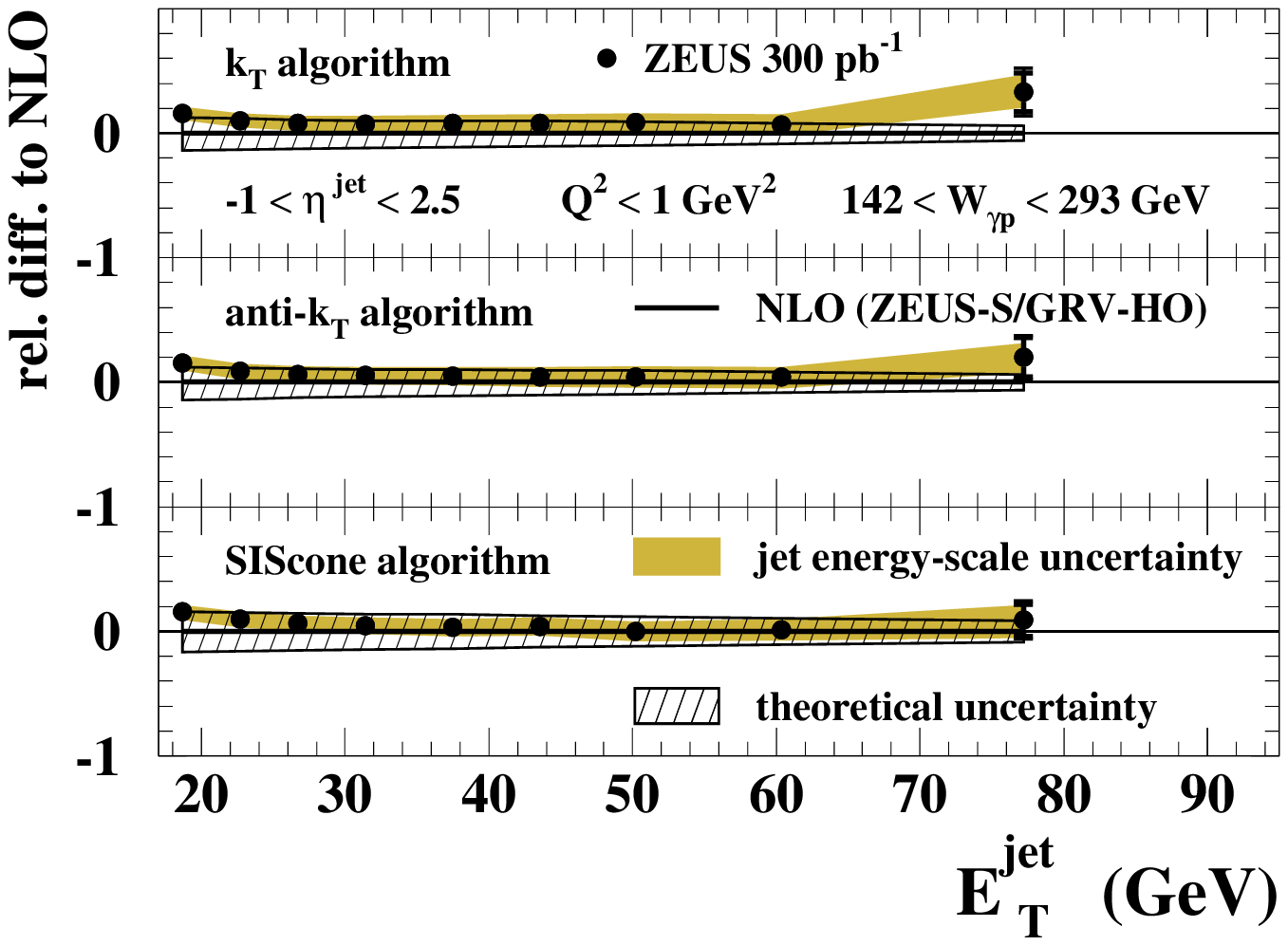,width=15cm}}}
\put (0.0,-6.5){\centerline{\epsfig{figure=\figdir 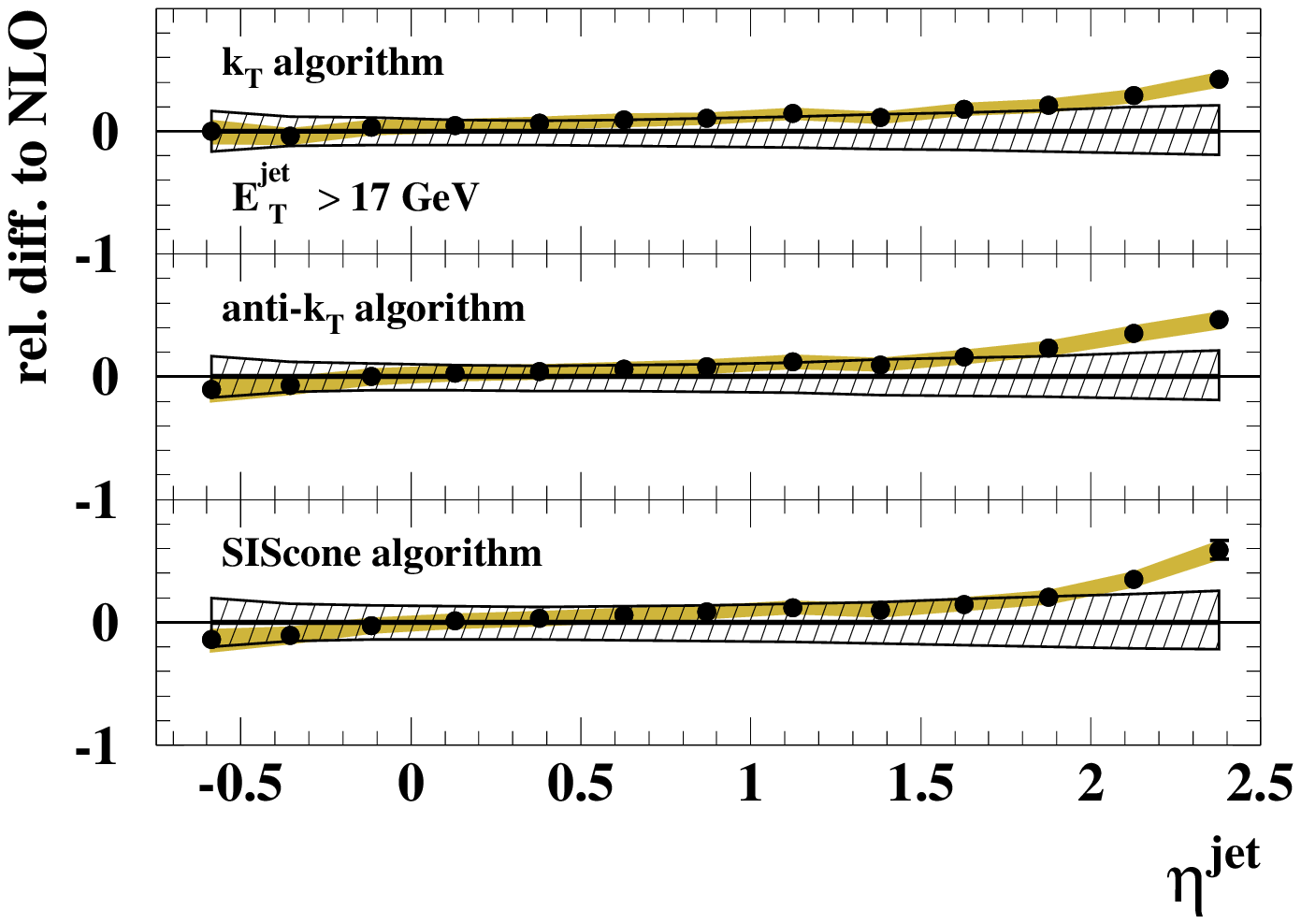,width=15cm}}}
\put (11.5,14.3){\bf\small (a)}
\put (11.5,6.6){\bf\small (b)}
\end{picture}
\caption
{\it 
The ratios between the measured cross-sections (a) $\set$ and (b)
$\seta$ and the NLO QCD calculations (dots) from Fig.~\ref{fig13}. 
Other details as in the caption to Fig.~\ref{fig2}.
}
\label{fig14}
\vfill
\end{figure}

The NLO QCD predictions are compared to the data in
Fig.~\ref{fig13}. The hadronisation correction factors applied to the
calculations and their uncertainties are also shown. It is seen that
the hadronisation correction factors are closest to (farthest from)
unity for the $\kt$ (SIScone) jet algorithm (see also
Tables~\ref{tab7} and \ref{tab8}). The ratios of the measured cross
sections to the NLO QCD calculations are shown in Fig.~\ref{fig14}
separately for each jet algorithm. The measured cross sections are well
reproduced by the calculations, except at high $\etajet$.

\begin{figure}[p]
\vfill
\setlength{\unitlength}{1.0cm}
\begin{picture} (18.0,15.0)
\put (0.0,5.5){\centerline{\epsfig{figure=\figdir DESY-12-045_0.eps,width=12cm}}}
\put (0.0,1.3){\centerline{\epsfig{figure=\figdir 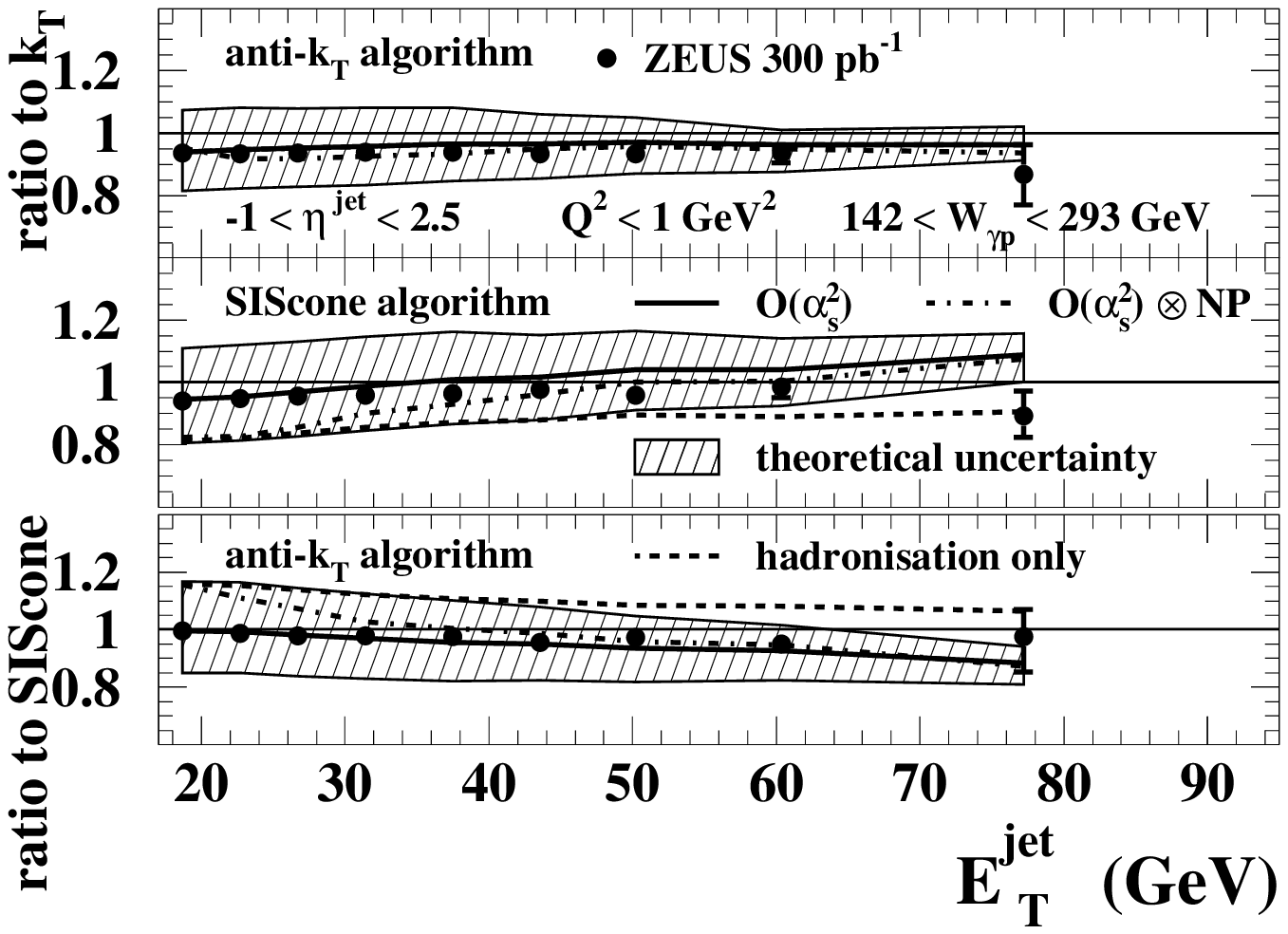,width=15cm}}}
\put (0.0,-6.5){\centerline{\epsfig{figure=\figdir 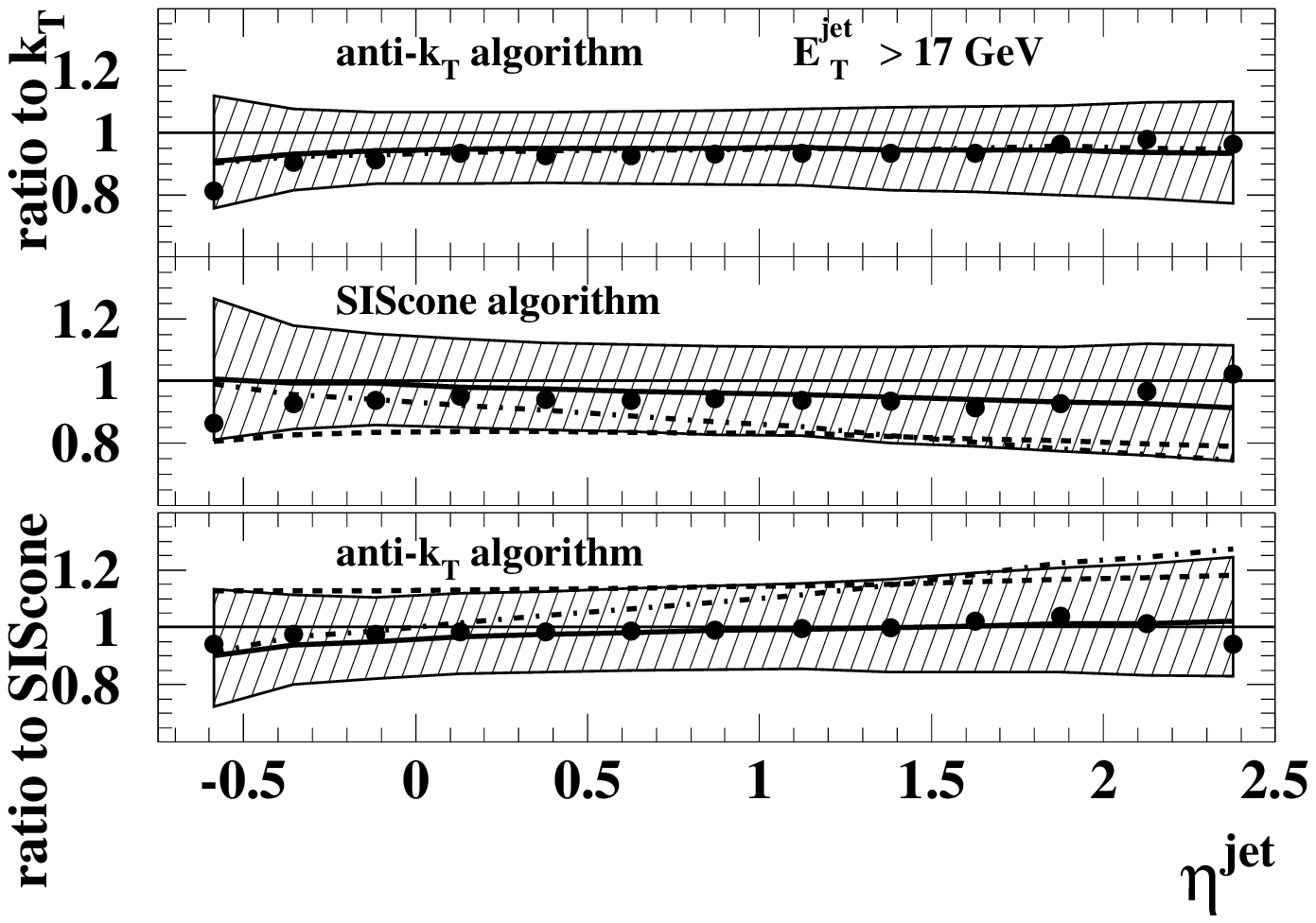,width=15cm}}}
\put (11.5,14.3){\bf\small (a)}
\put (11.5,6.6){\bf\small (b)}
\end{picture}
\caption
{\it 
The ratios of the measured cross-sections anti-$\kt$/$\kt$,
SIScone/$\kt$ and anti-$\kt$/SIScone (dots) as functions of
(a) $\etjet$ and (b) $\etajet$. In these plots, the outer error bars
also include the uncertainty on the absolute energy scale of the jets.
The predicted ratios based on calculations which include up to
$\oass$ terms are also shown (solid lines). The hatched bands display
the theoretical uncertainty on the ratio. The dashed lines indicate
the ratios of the hadronisation correction factors and the dash-dotted
lines represent the ratios of the NLO QCD calculations including an
estimation of non-perturbative effects (see text).
Other details as in the caption to Fig.~\ref{fig2}.
}
\label{fig15}
\vfill
\end{figure}

The ratios of the cross-sections anti-$\kt$/$\kt$, SIScone/$\kt$ and
anti-$\kt$/SIScone were studied to compare the jet algorithms in more
detail. These ratios allow, in particular, a stringent test of the
description of the differences between jet algorithms in terms of
parton radiation due to the partial cancellation of experimental and
theoretical uncertainties. The measured ratios are shown in
Fig.~\ref{fig15}. In these ratios, the statistical correlations among
the event samples as well as those among the jets in the same event
were taken into account in the estimation of the statistical
uncertainties; most of the systematic uncertainties, including that
due to the jet energy scale, cancel out. The measurements show that the
cross sections based on the anti-$\kt$ algorithm are similar in shape
to those based on the $\kt$ algorithm but $\approx 6\%$ lower and that
the cross sections based on the SIScone have a slightly different
shape than those based on the $\kt$ and anti-$\kt$ algorithms.

The pQCD predictions including up to $\oass$ terms for the ratios of
the cross sections are also shown in Fig.~\ref{fig15}. In the
estimation of the total theoretical uncertainty of the predicted
ratios, all the theoretical contributions were assumed to be
correlated except those due to terms beyond $\oass$ and to the
modelling of the QCD cascade and hadronisation. Figure~\ref{fig15}
also includes the ratio of the hadronisation correction factors
applied to the $\oass$ predictions. The predictions for the anti-$\kt$
and $\kt$ algorithms are identical at $\oass$ and, thus, the predicted
anti-$\kt$/$\kt$ ratio coincides with the ratio of the hadronisation
corrections for both algorithms. The predictions for the $\kt$ and
anti-$\kt$ algorithms are expected to start to differ at $\oasss$ and,
conservatively, the uncertainties due to higher-order terms were
assumed to be uncorrelated in the ratio; otherwise, a coherent
variation of the renormalisation scale in the ratio would yield an
unrealistic zero contribution. In the case of the SIScone/$\kt$ and
anti-$\kt$/SIScone, the $\oass$ predictions are different and the
ratios differ significantly from the ratios of the hadronisation
correction factors. The measured ratios are well described by the
calculations including terms up to $\oass$ within the experimental and
theoretical uncertainties, which are dominated by the uncertainty due
to higher orders. 

The differences in the influence of non-perturbative
effects not related to hadronisation for the different jet algorithms
were investigated by comparing the ratios of the NLO$\otimes$NP
predictions with $p_{T,{\rm min}}^{\rm sec}=1.5$~GeV, which are also
shown in Fig.~\ref{fig15}. For the 
anti-$\kt$/$\kt$ ratio, the NLO and the NLO$\otimes$NP predictions are
very similar, which indicates that the non-perturbative effects affect
the jets in the same way. However,
in the SIScone/$\kt$ and anti-$\kt$/SIScone ratios, the ratios of the
NLO$\otimes$NP predictions differ from the ratios of the NLO
calculations at low $\etjet$ and high $\etajet$; these differences,
which are located in the regions of phase space where the NLO
calculations fail to describe the data, show that there is a
dependence on the details of the jet reconstruction concerning
non-perturbative effects not related to hadronisation. These
differences are at most of the same size as the theoretical
uncertainties.

In summary, it is concluded that the measured inclusive-jet cross
sections based on the $\kt$, anti-$\kt$ and SIScone jet algorithms are
well described by the NLO QCD calculations, except at high
$\etajet$. The data and the calculations for the three jet algorithms
have a similar experimental and theoretical precision. Furthermore,
the measured ratios are well described by the predictions including up
to $\oass$ terms, demonstrating the ability of the pQCD calculations
with up to three partons in the final state to account adequately for
the differences between the SIScone and the $\kt$ or anti-$\kt$ jet
algorithms.

\subsection{Determination of {\boldmath $\asz$}}
The measured single-differential cross-sections $\set$ based on the
three jet algorithms were used to determine values of $\asz$ using the
method presented previously~\cite{pl:b547:164}. The NLO QCD
calculations were performed using five different sets of the ZEUS-S
proton PDFs which were determined from global fits assuming different
values of $\asz$, namely $\asz=0.115$, $0.117$, $0.119$, $0.121$ and
$0.123$. The GRV-HO set was used as default for the photon PDFs. The
value of $\asz$ used in each calculation was that associated with the
corresponding set of proton PDFs.

The $\asz$ dependence of the predicted cross sections in each bin $i$
of $\etjet$ was parameterised according to

$$\left [ d\sigma/d\etjet(\asz) \right ]_i=C_1^i\asz+C_2^i\as^2(\mz),$$

where $C_1^i$ and $C_2^i$ were determined from a $\chi^2$ fit to
the NLO QCD calculations. The value of $\asz$ was determined by a
$\chi^2$ fit to the measured $\set$ values. In the fitting procedure,
the running of $\as$ as predicted by QCD was assumed. Only the
measurements for $\etjet>21$ GeV were used in the fit to minimise the
effects of a possible non-perturbative contribution in addition to
that of hadronisation and the uncertainty coming from higher
orders. In addition, the fit was restricted to $\etjet<71$ GeV because
of the relatively large uncertainty coming from the proton PDFs for
higher $\etjet$ values.

The experimental uncertainties on the extracted values of $\asz$ were
evaluated by repeating the analysis for each systematic check
presented in Section~\ref{expunc}. The overall normalisation
uncertainty from the luminosity determination was also included. The
largest contribution comes from the uncertainty in the absolute
jet energy scale. The theoretical uncertainties were evaluated as 
described in Section~\ref{thun}. The largest contribution arises from
the terms beyond NLO, which was estimated by using the method of Jones
\etal~\cite{jhep:0312:007}. The uncertainty due to the photon PDFs is
of the same order as that arising from higher orders. The uncertainty
due to the proton PDFs and that arising from the hadronisation effects
were also estimated. All uncertainties are listed separately in
Table~\ref{tab9}.

As a cross-check, $\asz$ was determined by using NLO QCD calculations
based on the CTEQ6.1~\cite{jhep:0207:012,*jhep:0310:046}
(MSTW08) sets of proton PDFs. The values obtained are consistent
within $1.0\ (1.0),\ 0.9\ (0.9)$ and $0.6\ (0.8)\%$ with those based
on ZEUS-S for the $\kt$, anti-$\kt$ and SIScone determinations,
respectively. The uncertainty arising from the proton PDFs was
estimated to be $\pm 1.3\ (0.9,\ 0.5)\%$ for the $\kt$ and anti-$\kt$ and
$\pm 1.2\ (0.8,\ 0.4)\%$ for the SIScone determinations using the
results of the CTEQ6.1 (MSTW08nlo90cl, MSTW08nlo68cl) analysis.

The values of $\asz$ obtained from the measured $\set$ are

\begin{eqnarray}
\asz|_{\kt} &=& 0.1206\ ^{+0.0023}_{-0.0022}\ {\rm (exp.)}\ ^{+0.0042}_{-0.0035}\ {\rm (th.)},\nonumber\\
\asz|_{{\rm anti-}\kt} &=& 0.1198\ ^{+0.0023}_{-0.0022}\ {\rm (exp.)}\ ^{+0.0041}_{-0.0034}\ {\rm (th.)},\nonumber\\
\asz|_{\rm SIScone} &=& 0.1196\ ^{+0.0022}_{-0.0021}\ {\rm (exp.)}\ ^{+0.0046}_{-0.0043}\ {\rm (th.)}.\nonumber
\end{eqnarray}

\begin{figure}[p]
\vfill
\setlength{\unitlength}{1.0cm}
\begin{picture} (18.0,15.0)
\put (0.0,0.5){\centerline{\epsfig{figure=\figdir 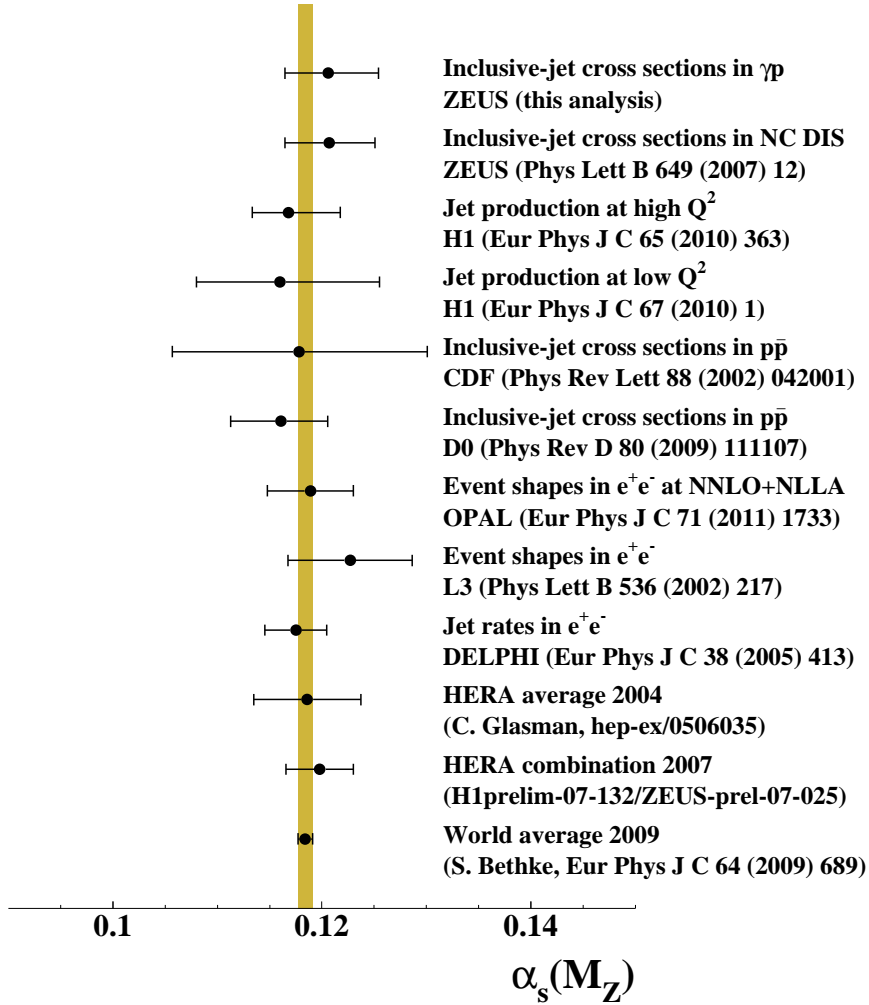,width=15cm}}}
\end{picture}
\vspace{-1.5cm}
\caption
{\it 
Extracted $\asz$ value from this analysis (upper dot). For comparison,
determinations from other experiments and reactions, the HERA
average 2004, the HERA combination 2007 and the world average 2009 are
also shown. The horizontal error bars represent the experimental and
theoretical uncertainties added in quadrature. The shaded band
represents the uncertainty of the world average.
}
\label{fig16}
\vfill
\end{figure}

The value of $\asz$ determined from the anti-$\kt$ (SIScone) measurements
is consistent with that obtained from the $\kt$ analysis within
$-0.7\ (-0.8)\%$. The total uncertainty from each determination, as
shown in Table~\ref{tab9}, indicates that the performance of the three jet
algorithms is similar. These determinations are consistent with
previous determinations in NC
DIS~\cite{pl:b649:12,epj:c65:363,epj:c67:1} at HERA, 
with the results obtained in $\pp$
collisions~\cite{prl:88:042001,*pr:d80:111107} and have a precision
comparable to those obtained in individual determinations from $\ele$
experiments~\cite{pl:b536:217,*epj:c38:413,*epj:c71:1733}. These
values are also consistent with the world average~\cite{jp:g26:r27},
as well as with the HERA 2004 average~\cite{proc:dis:2005:689}
and the HERA 2007 combined value~\cite{eps:2007:022013}.
Figure~\ref{fig16} shows the value of $\asz$ determined from the
$\kt$-based analysis together with determinations from other
experiments and the HERA and world averages.

\subsection{Energy-scale dependence of $\as$}
The energy-scale dependence of $\as$ was determined from a NLO QCD fit
to the measured $\set$ cross section. Values of $\as$ were extracted
at each mean value of measured $\etjet$ without assuming the running
of $\as$. The method employed was the same as described above, but
parameterising the $\as$ dependence of $\set$ in terms of
$\as(\langle\etjet\rangle)$ instead of $\asz$, where
$\langle\etjet\rangle$ is the average $\etjet$ of the data in each
bin.

\begin{figure}[p]
\vfill
\setlength{\unitlength}{1.0cm}
\begin{picture} (18.0,18.0)
\put (-1.0,0.0){\epsfig{figure=\figdir 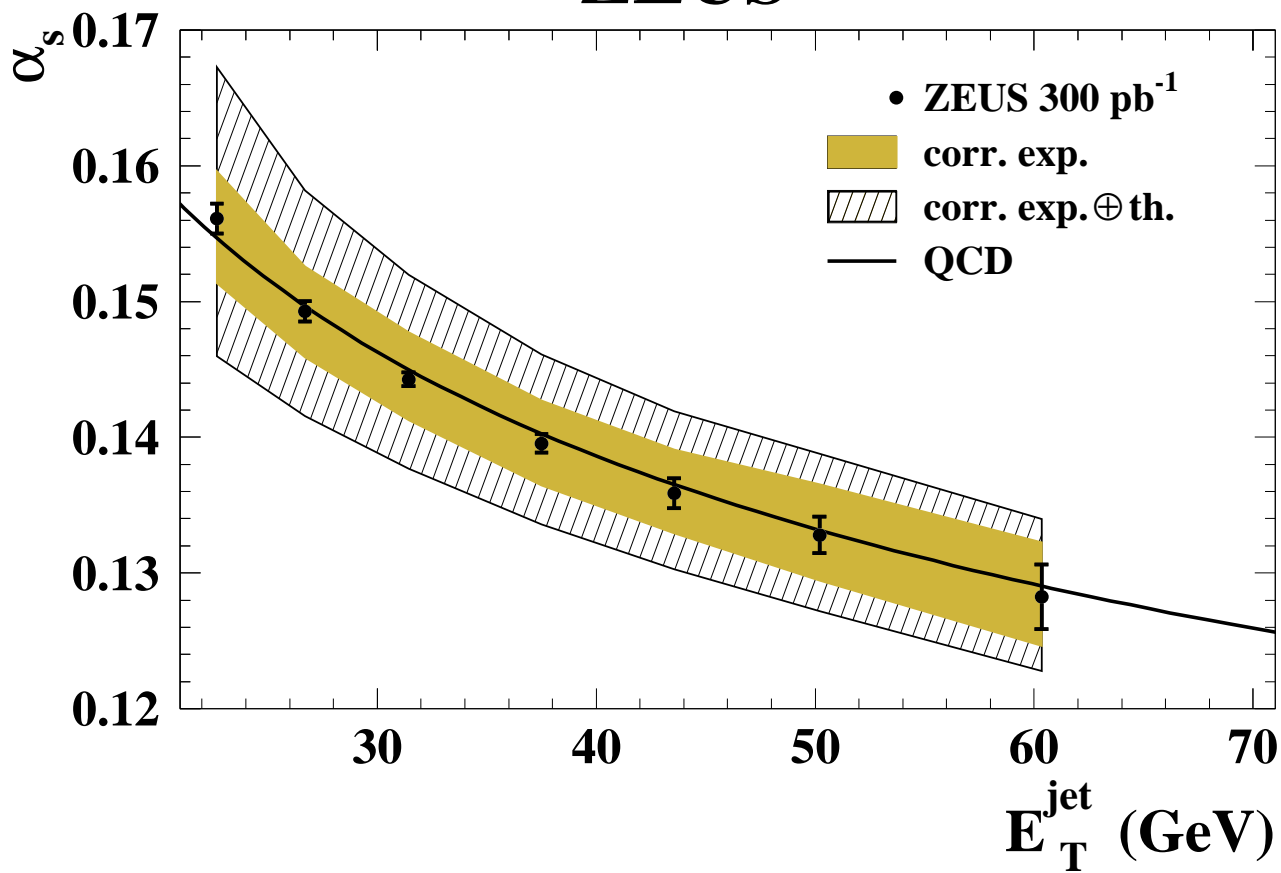,width=18cm}}
\end{picture}
\vspace{-1.5cm}
\caption
{\it 
The $\as$ values determined in each $\langle\etjet\rangle$ value from
the analysis of the measured $\set$ cross section based on the $\kt$
jet algorithm (open circles). The error bars represent the
uncorrelated experimental uncertainties; the shaded area represents
the correlated experimental uncertainties and the hatched area
represents the correlated experimental and theoretical uncertainties
added in quadrature. The solid line indicates the
renormalisation-group prediction at two loops obtained from the
corresponding $\asz$ value determined in this analysis.
}
\label{fig17}
\vfill
\end{figure}

The extracted values of $\as$ as a function of $\etjet$ are shown in
Fig.~\ref{fig17} and Table~\ref{tab10} for the analysis based on the
$\kt$ algorithm. The data demonstrate the running of $\as$ over a
large range in $\etjet$ from a single experiment. The predicted
running of the strong
coupling~\cite{prl:30:1343,*prl:30:1346,*pr:d8:3633,*prep:14:129} 
calculated at two loops is in good agreement with the data.
The values of $\as$ as functions of $\etjet$ determined
from the anti-$\kt$ and SIScone measurements are consistent with those
obtained from the $\kt$ analysis and have a similar precision.

\section{Summary and conclusions}
Measurements of differential cross sections for inclusive-jet
photoproduction at a centre-of-mass energy of 318~GeV using an
integrated luminosity of 300~\pb1\ collected by the ZEUS detector
have been presented. The cross sections refer to jets of hadrons of
$\etjet>17$~GeV and $-1<\etajet<2.5$ identified in the laboratory
frame with the $\kt$, anti-$\kt$ or SIScone jet algorithms with
jet radius $R=1$. The cross sections are given in the kinematic region
of $\q2<1$~\g2\ and \wrn.

Measurements of single-differential cross sections were presented as
functions of $\etjet$ and $\etajet$. The NLO QCD calculations provide
a good description of the measured cross sections, except at high
$\etajet$. Non-perturbative effects not related to hadronisation and
the influence of the photon PDFs were found to be most significant in
this region.

Cross-section measurements were performed as functions
of $\etjet$ in different regions of $\etajet$. The data are well
described by the NLO QCD predictions, except for jets with low
$\etjet$ and high $\etajet$. These cross sections are sensitive to the
parton densities in the proton and the photon in regions of phase
space where the theoretical uncertainties are small. The precision
measurements presented here are therefore of particular relevance for
improving the determination of the PDFs in future QCD fits.

A detailed comparison between the measurements for the three jet
algorithms was performed. The measured cross sections for the three
jet algorithms have a similar shape, normalisation and precision. The
NLO QCD calculations of inclusive-jet cross sections and their
uncertainties for the different jet algorithms were also compared: the
predictions describe the data well, except at high $\etajet$; the
calculations based on the SIScone algorithm are somewhat less precise
than those based on the $\kt$ or anti-$\kt$ due to a larger
contribution from terms beyond NLO. The ratios of the cross sections
based on the different jet algorithms were also presented. The
measured ratios are well reproduced by the $\oass$ predictions,
demonstrating the ability of the pQCD calculations including up to
three partons in the final state to account adequately for the details
of the differences between the SIScone and the $\kt$ or anti-$\kt$ jet
algorithms.

The measured cross sections were used to determine values of
$\asz$. QCD fits to the cross-section $\set$ for $21<\etjet<71$~GeV
based on the $\kt$ jet algorithm yielded

\begin{eqnarray}
\asz &=& 0.1206\ ^{+0.0023}_{-0.0022}\ {\rm (exp.)}\ ^{+0.0042}_{-0.0035}\ {\rm (th.)}.\nonumber
\end{eqnarray}

This value is in good agreement with the world and HERA averages. The
extracted values of $\as$ as a function of $\etjet$ are in good
agreement with the predicted running of the strong coupling over a
large range in $\etjet$.

\vspace{0.5cm}
\noindent {\Large\bf Acknowledgements}
\vspace{0.3cm}

We thank the DESY Directorate for their strong support and
encouragement. The remarkable achievements of the HERA machine group
were essential for the successful completion of this work and are
greatly appreciated. We are grateful for the support of the DESY
computing and network services. The design, construction, installation
and running of the ZEUS detector were made possible owing to the
ingenuity and effort of many people who are not listed as authors.

\providecommand{\etal}{et al.\xspace}
\providecommand{\coll}{Coll.\xspace}
\catcode`\@=11
\def\@bibitem#1{%
\ifmc@bstsupport
  \mc@iftail{#1}%
    {;\newline\ignorespaces}%
    {\ifmc@first\else.\fi\orig@bibitem{#1}}
  \mc@firstfalse
\else
  \mc@iftail{#1}%
    {\ignorespaces}%
    {\orig@bibitem{#1}}%
\fi}%
\catcode`\@=12

\clearpage
\newpage
\begin{table}
\begin{center}
    \begin{tabular}{|l||r|r|r|}
\hline
  
& $\kt$
& anti-$\kt$
& SIScone\\ \hline \hline

events           & 483328 & 444295  & 438906 \\ \hline
jets             & 613165 & 572865  & 566000 \\ \hline
one jet          & 355691 & 317509  & 313519 \\ \hline
two jets         & 125468 & 125016  & 123700 \\ \hline
three jets       &   2138 &   1756  &   1667 \\ \hline
four jets        &     31 &     14  &     20 \\ \hline
    \end{tabular}
 \caption{\it
Number of events and jets selected in data with $\etjet>17$~GeV
and $-1<\etajet<2.5$ in the kinematic region of $\q2<1$~\gev$^2$ and
\wrn\ for the $\kt$, anti-$\kt$ and SIScone jet algorithms. The
number of events with one, two, three and four jets are also
listed.
}
 \label{tab2}
\end{center}
\end{table}

\clearpage
\newpage
\begin{table}
\begin{center}
    \begin{tabular}{|c|c||cccc||c|}
\hline
  $\etjet$ bin
& $\langle\etjet\rangle$
& $d\sigma/d\etjet$
& $\delta_{\rm stat}$
& $\delta_{\rm syst}$
& $\delta_{\rm ES}$
& $C_{\rm had}$\\
  (GeV)
& (GeV)
& (pb/GeV)
&
&
&
& \\
\hline\hline
$17-21$ & $18.7$ & $295.84  $ & $\pm 0.52  $ & $_{-6.36}^{+6.37}    $ & $_{-11.93}^{+11.35}  $ & $0.99$ \\
$21-25$ & $22.7$ & $ 95.86  $ & $\pm 0.28  $ & $_{-1.55}^{+1.56}    $ & $_{-4.50}^{+4.19}    $ & $0.99$ \\
$25-29$ & $26.7$ & $ 36.88  $ & $\pm 0.18  $ & $_{-0.52}^{+0.52}    $ & $_{-1.80}^{+1.76}    $ & $0.98$ \\
$29-35$ & $31.4$ & $ 13.606 $ & $\pm 0.090 $ & $_{-0.142}^{+0.150}  $ & $_{-0.772}^{+0.687}  $ & $0.98$ \\
$35-41$ & $37.5$ & $  4.492 $ & $\pm 0.051 $ & $_{-0.102}^{+0.104}  $ & $_{-0.251}^{+0.250}  $ & $0.98$ \\
$41-47$ & $43.6$ & $  1.677 $ & $\pm 0.032 $ & $_{-0.033}^{+0.033}  $ & $_{-0.102}^{+0.098}  $ & $0.98$ \\
$47-55$ & $50.2$ & $  0.589 $ & $\pm 0.017 $ & $_{-0.015}^{+0.015}  $ & $_{-0.044}^{+0.038}  $ & $0.98$ \\
$55-71$ & $60.3$ & $  0.1216$ & $\pm 0.0057$ & $_{-0.0028}^{+0.0033}$ & $_{-0.0090}^{+0.0086}$ & $0.97$ \\
$71-95$ & $77.2$ & $  0.0121$ & $\pm 0.0014$ & $_{-0.0010}^{+0.0010}$ & $_{-0.0011}^{+0.0012}$ & $1.00$ \\
\hline
\multicolumn{7}{c}{ } \\
\hline
  $\etajet$ bin
& $\langle\etajet\rangle$
& $d\sigma/d\etajet$
& $\delta_{\rm stat}$
& $\delta_{\rm syst}$
& $\delta_{\rm ES}$
& $C_{\rm had}$\\

& 
& (pb)
&
&
&
& \\
\hline\hline
$-0.75 - -0.50$ & $-0.59$ & $80.65$ & $\pm 0.92$ & $_{-4.89}^{+5.12}$ & $_{-7.78}^{+6.86}$ & $0.82$ \\
$-0.50 - -0.25$ & $-0.36$ & $221.4$ & $\pm 1.6 $ & $_{-8.5}^{+8.7}  $ & $_{-15.3}^{+14.0}$ & $0.93$ \\
$-0.25 - +0.00$ & $-0.11$ & $404.4$ & $\pm 2.2 $ & $_{-9.9}^{+10.0} $ & $_{-23.1}^{+21.7}$ & $0.97$ \\
$+0.00 - +0.25$ & $+0.13$ & $555.2$ & $\pm 2.6 $ & $_{-7.9}^{+8.1}  $ & $_{-28.2}^{+25.9}$ & $0.98$ \\
$+0.25 - +0.50$ & $+0.38$ & $685.8$ & $\pm 2.9 $ & $_{-8.0}^{+8.2}  $ & $_{-32.5}^{+30.3}$ & $0.98$ \\
$+0.50 - +0.75$ & $+0.63$ & $779.6$ & $\pm 3.1 $ & $_{-12.2}^{+12.3}$ & $_{-34.4}^{+32.5}$ & $0.98$ \\
$+0.75 - +1.00$ & $+0.87$ & $803.0$ & $\pm 3.2 $ & $_{-19.3}^{+19.3}$ & $_{-32.8}^{+31.4}$ & $0.99$ \\
$+1.00 - +1.25$ & $+1.12$ & $784.9$ & $\pm 3.3 $ & $_{-19.6}^{+19.6}$ & $_{-30.7}^{+29.1}$ & $0.99$ \\
$+1.25 - +1.50$ & $+1.38$ & $694.9$ & $\pm 2.9 $ & $_{-24.0}^{+24.0}$ & $_{-27.0}^{+25.1}$ & $1.00$ \\
$+1.50 - +1.75$ & $+1.63$ & $654.8$ & $\pm 2.8 $ & $_{-23.8}^{+23.8}$ & $_{-25.3}^{+24.6}$ & $1.00$ \\
$+1.75 - +2.00$ & $+1.88$ & $592.1$ & $\pm 2.6 $ & $_{-16.3}^{+16.4}$ & $_{-23.2}^{+22.5}$ & $1.00$ \\
$+2.00 - +2.25$ & $+2.13$ & $547.3$ & $\pm 2.6 $ & $_{-13.4}^{+13.4}$ & $_{-21.1}^{+20.4}$ & $1.00$ \\
$+2.25 - +2.50$ & $+2.38$ & $507.2$ & $\pm 2.4 $ & $_{-20.2}^{+20.2}$ & $_{-19.9}^{+18.7}$ & $1.00$ \\
\hline
    \end{tabular}
 \caption{
The measured differential cross-sections $\set$ and $\seta$ based on
the $\kt$ jet algorithm for inclusive-jet photoproduction with
$-1<\etajet<2.5$ and $\etjet>17$~GeV in the kinematic region given by
$\q2<1$~\gev$^2$ and \wrn. The statistical ($\delta_{\rm stat}$), 
uncorrelated systematic ($\delta_{\rm syst}$) and jet-energy scale 
($\delta_{\rm ES}$) uncertainties are shown separately. The corrections for 
hadronisation effects to be applied to the parton-level NLO QCD calculations 
($C_{\rm had}$) are shown in the last column.}
 \label{tab3}
\end{center}
\end{table}

\clearpage
\newpage
\begin{table}
\begin{center}
    \begin{tabular}{|c|c||cccc||c|}
\hline
  $\etajet$ bin
& $\langle\etajet\rangle$
& $d\sigma/d\etajet$
& $\delta_{\rm stat}$
& $\delta_{\rm syst}$
& $\delta_{\rm ES}$
& $C_{\rm had}$\\

& 
& (pb)
&
&
&
& \\
\hline\hline
$-0.75 - -0.50$ & $-0.56$ & $  4.12$ & $\pm 0.13$ & $_{-0.49}^{+0.55}$ & $_{-0.62}^{+0.56}$ & $0.66$ \\
$-0.50 - -0.25$ & $-0.34$ & $ 38.12$ & $\pm 0.60$ & $_{-2.45}^{+2.51}$ & $_{-3.88}^{+3.47}$ & $0.87$ \\
$-0.25 - +0.00$ & $-0.11$ & $108.8 $ & $\pm 1.1 $ & $_{-3.8}^{+3.8}  $ & $_{-7.9}^{+7.3}  $ & $0.94$ \\
$+0.00 - +0.25$ & $+0.13$ & $176.2 $ & $\pm 1.4 $ & $_{-3.6}^{+3.8}  $ & $_{-10.6}^{+9.7} $ & $0.97$ \\
$+0.25 - +0.50$ & $+0.38$ & $238.3 $ & $\pm 1.7 $ & $_{-3.2}^{+3.3}  $ & $_{-12.7}^{+11.8}$ & $0.97$ \\
$+0.50 - +0.75$ & $+0.63$ & $282.0 $ & $\pm 1.8 $ & $_{-4.4}^{+4.6}  $ & $_{-14.1}^{+13.0}$ & $0.97$ \\
$+0.75 - +1.00$ & $+0.87$ & $309.5 $ & $\pm 2.0 $ & $_{-5.9}^{+5.9}  $ & $_{-14.5}^{+14.0}$ & $0.98$ \\
$+1.00 - +1.25$ & $+1.12$ & $314.5 $ & $\pm 2.0 $ & $_{-5.1}^{+5.1}  $ & $_{-14.4}^{+13.3}$ & $0.99$ \\
$+1.25 - +1.50$ & $+1.38$ & $281.6 $ & $\pm 1.8 $ & $_{-5.9}^{+5.9}  $ & $_{-12.2}^{+11.5}$ & $1.00$ \\
$+1.50 - +1.75$ & $+1.63$ & $262.9 $ & $\pm 1.7 $ & $_{-6.5}^{+6.5}  $ & $_{-11.4}^{+11.2}$ & $1.00$ \\
$+1.75 - +2.00$ & $+1.88$ & $225.3 $ & $\pm 1.6 $ & $_{-4.3}^{+4.3}  $ & $_{-9.9}^{+9.3}  $ & $1.00$ \\
$+2.00 - +2.25$ & $+2.13$ & $200.9 $ & $\pm 1.5 $ & $_{-2.6}^{+2.6}  $ & $_{-9.1}^{+8.4}  $ & $1.01$ \\
$+2.25 - +2.50$ & $+2.38$ & $173.2 $ & $\pm 1.3 $ & $_{-5.3}^{+5.3}  $ & $_{-8.3}^{+7.7}  $ & $1.01$ \\
\hline
    \end{tabular}
 \caption{
The measured differential cross-sections $\seta$ based on
the $\kt$ jet algorithm for inclusive-jet photoproduction with
$\etjet>21$~GeV in the kinematic region given by $\q2<1$~\gev$^2$ and
\wrn. Other details as in the caption to Table~\ref{tab3}.
}
 \label{tab4}
\end{center}
\end{table}

\clearpage
\newpage
\begin{table}
\vspace*{-3.cm}
\begin{center}
    \begin{tabular}{|c|c||cccc||c|}
\hline
  $\etjet$ bin
& $\langle\etjet\rangle$
& $d\sigma/d\etjet$
& $\delta_{\rm stat}$
& $\delta_{\rm syst}$
& $\delta_{\rm ES}$
& $C_{\rm had}$\\
  (GeV)
& (GeV)
& (pb/GeV)
&
&
&
& \\
\hline
\multicolumn{7}{c}{$-1<\etajet<0$} \\
\hline
$17-21$ & $18.6$ & $35.99  $ & $\pm 0.17  $ & $_{-1.14}^{+1.15}    $ & $_{-2.21}^{+2.04}    $ & $0.93$ \\
$21-25$ & $22.6$ & $ 7.522 $ & $\pm 0.071 $ & $_{-0.331}^{+0.337}  $ & $_{-0.592}^{+0.543}  $ & $0.91$ \\
$25-29$ & $26.5$ & $ 1.695 $ & $\pm 0.032 $ & $_{-0.109}^{+0.115}  $ & $_{-0.167}^{+0.157}  $ & $0.89$ \\
$29-35$ & $30.9$ & $ 0.268 $ & $\pm 0.010 $ & $_{-0.024}^{+0.024}  $ & $_{-0.039}^{+0.032}  $ & $0.85$ \\
$35-41$ & $37.0$ & $ 0.0138$ & $\pm 0.0018$ & $_{-0.0021}^{+0.0026}$ & $_{-0.0031}^{+0.0023}$ & $0.80$ \\
\hline
\multicolumn{7}{c}{$0<\etajet<1$} \\
\hline
$17-21$ & $18.7$ & $113.61   $ & $\pm 0.31   $ & $_{-1.67}^{+1.68}      $ & $_{-4.72}^{+4.45}      $ & $0.99$ \\
$21-25$ & $22.7$ & $ 37.73   $ & $\pm 0.17   $ & $_{-0.53}^{+0.54}      $ & $_{-1.83}^{+1.70}      $ & $0.98$ \\
$25-29$ & $26.7$ & $ 14.27   $ & $\pm 0.11   $ & $_{-0.24}^{+0.24}      $ & $_{-0.74}^{+0.72}      $ & $0.98$ \\
$29-35$ & $31.4$ & $  5.034  $ & $\pm 0.052  $ & $_{-0.087}^{+0.093}    $ & $_{-0.300}^{+0.261}    $ & $0.97$ \\
$35-41$ & $37.5$ & $  1.490  $ & $\pm 0.027  $ & $_{-0.039}^{+0.040}    $ & $_{-0.098}^{+0.090}    $ & $0.96$ \\
$41-47$ & $43.4$ & $  0.485  $ & $\pm 0.015  $ & $_{-0.015}^{+0.016}    $ & $_{-0.032}^{+0.031}    $ & $0.96$ \\
$47-55$ & $50.1$ & $  0.1356 $ & $\pm 0.0068 $ & $_{-0.0056}^{+0.0059}  $ & $_{-0.0110}^{+0.0099}  $ & $0.95$ \\
$55-71$ & $59.8$ & $  0.0220 $ & $\pm 0.0017 $ & $_{-0.0013}^{+0.0014}  $ & $_{-0.0019}^{+0.0019}  $ & $0.93$ \\
$71-95$ & $76.5$ & $  0.00075$ & $\pm 0.00022$ & $_{-0.00008}^{+0.00011}$ & $_{-0.00015}^{+0.00010}$ & $0.93$ \\
\hline
\multicolumn{7}{c}{$1<\etajet<1.5$} \\
\hline
$17-21$ & $18.7$ & $55.26   $ & $\pm 0.21   $ & $_{-2.07}^{+2.07}      $ & $_{-1.92}^{+1.81}      $ & $1.00$ \\
$21-25$ & $22.7$ & $20.17   $ & $\pm 0.12   $ & $_{-0.43}^{+0.43}      $ & $_{-0.83}^{+0.77}      $ & $1.00$ \\
$25-29$ & $26.7$ & $ 8.723  $ & $\pm 0.082  $ & $_{-0.133}^{+0.133}    $ & $_{-0.387}^{+0.355}    $ & $0.99$ \\
$29-35$ & $31.5$ & $ 3.461  $ & $\pm 0.044  $ & $_{-0.060}^{+0.062}    $ & $_{-0.184}^{+0.173}    $ & $0.99$ \\
$35-41$ & $37.6$ & $ 1.209  $ & $\pm 0.026  $ & $_{-0.031}^{+0.031}    $ & $_{-0.058}^{+0.060}    $ & $0.97$ \\
$41-47$ & $43.6$ & $ 0.486  $ & $\pm 0.016  $ & $_{-0.013}^{+0.013}    $ & $_{-0.027}^{+0.026}    $ & $0.99$ \\
$47-55$ & $50.3$ & $ 0.1911 $ & $\pm 0.0089 $ & $_{-0.0073}^{+0.0074}  $ & $_{-0.0130}^{+0.0117}  $ & $0.97$ \\
$55-71$ & $60.5$ & $ 0.0454 $ & $\pm 0.0031 $ & $_{-0.0023}^{+0.0023}  $ & $_{-0.0031}^{+0.0028}  $ & $0.96$ \\
$71-95$ & $77.5$ & $ 0.00508$ & $\pm 0.00083$ & $_{-0.00083}^{+0.00083}$ & $_{-0.00038}^{+0.00041}$ & $0.99$ \\
\hline
    \end{tabular}
 \caption{
The measured differential cross-sections $\set$ based on the $\kt$ jet
algorithm for inclusive-jet photoproduction with $\etjet>17$~GeV in
different regions of $\etajet$ in the kinematic region given by
$\q2<1$~\gev$^2$ and \wrn. Other details as in the caption to 
Table~\ref{tab3}. 
}
\label{tab5}
\end{center}
\end{table}

\clearpage
\newpage
\begin{table}
\vspace*{-3.cm}
\begin{center}
    \begin{tabular}{|c|c||cccc||c|}
\hline
  $\etjet$ bin
& $\langle\etjet\rangle$
& $d\sigma/d\etjet$
& $\delta_{\rm stat}$
& $\delta_{\rm syst}$
& $\delta_{\rm ES}$
& $C_{\rm had}$\\
  (GeV)
& (GeV)
& (pb/GeV)
&
&
&
& \\
\hline
\multicolumn{7}{c}{$1.5<\etajet<2$} \\
\hline
$17-21$ & $18.7$ & $47.51   $ & $\pm 0.19   $ & $_{-1.87}^{+1.86}      $ & $_{-1.68}^{+1.64}      $ & $0.99$ \\
$21-25$ & $22.7$ & $16.70   $ & $\pm 0.11   $ & $_{-0.48}^{+0.48}      $ & $_{-0.68}^{+0.66}      $ & $1.00$ \\
$25-29$ & $26.7$ & $ 6.844  $ & $\pm 0.069  $ & $_{-0.134}^{+0.134}    $ & $_{-0.294}^{+0.285}    $ & $0.99$ \\
$29-35$ & $31.5$ & $ 2.818  $ & $\pm 0.037  $ & $_{-0.035}^{+0.035}    $ & $_{-0.138}^{+0.129}    $ & $0.99$ \\
$35-41$ & $37.5$ & $ 1.055  $ & $\pm 0.023  $ & $_{-0.020}^{+0.020}    $ & $_{-0.052}^{+0.056}    $ & $0.99$ \\
$41-47$ & $43.7$ & $ 0.432  $ & $\pm 0.015  $ & $_{-0.006}^{+0.006}    $ & $_{-0.026}^{+0.023}    $ & $0.99$ \\
$47-55$ & $50.2$ & $ 0.1703 $ & $\pm 0.0086 $ & $_{-0.0022}^{+0.0020}  $ & $_{-0.0119}^{+0.0099}  $ & $0.99$ \\
$55-71$ & $60.6$ & $ 0.0334 $ & $\pm 0.0027 $ & $_{-0.0011}^{+0.0012}  $ & $_{-0.0025}^{+0.0021}  $ & $1.00$ \\
$71-95$ & $78.1$ & $ 0.00393$ & $\pm 0.00081$ & $_{-0.00028}^{+0.00027}$ & $_{-0.00029}^{+0.00041}$ & $0.98$ \\
\hline
\multicolumn{7}{c}{$2<\etajet<2.5$} \\
\hline
$17-21$ & $18.6$ & $42.88   $ & $\pm 0.18   $ & $_{-1.63}^{+1.63}      $ & $_{-1.43}^{+1.40}      $ & $1.00$ \\
$21-25$ & $22.7$ & $13.479  $ & $\pm 0.097  $ & $_{-0.325}^{+0.325}    $ & $_{-0.591}^{+0.523}    $ & $1.01$ \\
$25-29$ & $26.7$ & $ 5.223  $ & $\pm 0.058  $ & $_{-0.134}^{+0.134}    $ & $_{-0.222}^{+0.240}    $ & $1.01$ \\
$29-35$ & $31.5$ & $ 1.977  $ & $\pm 0.029  $ & $_{-0.027}^{+0.027}    $ & $_{-0.114}^{+0.091}    $ & $1.00$ \\
$35-41$ & $37.6$ & $ 0.708  $ & $\pm 0.018  $ & $_{-0.015}^{+0.016}    $ & $_{-0.038}^{+0.039}    $ & $1.00$ \\
$41-47$ & $43.6$ & $ 0.268  $ & $\pm 0.011  $ & $_{-0.004}^{+0.003}    $ & $_{-0.017}^{+0.017}    $ & $1.00$ \\
$47-55$ & $50.2$ & $ 0.0928 $ & $\pm 0.0059 $ & $_{-0.0065}^{+0.0065}  $ & $_{-0.0082}^{+0.0061}  $ & $1.00$ \\
$55-71$ & $60.3$ & $ 0.0192 $ & $\pm 0.0019 $ & $_{-0.0007}^{+0.0007}  $ & $_{-0.0013}^{+0.0016}  $ & $0.97$ \\
$71-95$ & $74.9$ & $ 0.00238$ & $\pm 0.00065$ & $_{-0.00023}^{+0.00023}$ & $_{-0.00021}^{+0.00024}$ & $1.11$ \\
\hline
    \end{tabular}
 \caption{Continuation of Table~\ref{tab5}. 
}
\label{tab6}
\end{center}
\end{table}

\clearpage
\newpage
\begin{table}
\begin{center}
    \begin{tabular}{|c|c||cccc||c|}
\hline
  $\etjet$ bin
& $\langle\etjet\rangle$
& $d\sigma/d\etjet$
& $\delta_{\rm stat}$
& $\delta_{\rm syst}$
& $\delta_{\rm ES}$
& $C_{\rm had}$\\
  (GeV)
& (GeV)
& (pb/GeV)
&
&
&
& \\
\hline
\multicolumn{7}{c}{anti-$\kt$} \\
\hline
$17-21$ & $18.6$ & $276.84  $ & $\pm 0.50  $ & $_{-5.55}^{+5.55}    $ & $_{-11.75}^{+11.09}  $ & $0.93$ \\
$21-25$ & $22.7$ & $ 89.51  $ & $\pm 0.28  $ & $_{-1.35}^{+1.36}    $ & $_{-4.30}^{+4.03}    $ & $0.93$ \\
$25-29$ & $26.7$ & $ 34.54  $ & $\pm 0.18  $ & $_{-0.49}^{+0.49}    $ & $_{-1.76}^{+1.68}    $ & $0.94$ \\
$29-35$ & $31.4$ & $ 12.771 $ & $\pm 0.091 $ & $_{-0.170}^{+0.180}  $ & $_{-0.714}^{+0.640}  $ & $0.94$ \\
$35-41$ & $37.5$ & $  4.218 $ & $\pm 0.052 $ & $_{-0.087}^{+0.090}  $ & $_{-0.243}^{+0.235}  $ & $0.94$ \\
$41-47$ & $43.5$ & $  1.567 $ & $\pm 0.032 $ & $_{-0.028}^{+0.030}  $ & $_{-0.101}^{+0.093}  $ & $0.95$ \\
$47-55$ & $50.2$ & $  0.550 $ & $\pm 0.017 $ & $_{-0.018}^{+0.018}  $ & $_{-0.039}^{+0.037}  $ & $0.95$ \\
$55-71$ & $60.3$ & $  0.1139$ & $\pm 0.0058$ & $_{-0.0034}^{+0.0039}$ & $_{-0.0081}^{+0.0074}$ & $0.93$ \\
$71-95$ & $77.5$ & $  0.0105$ & $\pm 0.0014$ & $_{-0.0007}^{+0.0006}$ & $_{-0.0010}^{+0.0009}$ & $0.96$ \\
\hline
\multicolumn{7}{c}{SIScone} \\
\hline
$17-21$ & $18.7$ & $278.01  $ & $\pm 0.51  $ & $_{-4.30}^{+4.31}    $ & $_{-11.52}^{+10.89}  $ & $0.80$ \\
$21-25$ & $22.7$ & $ 90.82  $ & $\pm 0.28  $ & $_{-1.21}^{+1.22}    $ & $_{-4.27}^{+3.96}    $ & $0.81$ \\
$25-29$ & $26.7$ & $ 35.27  $ & $\pm 0.18  $ & $_{-0.48}^{+0.49}    $ & $_{-1.75}^{+1.65}    $ & $0.82$ \\
$29-35$ & $31.4$ & $ 13.059 $ & $\pm 0.090 $ & $_{-0.179}^{+0.185}  $ & $_{-0.695}^{+0.651}  $ & $0.84$ \\
$35-41$ & $37.5$ & $  4.330 $ & $\pm 0.051 $ & $_{-0.107}^{+0.108}  $ & $_{-0.252}^{+0.233}  $ & $0.85$ \\
$41-47$ & $43.6$ & $  1.639 $ & $\pm 0.031 $ & $_{-0.029}^{+0.031}  $ & $_{-0.100}^{+0.090}  $ & $0.86$ \\
$47-55$ & $50.3$ & $  0.565 $ & $\pm 0.016 $ & $_{-0.020}^{+0.020}  $ & $_{-0.039}^{+0.034}  $ & $0.87$ \\
$55-71$ & $60.5$ & $  0.1199$ & $\pm 0.0055$ & $_{-0.0042}^{+0.0045}$ & $_{-0.0085}^{+0.0085}$ & $0.86$ \\
$71-95$ & $77.7$ & $  0.0108$ & $\pm 0.0013$ & $_{-0.0008}^{+0.0008}$ & $_{-0.0013}^{+0.0012}$ & $0.90$ \\
\hline
    \end{tabular}
 \caption{
The measured differential cross-sections $\set$
based on different jet algorithms for inclusive-jet photoproduction
with $\etar$ in the kinematic region given
by $\q2<1$~\gev$^2$ and \wrn.
Other details as in the caption to Table~\ref{tab3}.}
 \label{tab7}
\end{center}
\end{table}

\clearpage
\newpage
\begin{table}
\vspace*{-3.cm}
\begin{center}
    \begin{tabular}{|c|c||cccc||c|}
\hline
  $\etajet$ bin
& $\langle\etajet\rangle$
& $d\sigma/d\etajet$
& $\delta_{\rm stat}$
& $\delta_{\rm syst}$
& $\delta_{\rm ES}$
& $C_{\rm had}$\\

& 
& (pb)
&
&
&
& \\
\hline
\multicolumn{7}{c}{anti-$\kt$} \\
\hline
$-0.75 - -0.50$ & $-0.58$ & $65.63$ & $\pm 0.84$ & $_{-3.88}^{+4.16}$ & $_{-6.51}^{+5.83}$ & $0.74$ \\
$-0.50 - -0.25$ & $-0.35$ & $200.3$ & $\pm 1.6$ & $_{-7.9}^{+8.1}  $ & $_{-14.1}^{+13.3}$ & $0.87$ \\
$-0.25 - +0.00$ & $-0.12$ & $369.5$ & $\pm 2.1 $ & $_{-9.9}^{+10.1} $ & $_{-21.4}^{+20.3}$ & $0.92$ \\
$+0.00 - +0.25$ & $+0.13$ & $518.6$ & $\pm 2.5$ & $_{-7.3}^{+7.5}  $ & $_{-26.3}^{+25.5}$ & $0.93$ \\
$+0.25 - +0.50$ & $+0.38$ & $634.6$ & $\pm 2.8 $ & $_{-6.6}^{+6.9}  $ & $_{-30.3}^{+28.3}$ & $0.93$ \\
$+0.50 - +0.75$ & $+0.63$ & $722.2$ & $\pm 3.0 $ & $_{-9.5}^{+9.6}$ & $_{-32.7}^{+30.8}$ & $0.93$ \\
$+0.75 - +1.00$ & $+0.87$ & $748.2$ & $\pm 3.1 $ & $_{-15.5}^{+15.5}$ & $_{-32.1}^{+30.0}$ & $0.94$ \\
$+1.00 - +1.25$ & $+1.12$ & $732.3$ & $\pm 3.2 $ & $_{-18.3}^{+18.3}$ & $_{-29.8}^{+28.4}$ & $0.95$ \\
$+1.25 - +1.50$ & $+1.38$ & $649.7$ & $\pm 2.8 $ & $_{-19.1}^{+19.1}$ & $_{-25.8}^{+24.0}$ & $0.94$ \\
$+1.50 - +1.75$ & $+1.63$ & $610.9$ & $\pm 2.7 $ & $_{-20.2}^{+20.1}$ & $_{-25.0}^{+23.4}$ & $0.94$ \\
$+1.75 - +2.00$ & $+1.87$ & $569.5$ & $\pm 2.6 $ & $_{-16.2}^{+16.2}$ & $_{-24.1}^{+22.4}$ & $0.94$ \\
$+2.00 - +2.25$ & $+2.13$ & $536.2$ & $\pm 2.6 $ & $_{-16.1}^{+16.1}$ & $_{-22.8}^{+21.3}$ & $0.94$ \\
$+2.25 - +2.50$ & $+2.38$ & $488.6$ & $\pm 2.5 $ & $_{-18.5}^{+18.5}$ & $_{-21.2}^{+18.7}$ & $0.94$ \\
\hline
\multicolumn{7}{c}{SIScone} \\
\hline
$-0.75 - -0.50$ & $-0.58$ & $69.65$ & $\pm 0.87$ & $_{-4.36}^{+4.67}$ & $_{-7.06}^{+6.21}$ & $0.66$ \\
$-0.50 - -0.25$ & $-0.35$ & $205.3$ & $\pm 1.6 $ & $_{-8.7}^{+8.7}  $ & $_{-14.7}^{+13.5}$ & $0.77$ \\
$-0.25 - +0.00$ & $-0.12$ & $379.3$ & $\pm 2.2 $ & $_{-10.9}^{+11.0} $ & $_{-22.1}^{+20.4}$ & $0.81$ \\
$+0.00 - +0.25$ & $+0.13$ & $527.7$ & $\pm 2.6 $ & $_{-8.0}^{+8.3}  $ & $_{-27.1}^{+25.6}$ & $0.82$ \\
$+0.25 - +0.50$ & $+0.38$ & $645.1$ & $\pm 2.9 $ & $_{-6.7}^{+6.8}  $ & $_{-30.6}^{+28.1}$ & $0.82$ \\
$+0.50 - +0.75$ & $+0.63$ & $731.6$ & $\pm 3.1 $ & $_{-9.0}^{+9.1}$ & $_{-32.5}^{+30.1}$ & $0.82$ \\
$+0.75 - +1.00$ & $+0.87$ & $756.8$ & $\pm 3.2 $ & $_{-13.8}^{+13.8}$ & $_{-31.8}^{+29.9}$ & $0.83$ \\
$+1.00 - +1.25$ & $+1.12$ & $735.0$ & $\pm 3.3 $ & $_{-14.6}^{+14.7}$ & $_{-29.3}^{+27.7}$ & $0.83$ \\
$+1.25 - +1.50$ & $+1.38$ & $650.0$ & $\pm 2.9 $ & $_{-15.3}^{+15.2}$ & $_{-24.3}^{+23.3}$ & $0.82$ \\
$+1.50 - +1.75$ & $+1.63$ & $598.7$ & $\pm 2.7 $ & $_{-14.3}^{+14.3}$ & $_{-23.1}^{+22.3}$ & $0.81$ \\
$+1.75 - +2.00$ & $+1.87$ & $549.0$ & $\pm 2.6 $ & $_{-10.9}^{+10.8}$ & $_{-22.2}^{+20.9}$ & $0.81$ \\
$+2.00 - +2.25$ & $+2.13$ & $529.8$ & $\pm 2.5 $ & $_{-14.1}^{+14.1}$ & $_{-22.0}^{+20.5}$ & $0.80$ \\
$+2.25 - +2.50$ & $+2.38$ & $518.5$ & $\pm 2.5 $ & $_{-24.4}^{+24.3}$ & $_{-21.6}^{+20.1}$ & $0.79$ \\
\hline
    \end{tabular}
 \caption{
The measured differential cross-sections $\seta$
based on different jet algorithms for inclusive-jet photoproduction
with $\etjet>17$~GeV in the kinematic region given
by $\q2<1$~\gev$^2$ and \wrn.
Other details as in the caption to Table~\ref{tab3}.}
 \label{tab8}
\end{center}
\end{table}

\clearpage
\newpage
\begin{table}
\begin{center}
    \begin{tabular}{|l||c|c|c|}
\hline
  
& $\kt$
& anti-$\kt$
& SIScone\\ \hline
\multicolumn{4}{|c|}{experimental uncertainties} \\ \hline
jet energy scale & $_{-1.7}^{+1.8}\%$ & $_{-1.8}^{+1.8}\%$ & $_{-1.6}^{+1.7}\%$ \\ \hline
luminosity       & $\pm 0.6\%$        & $\pm 0.6\%$        & $\pm 0.6\%$        \\ \hline
uncorrelated     & $_{-0.4}^{+0.3}\%$ & $_{-0.4}^{+0.3}\%$ & $_{-0.4}^{+0.3}\%$ \\ \hline
statistical      & $\pm 0.2\%$        & $\pm 0.2\%$        & $\pm 0.2\%$        \\ \hline
\multicolumn{4}{|c|}{theoretical uncertainties} \\ \hline
terms beyond NLO & $_{-2.5}^{+2.4}\%$ & $_{-2.4}^{+2.3}\%$ & $_{-3.3}^{+3.2}\%$ \\ \hline
photon PDFs      & $_{-0.9}^{+2.3}\%$ & $_{-0.9}^{+2.2}\%$ & $_{-0.9}^{+1.9}\%$ \\ \hline
proton PDFs      & $\pm 1.0\%$        & $\pm 1.0\%$        & $\pm 1.0\%$        \\ \hline
hadronisation    & $\pm 0.4\%$        & $\pm 0.4\%$        & $\pm 0.2\%$        \\ \hline
\multicolumn{4}{|c|}{total uncertainty} \\ \hline
                 & $_{-3.4}^{+4.0}\%$ & $_{-3.4}^{+3.9}\%$ & $_{-4.0}^{+4.2}\%$ \\ \hline
    \end{tabular}
 \caption{\it
Experimental, theoretical and total uncertainties in the determination of $\asz$ from the 
$\kt$, anti-$\kt$ and SIScone analyses.
}
 \label{tab9}
\end{center}
\end{table}

\begin{table}
\begin{center}
    \begin{tabular}{|c||cccc|}
\hline
  $\langle\etjet\rangle$
& $\as$
& $\delta_{\rm uncorr}$
& $\delta_{\rm corr}$
& $\delta_{\rm th}$\\
  (GeV)
&
&
&
& \\
\hline
$22.7$ & $0.1561$ & $\pm 0.0011$ & $_{-0.0048}^{+0.0035}$ & $_{-0.0089}^{+0.0106}$ \\
$26.7$ & $0.1493$ & $\pm 0.0007$ & $_{-0.0034}^{+0.0033}$ & $_{-0.0070}^{+0.0083}$ \\
$31.4$ & $0.1443$ & $\pm 0.0005$ & $_{-0.0030}^{+0.0035}$ & $_{-0.0059}^{+0.0069}$ \\
$37.5$ & $0.1396$ & $\pm 0.0007$ & $_{-0.0031}^{+0.0032}$ & $_{-0.0051}^{+0.0057}$ \\
$43.6$ & $0.1359$ & $\pm 0.0011$ & $_{-0.0030}^{+0.0032}$ & $_{-0.0047}^{+0.0051}$ \\
$50.2$ & $0.1328$ & $\pm 0.0014$ & $_{-0.0034}^{+0.0037}$ & $_{-0.0045}^{+0.0047}$ \\
$60.3$ & $0.1283$ & $\pm 0.0024$ & $_{-0.0036}^{+0.0040}$ & $_{-0.0041}^{+0.0041}$ \\
\hline
    \end{tabular}
 \caption{\it
The $\as$ values determined in each $\langle\etjet\rangle$ value from the analysis of the
measured $\set$ cross section based on the $\kt$ jet algorithm.
The uncorrelated ($\delta_{\rm uncorr}$) and correlated ($\delta_{\rm corr}$) experimental 
and theoretical ($\delta_{\rm th}$) uncertainties are listed separately.
}
 \label{tab10}
\end{center}
\end{table}

\newpage
\clearpage
\newpage
\clearpage

\newpage
\clearpage

\newpage
\clearpage

\newpage
\clearpage

\newpage
\clearpage

\newpage
\clearpage

\newpage
\clearpage

\newpage
\clearpage

\newpage
\clearpage

\newpage
\clearpage

\newpage
\clearpage

\newpage
\clearpage

\newpage
\clearpage

\newpage
\clearpage

\newpage
\clearpage

\newpage
\clearpage

\end{document}

%% file: zeus_def.tex
\chardef\usc=95
\chardef\til=126
\catcode`\@=11 
\DeclareRobustCommand\xdotspace{\futurelet\@let@token\@xdotspace}
\def\@xdotspace{%
  \ifx\@let@token.\else
  \ifx\@let@token\bgroup.\else
  \ifx\@let@token\egroup.\else
  \ifx\@let@token\/.\else
  \ifx\@let@token\ .\else
  \ifx\@let@token~.\else
  \ifx\@let@token!.\else
  \ifx\@let@token,.\else
  \ifx\@let@token:.\else
  \ifx\@let@token;.\else
  \ifx\@let@token?.\else
  \ifx\@let@token/.\else
  \ifx\@let@token'.\else
  \ifx\@let@token).\else
  \ifx\@let@token-.\else
  \ifx\@let@token\@xobeysp.\else
  \ifx\@let@token\space.\else
  \ifx\@let@token\@sptoken.\else
   .\space
   \fi\fi\fi\fi\fi\fi\fi\fi\fi\fi\fi\fi\fi\fi\fi\fi\fi\fi}
\catcode`\@=12 

\newcommand{\stru}[2]{%
   \relax\ifmmode\hbox{\vrule height#1 depth#2 width0pt}%
   \else\vrule height#1 depth#2 width0pt\fi}

\newcommand{\Ronum}[1]{\uppercase\expandafter{\romannumeral#1}}
\newcommand{\ronum}[1]{\expandafter{\romannumeral#1}}
\DeclareRobustCommand{\LaTeXZ}{%
  \LaTeX\kern-.05em4\kern-.1em
  {\raisebox{-0.2ex}{$\scriptstyle\text{ZEUS}$}}\xspace}

\DeclareMathAlphabet{\mathbf}{OT1}{cmr}{bx}{sl}
\newcommand{\eVdist}{\kern-0.06667em}

\newcommand{\gev}{{\,\text{Ge}\eVdist\text{V\/}}}

\newcommand{\pb}{\,\text{pb}}
\newcommand{\fb}{\,\text{fb}}

\newcommand{\Tesla}{\,\text{T}}

\newcommand{\slashfrac}[2]{%
  \raisebox{0.5ex}{\ensuremath #1}\kern-0.12em/\kern-0.08em
  \raisebox{-.8ex}{\ensuremath #2}}

\newcommand{\sqr}[3]{%
    {\vcenter{\hrule height.#3ex\hbox{\vrule width.#2ex height#1ex
     \kern#1ex\vrule width.#3ex}\hrule height.#2ex}}}

\catcode`\@=11 
\newcommand{\parenbar}{\mathpalette\p@renb@r}
\def\p@renb@r#1#2{\vbox{%
  \ifx#1\scriptscriptstyle \dimen@.7em\dimen@ii.2em\else
  \ifx#1\scriptstyle \dimen@.8em\dimen@ii.25em\else
  \dimen@1em\dimen@ii.4em\fi\fi \offinterlineskip
  \ialign{\hfill##\hfill\cr
    \vbox{\hrule width\dimen@ii}\cr
    \noalign{\vskip-.3ex}%
    \hbox to\dimen@{$\mathchar300\hfil\mathchar301$}\cr
    \noalign{\vskip-.3ex}%
    $#1#2$\cr}}}
\catcode`\@=12 

\newcommand{\IP}{{\rm I$\kern-0.01667em$P}\xspace}

\newcommand{\F}{{\cal F}}

\mathchardef\qsm=63
\mathchardef\pls=43
\mathchardef\mns=512
\mathchardef\plm=518
\mathchardef\eql=61
\mathchardef\smallleft=300
\mathchardef\smallright=301
\mathchardef\les=316
\mathchardef\gre=318
\mathchardef\leq=532
\mathchardef\grq=533

\catcode`\@=11 
\newcounter{pict@width}
\newcounter{pict@height}
\newlength{\pict@scale}
\newcommand{\psfigadd}[4]{%
\setcounter{pict@width}{1*\ratio{#2+\pict@scale/2}{\pict@scale}}
\setcounter{pict@height}{1*\ratio{#3+\pict@scale/2}{\pict@scale}}
\setlength{\unitlength}{\pict@scale}
\hbox to #2{\hspace{-\fill}\begin{picture}(\thepict@width,\thepict@height)
\put(0,0){\psfig{figure=#1,width=#2,height=#3,clip=}}
\SetScale{0.283466457}
\SetWidth{1.763889}
{#4}
\end{picture}}
}
\newcounter{pict@widthfst}
\newcounter{pict@widthscd}
\newcounter{pict@widthtot}
\newcommand{\psfigaddtwo}[7]{%
\setcounter{pict@widthfst}{1*\ratio{#2+\pict@scale/2}{\pict@scale}}
\setcounter{pict@widthscd}{1*\ratio{#2+#4+\pict@scale/2}{\pict@scale}}
\setcounter{pict@widthtot}{1*\ratio{#2+#4+#6+\pict@scale/2}{\pict@scale}}
\setcounter{pict@height}{1*\ratio{#3+\pict@scale/2}{\pict@scale}}
\setlength{\unitlength}{\pict@scale}
\hbox{\hspace{-\fill}\begin{picture}(\thepict@widthtot,\thepict@height)
\put(0,0){\psfig{figure=#1,width=#2,height=#3,clip=}}
\put(\thepict@widthscd,0){\psfig{figure=#5,width=#6,height=#3,clip=}}
\SetScale{0.283466457}
\SetWidth{1.763889}
{#7}
\end{picture}}
}
\newcommand{\psfigror}[4]{%
\setcounter{pict@width}{1*\ratio{#2+\pict@scale/2}{\pict@scale}}
\setcounter{pict@height}{1*\ratio{#3+\pict@scale/2}{\pict@scale}}
\setlength{\unitlength}{\pict@scale}
\hbox{\begin{picture}(\thepict@width,\thepict@height)
\put(0,\thepict@height){\psfig{figure=#1,width=#3,height=#2,clip=,angle=270}}
\SetScale{0.283466457}
\SetWidth{1.763889}
{#4}
\end{picture}}
}
\newcommand{\psfigrol}[4]{%
\setcounter{pict@width}{1*\ratio{#2+\pict@scale/2}{\pict@scale}}
\setcounter{pict@height}{1*\ratio{#3+\pict@scale/2}{\pict@scale}}
\setlength{\unitlength}{\pict@scale}
\hbox{\begin{picture}(\thepict@width,\thepict@height)
\put(0,0){\psfig{figure=#1,width=#3,height=#2,clip=,angle=90}}
\SetScale{0.283466457}
\SetWidth{1.763889}
{#4}
\end{picture}}
}
\catcode`\@=12 
\newlength\listtextwidth

\catcode`\@=11 
\newlength{\@tabfninsert}
\newlength{\@tabfnwidth}
\newcommand{\tabfootnote}[2]{%
  \setlength{\@tabfninsert}{0.8em}
  \setlength{\@tabfnwidth}{\textwidth}
  \addtolength{\@tabfnwidth}{-\@tabfninsert}
  \addtolength{\@tabfnwidth}{-0.4em}
  \noindent\makebox[\@tabfninsert][r]{\footnotesize$^{#1}$\hfil}\hfill%
  \parbox[t]{\@tabfnwidth}{\footnotesize #2\hfill}}
\catcode`\@=12 

%% file: definitions.tex
\def\JHEP{JHEP}

\def\etjet{E_T^{\rm jet}}
\def\etajet{\eta^{\rm jet}}

\def\etcal{E_{T,{\rm cal}}^{\rm jet}}
\def\etacal{\eta_{\rm cal}^{\rm jet}}
\def\phical{\phi_{\rm cal}^{\rm jet}}

\def\etaphi{\eta-\phi}

\def\etar{-1<\etajet<2.5}

\def\etacr{-1<\etacal<2.5}

\def\qg2{$\q2>125$~\g2}

\def\seta{d\sigma/d\etajet}

\def\set{d\sigma/d\etjet}

\def\sq2{d\sigma/d\q2}

\def\q2{Q^2}

\def\pb1{pb$^{-1}$}
\def\fb1{fb$^{-1}$}
\def\gp{\gamma p}

\def\g2{GeV$^2$}

\def\F2g{F_2^{\gamma}}
\def\f2gv{F_2^{\gamma^*}}

\def\rr1{R=1}
\def\r7{R=0.7}
\def\R71{R=0.7\ {\rm and}\ 1}

\def\m3j{M^{\rm 3j}}

\def\kt{k_T}

\def\lq2{\log_{10}(\q2)}

\def\ele{e^+e^-}
\def\pp{p\bar p}

\def\colab#1{#1 Coll.}

\def\ptmis{p_T^{\rm miss}}

\def\z0{Z^0}
\def\mz{M_Z}

\def\as{\alpha_s}
\def\oalphas2{{\cal O}(\alpha\as^2)}

\def\oass{{\cal O}(\as^2)}
\def\oasss{{\cal O}(\as^3)}
\def\asz{\as(\mz)}

\def\p2{P^2}

\def\mr2{\mu_R^2}
\def\mf2{\mu_F^2}

\def\etal{et al.}

\def\eprn{ep\rightarrow e+{\rm jet}+{\rm X}}

\def\wgp{W_{\gp}}

\def\wrn{$142<\wgp<293$ GeV}

\def\bet0#1#2#3#4#5#6{\beta_0 = #1\pm #2\ {\rm (stat.)}\ ^{+#4}_{-#3}\ {\rm (exp.)}\ ^{+#6}_{-#5}\ {\rm (th.)}}

\def\a34{\alpha_{23}}

\def\k0s{K_S^0}

%% file: DESY-12-045.bbl
\begin{mcbibliography}{10}

\bibitem{epj:c19:289}
\colab{H1}, C. Adloff \etal,
\newblock Eur.\ Phys.\ J.{} C~19~(2001)~289\relax
\relax
\bibitem{pl:b547:164}
\colab{ZEUS}, S. Chekanov \etal,
\newblock Phys.\ Lett.{} B~547~(2002)~164\relax
\relax
\bibitem{pl:b551:226}
\colab{ZEUS}, S. Chekanov \etal,
\newblock Phys.\ Lett.{} B~551~(2003)~226\relax
\relax
\bibitem{np:b765:1}
\colab{ZEUS}, S.~Chekanov \etal,
\newblock Nucl.\ Phys.{} B~765~(2007)~1\relax
\relax
\bibitem{pl:b649:12}
\colab{ZEUS}, S. Chekanov \etal,
\newblock Phys.\ Lett.{} B~649~(2007)~12\relax
\relax
\bibitem{epj:c65:363}
\colab{H1}, F.D. Aaron \etal,
\newblock Eur.\ Phys.\ J.{} C~65~(2010)~363\relax
\relax
\bibitem{epj:c67:1}
\colab{H1}, F.D. Aaron \etal,
\newblock Eur.\ Phys.\ J.{} C~67~(2010)~1\relax
\relax
\bibitem{pl:b507:70}
\colab{ZEUS}, J. Breitweg \etal,
\newblock Phys.\ Lett.{} B~507~(2001)~70\relax
\relax
\bibitem{epj:c23:13}
\colab{ZEUS}, S. Chekanov \etal,
\newblock Eur.\ Phys.\ J.{} C~23~(2002)~13\relax
\relax
\bibitem{pl:b515:17}
\colab{H1}, C. Adloff \etal,
\newblock Phys.\ Lett.{} B~515~(2001)~17\relax
\relax
\bibitem{epj:c44:183}
\colab{ZEUS}, S. Chekanov \etal,
\newblock Eur.\ Phys.\ J.{} C~44~(2005)~183\relax
\relax
\bibitem{pr:d85:052008}
\colab{ZEUS}, H. Abramowicz \etal,
\newblock Phys.\ Rev.{} D~85~(2012)~052008\relax
\relax
\bibitem{pl:b560:7}
\colab{ZEUS}, S. Chekanov \etal,
\newblock Phys.\ Lett.{} B~560~(2003)~7\relax
\relax
\bibitem{epj:c29:497}
\colab{H1}, C. Adloff \etal,
\newblock Eur.\ Phys.\ J.{} C~29~(2003)~497\relax
\relax
\bibitem{epj:c11:35}
\colab{ZEUS}, J.~Breitweg \etal,
\newblock Eur.\ Phys.\ J.{} C~11~(1999)~35\relax
\relax
\bibitem{epj:c23:615}
\colab{ZEUS}, S. Chekanov \etal,
\newblock Eur.\ Phys.\ J.{} C~23~(2002)~615\relax
\relax
\bibitem{pl:b531:9}
\colab{ZEUS}, S. Chekanov \etal,
\newblock Phys.\ Lett.{} B~531~(2002)~9\relax
\relax
\bibitem{epj:c25:13}
\colab{H1}, C.~Adloff \etal,
\newblock Eur.\ Phys.\ J.{} C~25~(2002)~13\relax
\relax
\bibitem{pl:b639:21}
\colab{H1}, A. Aktas \etal,
\newblock Phys.\ Lett.{} B~639~(2006)~21\relax
\relax
\bibitem{pr:d76:072011}
\colab{ZEUS}, S. Chekanov \etal,
\newblock Phys.\ Rev.{} D~76~(2007)~072011\relax
\relax
\bibitem{pl:b443:394}
\colab{ZEUS}, J.~Breitweg \etal,
\newblock Phys.\ Lett.{} B~443~(1998)~394\relax
\relax
\bibitem{np:b792:1}
\colab{ZEUS}, S. Chekanov \etal,
\newblock Nucl.\ Phys.{} B~792~(2008)~1\relax
\relax
\bibitem{np:b545:3}
\colab{H1}, C. Adloff \etal,
\newblock Nucl.\ Phys.{} B~545~(1999)~3\relax
\relax
\bibitem{pl:b558:41}
\colab{ZEUS}, S. Chekanov \etal,
\newblock Phys.\ Lett.{} B~558~(2003)~41\relax
\relax
\bibitem{np:b700:3}
\colab{ZEUS}, S. Chekanov \etal,
\newblock Nucl.\ Phys.{} B~700~(2004)~3\relax
\relax
\bibitem{epj:c31:149}
\colab{ZEUS}, S.~Chekanov \etal,
\newblock Eur.\ Phys.\ J.{} C~31~(2003)~149\relax
\relax
\bibitem{pr:d78:032004}
\colab{ZEUS}, S.~Chekanov \etal,
\newblock Phys.\ Rev.{} D~78~(2008)~032004\relax
\relax
\bibitem{np:b406:187}
S. Catani \etal,
\newblock Nucl.\ Phys.{} B~406~(1993)~187\relax
\relax
\bibitem{pr:d48:3160}
S.D. Ellis and D.E. Soper,
\newblock Phys.\ Rev.{} D~48~(1993)~3160\relax
\relax
\bibitem{epj:c42:1}
\colab{ZEUS}, S. Chekanov \etal,
\newblock Eur.\ Phys.\ J.{} C~42~(2005)~1\relax
\relax
\bibitem{jhep:04:063}
M. Cacciari, G.P. Salam and G. Soyez,
\newblock \JHEP{} 0804~(2008)~063\relax
\relax
\bibitem{jhep:05:086}
G.P. Salam and G. Soyez,
\newblock \JHEP{} 0705~(2007)~086\relax
\relax
\bibitem{pl:b691:127}
\colab{ZEUS}, H. Abramowicz \etal,
\newblock Phys.\ Lett.{} B~691~(2010)~127\relax
\relax
\bibitem{pl:b79:83}
C.H. Llewellyn Smith,
\newblock Phys.\ Lett.{} B~79~(1978)~83\relax
\relax
\bibitem{np:b166:413}
I. Kang and C.H. Llewellyn Smith,
\newblock Nucl.\ Phys.{} B~166~(1980)~413\relax
\relax
\bibitem{pr:d21:54}
J.F. Owens,
\newblock Phys.\ Rev.{} D~21~(1980)~54\relax
\relax
\bibitem{zfp:c6:241}
M. Fontannaz, A. Mantrach and D. Schiff,
\newblock Z.\ Phys.{} C~6~(1980)~241\relax
\relax
\bibitem{proc:hera:1987:331}
W.J. Stirling and Z. Kunszt,
\newblock {\em Proc. HERA Workshop}, R.D. Peccei~(ed.), Vol.~2, p.~331.
\newblock DESY, Hamburg, Germany (1987)\relax
\relax
\bibitem{prl:61:275}
M. Drees and F. Halzen,
\newblock Phys.\ Rev.\ Lett.{} 61~(1988)~275\relax
\relax
\bibitem{prl:61:682}
M. Drees and R.M. Godbole,
\newblock Phys.\ Rev.\ Lett.{} 61~(1988)~682\relax
\relax
\bibitem{pr:d39:169}
M. Drees and R.M. Godbole,
\newblock Phys.\ Rev.{} D~39~(1989)~169\relax
\relax
\bibitem{zfp:c42:657}
H. Baer, J. Ohnemus and J.F. Owens,
\newblock Z.\ Phys.{} C~42~(1989)~657\relax
\relax
\bibitem{pr:d40:2844}
H. Baer, J. Ohnemus and J.F. Owens,
\newblock Phys.\ Rev.{} D~40~(1989)~2844\relax
\relax
\bibitem{pl:b293:465}
\colab{ZEUS}, M.~Derrick \etal,
\newblock Phys.\ Lett.{} B~293~(1992)~465\relax
\relax
\bibitem{zeus:1993:bluebook}
\colab{ZEUS}, U.~Holm~(ed.),
\newblock {\em The {ZEUS} Detector}.
\newblock Status Report (unpublished), DESY (1993),
\newblock available on
  \texttt{http://www-zeus.desy.de/bluebook/bluebook.html}\relax
\relax
\bibitem{nim:a279:290}
N.~Harnew \etal,
\newblock Nucl.\ Instr.\ Meth.{} A~279~(1989)~290\relax
\relax
\bibitem{npps:b32:181}
B.~Foster \etal,
\newblock Nucl.\ Phys.\ Proc.\ Suppl.{} B~32~(1993)~181\relax
\relax
\bibitem{nim:a338:254}
B.~Foster \etal,
\newblock Nucl.\ Instr.\ Meth.{} A~338~(1994)~254\relax
\relax
\bibitem{nim:a581:656}
A. Polini \etal,
\newblock Nucl.\ Instr.\ Meth.{} A~581~(2007)~656\relax
\relax
\bibitem{nim:a309:77}
M.~Derrick \etal,
\newblock Nucl.\ Instr.\ Meth.{} A~309~(1991)~77\relax
\relax
\bibitem{nim:a309:101}
A.~Andresen \etal,
\newblock Nucl.\ Instr.\ Meth.{} A~309~(1991)~101\relax
\relax
\bibitem{nim:a321:356}
A.~Caldwell \etal,
\newblock Nucl.\ Instr.\ Meth.{} A~321~(1992)~356\relax
\relax
\bibitem{nim:a336:23}
A.~Bernstein \etal,
\newblock Nucl.\ Instr.\ Meth.{} A~336~(1993)~23\relax
\relax
\bibitem{desy-92-066}
J.~Andruszk\'ow \etal,
\newblock Preprint \mbox{DESY-92-066}, DESY, 1992\relax
\relax
\bibitem{zfp:c63:391}
\colab{ZEUS}, M.~Derrick \etal,
\newblock Z.\ Phys.{} C~63~(1994)~391\relax
\relax
\bibitem{acpp:b32:2025}
J.~Andruszk\'ow \etal,
\newblock Acta Phys.\ Pol.{} B~32~(2001)~2025\relax
\relax
\bibitem{nim:a565:572}
M.~Helbich \etal,
\newblock Nucl.\ Instr.\ Meth.{} A~565~(2006)~572\relax
\relax
\bibitem{proc:chep:1992:222}
W.H.~Smith, K.~Tokushuku and L.W.~Wiggers,
\newblock {\em Proc.\ Computing in High-Energy Physics (CHEP), Annecy, France,
  Sept.~1992}, C.~Verkerk and W.~Wojcik~(eds.), p.~222.
\newblock CERN, Geneva, Switzerland (1992).
\newblock Also in preprint \mbox{DESY 92-150B}\relax
\relax
\bibitem{proc:snowmass:1990:134}
J.E. Huth \etal,
\newblock {\em Research Directions for the Decade. Proc. of Summer Study on
  High Energy Physics, 1990}, E.L. Berger~(ed.), p.~134.
\newblock World Scientific (1992).
\newblock Also in preprint \mbox{FERMILAB-CONF-90-249-E}\relax
\relax
\bibitem{pl:b641:57}
M. Cacciari and G.P. Salam,
\newblock Phys.\ Lett.{} B~641~(2006)~57\relax
\relax
\bibitem{cpc:82:74}
T. Sj\"ostrand,
\newblock Comput.\ Phys.\ Comm.{} 82~(1994)~74\relax
\relax
\bibitem{cpc:67:465}
G. Marchesini \etal,
\newblock Comput.\ Phys.\ Comm.{} 67~(1992)~465\relax
\relax
\bibitem{jhep:0101:010}
G. Corcella \etal,
\newblock \JHEP{} 0101~(2001)~010\relax
\relax
\bibitem{prep:97:31}
B. Andersson \etal,
\newblock Phys.\ Rep.{} 97~(1983)~31\relax
\relax
\bibitem{cpc:39:347}
T. Sj\"ostrand,
\newblock Comput.\ Phys.\ Comm.{} 39~(1986)~347\relax
\relax
\bibitem{cpc:43:367}
T. Sj\"ostrand and M. Bengtsson,
\newblock Comput.\ Phys.\ Comm.{} 43~(1987)~367\relax
\relax
\bibitem{np:b238:492}
B.R. Webber,
\newblock Nucl.\ Phys.{} B~238~(1984)~492\relax
\relax
\bibitem{pr:d55:1280}
H.L. Lai \etal,
\newblock Phys.\ Rev.{} D~55~(1997)~1280\relax
\relax
\bibitem{pr:d45:3986}
M. Gl\"uck, E. Reya and A. Vogt,
\newblock Phys.\ Rev.{} D~45~(1992)~3986\relax
\relax
\bibitem{pr:d46:1973}
M. Gl\"uck, E. Reya and A. Vogt,
\newblock Phys.\ Rev.{} D~46~(1992)~1973\relax
\relax
\bibitem{pr:d36:2019}
T. Sj\"ostrand and M. van Zijl,
\newblock Phys.\ Rev.{} D~36~(1987)~2019\relax
\relax
\bibitem{tech:cern-dd-ee-84-1}
R.~Brun et al.,
\newblock {\em {\sc geant3}},
\newblock Technical Report CERN-DD/EE/84-1, CERN, 1987\relax
\relax
\bibitem{hep-ex-0206036}
M. Wing (on behalf of the \colab{ZEUS}), {\it Proc. of the $10th$ International
  Conference on Calorimetry in High Energy Physics},
\newblock Preprint \mbox{hep-ex/0206036}, 2002\relax
\relax
\bibitem{epj:c4:463}
A.D. Martin \etal,
\newblock Eur.\ Phys.\ J.{} C~4~(1998)~463\relax
\relax
\bibitem{zp:c64:621}
P. Aurenche, J.-P. Guillet and M. Fontannaz,
\newblock Z.\ Phys.{} C~64~(1994)~621\relax
\relax
\bibitem{epjdirectcbg:c1:1}
M. Klasen, T. Kleinwort and G. Kramer,
\newblock Eur.\ Phys.\ J.\ Direct{} C~1~(1998)~1\relax
\relax
\bibitem{kramer:1984:slicing}
G. Kramer,
\newblock {\em Theory of Jets in Electron-Positron Annihilation}.
\newblock Springer, Berlin, (1984)\relax
\relax
\bibitem{pr:d67:012007}
\colab{ZEUS}, S.~Chekanov \etal,
\newblock Phys.\ Rev.{} D~67~(2003)~012007\relax
\relax
\bibitem{hep-ph:0508110}
M.R. Whalley, D. Bourilkov and R.C. Group,
\newblock Preprint \mbox{hep-ph/0508110}, 2005\relax
\relax
\bibitem{jp:g26:r27}
S. Bethke,
\newblock J.\ Phys.{} G~26~(2000)~R27.
\newblock Updated in S. Bethke, Eur. Phys. J. {\bf C 64} (2009) 689\relax
\relax
\bibitem{epj:c44:395}
P. Aurenche, M. Fontannaz and J.Ph. Guillet,
\newblock Eur.\ Phys.\ J.{} C~44~(2005)~395\relax
\relax
\bibitem{pr:d70:093004}
F. Cornet, P. Jankowski and M. Krawczyk,
\newblock Phys.\ Rev.{} D~70~(2004)~093004\relax
\relax
\bibitem{epj:c63:189}
A.D. Martin \etal,
\newblock Eur.\ Phys.\ J.{} C~63~(2009)~189\relax
\relax
\bibitem{jhep:1001:109}
H1 and ZEUS Collaborations, F.D. Aaron \etal,
\newblock \JHEP{} 1001~(2010)~109\relax
\relax
\bibitem{jhep:0312:007}
R.W.L. Jones \etal,
\newblock \JHEP{} 0312~(2003)~007\relax
\relax
\bibitem{jhep:0207:012}
J. Pumplin \etal,
\newblock \JHEP{} 0207~(2002)~012\relax
\relax
\bibitem{jhep:0310:046}
D. Stump \etal,
\newblock \JHEP{} 0310~(2003)~046\relax
\relax
\bibitem{prl:88:042001}
\colab{CDF}, T. Affolder \etal,
\newblock Phys.\ Rev.\ Lett.{} 88~(2002)~042001\relax
\relax
\bibitem{pr:d80:111107}
\colab{D\O}, V.M. Abazov \etal,
\newblock Phys.\ Rev.{} D~80~(2009)~111107\relax
\relax
\bibitem{pl:b536:217}
\colab{L3}, P. Achard \etal,
\newblock Phys.\ Lett.{} B~536~(2002)~217\relax
\relax
\bibitem{epj:c38:413}
\colab{DELPHI}, J. Abdallah \etal,
\newblock Eur.\ Phys.\ J.{} C~38~(2005)~413\relax
\relax
\bibitem{epj:c71:1733}
\colab{OPAL}, G. Abbiendi \etal,
\newblock Eur.\ Phys.\ J.{} C~71~(2011)~1733\relax
\relax
\bibitem{proc:dis:2005:689}
C. Glasman,
\newblock {\em Proc. of the 13th International Workshop on Deep Inelastic
  Scattering}, S.R. Dasu and W.H. Smith~(eds.), p.~689.
\newblock Madison, USA (2005).
\newblock Also in preprint \mbox{hep-ex/0506035}\relax
\relax
\bibitem{eps:2007:022013}
C. Glasman, {\it Proc. of the HEP2007 International Europhysics Conference on
  High Energy Physics},
\newblock J. Phys. Conf. Ser.{} 110~(2008)~022013. Also in preprint
  arXiv:0709.4426\relax
\relax
\bibitem{prl:30:1343}
D.J. Gross and F. Wilczek,
\newblock Phys.\ Rev.\ Lett.{} 30~(1973)~1343\relax
\relax
\bibitem{prl:30:1346}
H.D. Politzer,
\newblock Phys.\ Rev.\ Lett.{} 30~(1973)~1346\relax
\relax
\bibitem{pr:d8:3633}
D.J. Gross and F. Wilczek,
\newblock Phys.\ Rev.{} D~8~(1973)~3633\relax
\relax
\bibitem{prep:14:129}
H.D. Politzer,
\newblock Phys.\ Rep.{} 14~(1974)~129\relax
\relax
\end{mcbibliography}
